\documentclass[preprint,12pt]{elsarticle}
\usepackage{amssymb}
\usepackage{amsmath}
\usepackage{epigraph}
\journal{Phys. Rep.}

\begin{document}

\begin{frontmatter}



\title{Temperature in and out of equilibrium: a  review of concepts, tools and attempts}


\author[cnr,roma1]{A. Puglisi}
\author[cnr,roma1]{A. Sarracino}
\author[roma1,cnr,lincei]{A. Vulpiani}

\address[cnr]{CNR-ISC, Roma}
\address[roma1]{Dipartimento di Fisica, Universit\`a Sapienza, P.le A. Moro 2, I-00185 Roma}
\address[lincei]{Centro Linceo Interdisciplinare  ``Beniamino Segre'', Accademia dei Lincei}

\begin{abstract}
We review the general aspects of the concept of temperature in
equilibrium and non-equilibrium statistical mechanics. Although
temperature is an old and well-established notion, it still presents
controversial facets. After a short historical survey of the key role
of temperature in thermodynamics and statistical mechanics, we tackle
a series of issues which have been recently reconsidered. In
particular, we discuss different definitions and their relevance
for energy fluctuations.  The interest in such a topic has been
triggered by the recent observation of negative temperatures in
condensed matter experiments. Moreover, the ability to manipulate
systems at the micro and nano-scale urges to understand and clarify
some aspects related to the statistical properties of small systems
(as the issue of temperature's ``fluctuations'').  We also discuss the
notion of temperature in a dynamical context, within the theory of
linear response for Hamiltonian systems at equilibrium and stochastic
models with detailed balance, and the generalised fluctuation-response
relations, which provide a hint for an extension of the definition of
temperature in far-from-equilibrium systems. To conclude we consider
non-Hamiltonian systems, such as granular materials, turbulence and
active matter, where a general theoretical framework is still lacking.
\end{abstract}


\end{frontmatter}


\tableofcontents

\newpage

\epigraph{The meaning of the world is the separation of wish and fact.}{Kurt G\"odel}

\section{Introduction}
\label{sec:intro}
\subsection{Introductory remarks and plan of the paper}

Temperature is surely one of the central concepts in thermodynamics
and statistical mechanics.  Since we study such a topic in elementary
courses, at first glance, a review on this argument can appear of mere
scholarly interest, without particular relevance for the present
research.

The recent developements in statistical mechanics of small systems, as
well as for non-equilibrium and/or non-Hamiltonian systems, are
becoming more and more important due to the theoretical and
technological challenges of micro, and nano physics.  In the treatment
of situations far from the ``traditional ones'', i.e. equilibrium
Hamiltonian systems with many degrees of freedom, the need to
reconsider in a careful way the basic aspects of the notion of
temperature appears in a natural way.  Such a renewed interest in the
foundations of statistical mechanics has shown that temperature is a
rather difficult and subtle issue, and a clear understanding of its
role in thermodynamics and statistical mechanics goes beyond the mere
academic level.

The paper is organised as follows.  In this Section we give a brief
historical overview on the birth of the concept of temperature,
focusing in particular on its connection with the fluctuations theory.  Section 2 is devoted to some general aspects of
temperature.  In particular we review the role of temperature for the
energy fluctuations and its connection with the ergodic hypothesis in
the buiding of consistent statistical mechanics theory.  In addition
we discuss the link between entropy and large deviations, and how
temperature can be computed as time average of a suitable
observable. Such an approach allows one to determine the temperature
for any Hamiltonian system.  In Section 3 we treat several subtle
points, namely the (non existing) temperature fluctuations and systems
with negative temperature.  Morevover we discuss some statistical
features of small systems.  In Section 4 we review the role of the
temperature in the context of the Response Theory.  The link between
relaxation and fluctuations allows for a bridge from equilibrium to
non equilibrium statistical mechanics, and in addition it suggests a
possible path for the introduction of temperature out of equilibrium.
The statistical mechanics approach, and the possibility to introduce a
concept of temperature for non Hamiltonian systems is discusses in
Section 5.  We shortly review such a challenge in topics as granular
media, chaotic systems, fluids and active matter.  In Section 6 we
present some general considerations and final remarks.


\subsection{Historical overview}

\begin{figure}
  \begin{center}
    \includegraphics[width=5cm]{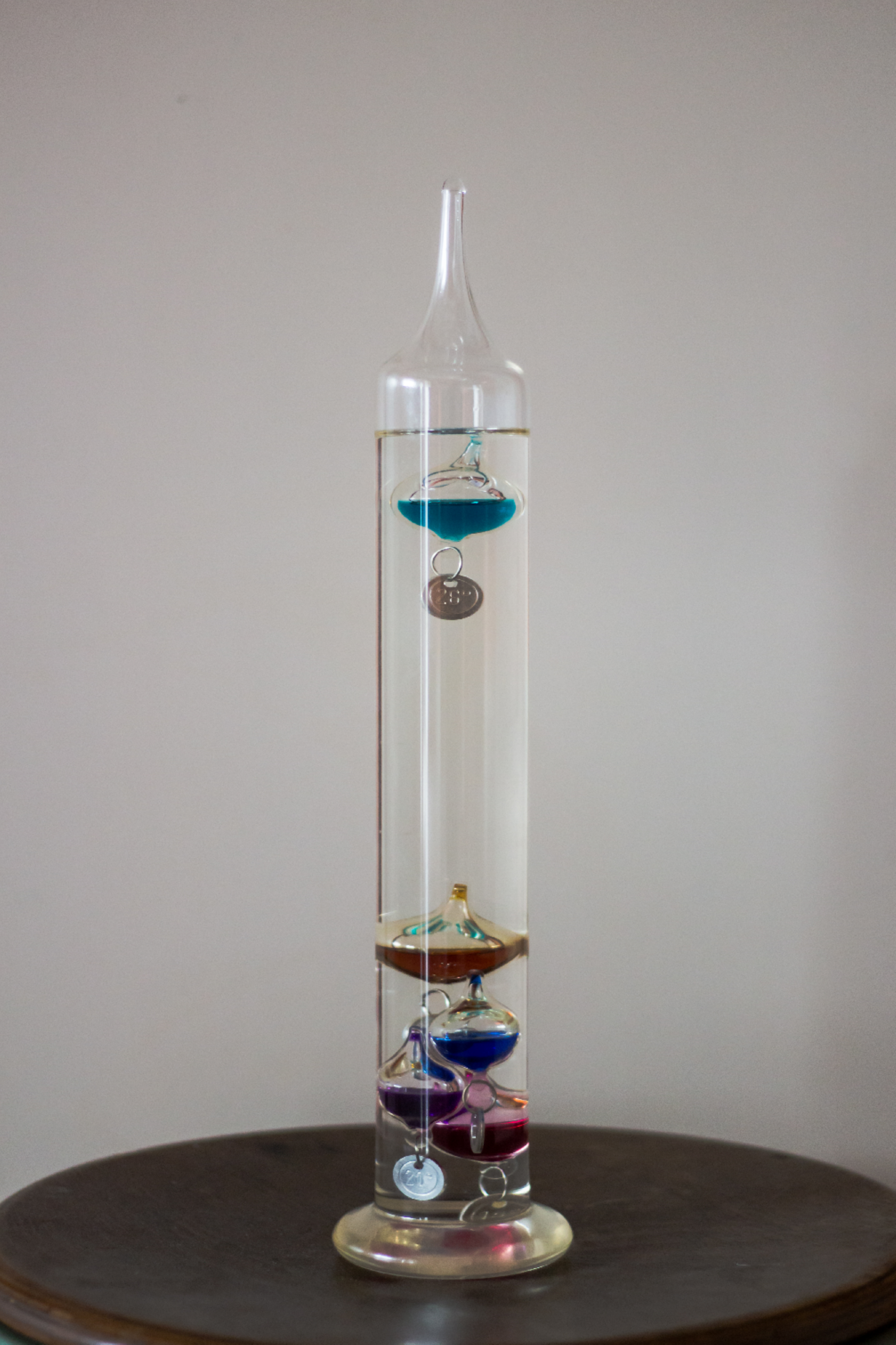}
    \end{center}
  \caption{\label{term} An example of Galileo's thermometer: the room temperature has an uncertainty of $1^oC$ (Courtesy of Andrea Baldassarri).}
  \end{figure}

Let us start this review with a brief, non systematic, historical
outlook. We believe to have an intuitive idea of temperature because
of our physiological sense of hot and cold. Actually our sense of
temperature depends also upon the heat flux
across the skin.  For instance, if we touch a piece of metal and a
piece of wood at the same temperature we have different perceptions:
the metal feels colder than the wood.  Although one can find some
ideas about the concept of temperature in the ancient Greek and
Islamic science, the first attempts to give a coherent definition
started in the 16-th century~\cite{muller07,chang04}. Galileo had a
crucial role in the development of the concept of temperature,
starting several studies which led to the construction of the first
thermometer, named after him: a closed glass cylinder containing a
clear liquid and several glass vessels of varying densities, see
Fig.~\ref{term}.  Actually such an instrument, which now is used as a
decorative object, has been invented by a group of academics and
technicians who included E. Torricelli and V. Viviani~\cite{loyson}.

Using Galileo's thermometer, G. Sagredo had been
able to understand the distinction between our
physiological senses and physical properties.  On February 7th, 1615,
in a letter to Galileo he wrote \\ \\ {\it ... Con questi istrumenti
  ho chiaramente veduto, essere molto pi\'u freda l' aqua de' nostri
  pozzi il verno che l'estate...., ancorch\'e il senso nostro giudichi
  diversamente.}  [... With these instruments I have clearly seen that
  water of our wells is colder in winter than in summer ..., although our
  senses tell differently.]~\cite{muller07} \\ \\ 

The introduction of the temperature at quantitative level needs a
thermometric scale.  For instance one can divide the interval between
the values of two suitable temperatures into a given number of
subdivisions.  In the Celsius scale, $0 ^oC$ corresponds to the
equilibrium between liquid water and ice, while $100 ^oC$ is the
temperature at the equilibrium between liquid water and vapour, both
at pressure of $1$ atmosphere.  For the Fahrenheit scale one has a
rather similar procedure.  A very important step has been the
observation that in all the gases, if the pressure $p$ is sufficiently
low (ideal gases), the so-called Boyle-Charles-Gay-Lussac law holds:
\begin{equation}
pV=n R ( t + 273.15 ^oC),
\label{I1}
\end{equation}
where $V$ is the volume, $n$ the number of moles, $R$ a universal
constant and $t$ the temperature in Celsius degrees.

\subsubsection{The zero law implies that  temperature is a consistent concept}

Before going on in the attempt to define temperature in a systematic
way, even in systems different from ideal gases, it is necessary to be
sure that the temperature ``really exists'', i.e.  one has to show
that it is possible to introduce such a concept in a non ambiguous way.

The key physical point for such an aim is the zero law of
thermodynamics.  If two systems $A$ and $B$ are each in thermal
equilibrium with a third, then they are in thermal equilibrium with
each other; in other words thermal equilibrium between two systems is
a transitive relation.  Such an empirical (and intuitive) law is the
physical basis which allows us to establish the concept of
temperature, as shown by the following simple and elegant argument,
see~\cite{miller52,pippard95}.  Consider two systems $A$ and $B$ whose
volumes and pressures are $(V_A, P_A)$ and $(V_B, P_B)$
respectively. We can say that there exists a function of the variables
of state such that, when the two systems are in equilibrium, one has:
\begin{equation}
F_1(P_A, V_A, P_B, V_B)=0.
\label{I2}
\end{equation}
Of course the form of the function depends on the considered system. In the following we show that, assuming the zero law, one has
\begin{equation}
\phi_1(P_A,V_A)=\phi_2(P_B,V_B),
\label{I3}
\end{equation}
from which one can derive the existence of a quantity (the temperature) 
common to the two systems.

Let us consider three systems $A$, $B$ and $C$;
if $A$ and $C$ are in equilibrium we have:
\begin{equation}
F_2(P_A, V_A, P_C, V_C)=0,
\label{I4}
\end{equation}
from which one can obtain $P_C$:
\begin{equation}
P_C=f_1(P_A,V_A,V_C).
\label{I5}
\end{equation}
If $B$ and $C$ are in equilibrium we can repeat the previous reasoning:
\begin{equation}
F_3(P_B, V_B, P_C, V_C)=0,
\label{I6}
\end{equation}
and
\begin{equation}
P_C=f_2(P_B,V_B,V_C),
\label{I7}
\end{equation}
and therefore
\begin{equation}
f_2(P_B,V_B,V_C)=f_1(P_A,V_A,V_C).
\label{I8}
\end{equation}
Due to  the zero law $A$ is in equilibrium with $B$, and therefore~(\ref{I2}) holds: however $V_C$ appears in~(\ref{I8}) but not  in~(\ref{I2}),
therefore   $f_1$ and $f_2$ must depend on $V_C$ in a proper way.
For instance the form
\begin{equation}
\label{I9}
f_1(P_A,V_A,V_C)=\phi_1(P_A,V_A)\zeta(V_C)+\eta(V_C)
\end{equation}
satisfies~(\ref{I8}) and implies~(\ref{I3});
in addition, see~\cite{miller52}, it is possible to show that not only~(\ref{I9}) is sufficient to have~(\ref{I3}),
but it is also necessary.

From the above reasoning one has that,  for systems in equilibrium,
there exists a quantity (empirical temperature $\theta$) depending on  $P$ and $V$
$$
\theta=\phi (P,V),
$$
which defines the equation of state.
For the ideal gases we have the  equation 
$$
PV=f(\theta),
$$
and the form of $f(\theta)$ is determined by the  empirical scale.
Being $t$ the temperature in degrees Celsius, the experimental result~(\ref{I1}) suggests a linear shape:  $f(\theta)=n R \theta$ with $\theta= 273.15 +t$.

\subsubsection{Absolute temperature in thermodynamics}

Using Carnot's theorem  on the efficiency of reversible engines working between  two heat reservoirs,
it is possible to go beyond the empirical  temperature and introduce the absolute temperature.
Let us recall Kelvin's idea~\cite{pippard95}:
denote  $Q_1$ the heat absorbed by the engine from the hotter reservoir (with empirical temperature $\theta_1$), and $Q_2$ the heat
delivered by the engine to the colder reservoir (with empirical temperature $\theta_2$),
in one cycle;  from Carnot's theorems 
one has that $Q_2/Q_1$ only depends  on the empirical temperatures $\theta_1$ and $\theta_2$:
$$
{Q_2 \over Q_1}= G(\theta_1, \theta_2).
$$
It is easy to show that~\footnote{Considering the cycles between $\theta_1$ and $\theta_3$, $\theta_2$ and $\theta_3$ and $\theta_1$ and $\theta_2$, we have 
  $$
  \frac{Q_1}{Q_3}=G(\theta_1,\theta_3), \;\;\;\;\;
  \frac{Q_2}{Q_3}=G(\theta_2,\theta_3), \;\;\;\;\;
  \frac{Q_1}{Q_2}=G(\theta_1,\theta_2),
  $$
  therefore we have 
  $$
  G(\theta_2,\theta_3)=\frac{Q_2}{Q_1}\frac{Q_1}{Q_3}= \frac{G(\theta_1,\theta_3)}{G(\theta_1,\theta_2)}=\frac{g(\theta_2)}{g(\theta_3)},
  $$
  proving Eq.~\eqref{carnot}.
} the function $G(\theta_1, \theta_2)$  must be independent of the details of the reservoirs and the engine,
one can write
\begin{equation} \label{carnot}
  {Q_2 \over Q_1}= {g(\theta_2) \over g(\theta_1)},
  \end{equation}
being $g(\theta)$ a universal function.
Therefore, using the result~(\ref{I1})  for the ideal gases,
 $g(\theta)$ can be identified, apart from a multiplicative constant, with the absolute temperature.

\subsection{Temperature in statistical mechanics}

Due to its intrinsic phenomenological character it is difficult to
associate the thermodynamic temperature with any specific dynamical
property of the considered system.  However once it was clear that
macroscopic systems are made of particles (atoms, molecules) ruled by
the mechanical laws, it was mandatory an attempt to find a link
between the mechanical world, on the one hand, and thermodynamics on
the other.

The first approach, dating back to Daniel Bernoulli, has been for
ideal gases considered to be point-like particles of mass $m$ that
collide with the walls of the container~\cite{brush03}.  Later
Clausius had been able to show that in such a model of matter,
Eq.~(\ref{I1}) holds, and the temperature $T$ is proportional to the
mean kinetic energy:
$$
m \langle v^2 \rangle= k_B T,
$$ 
being $v$ one component of the velocity, and $k_B$ the Boltzmann's
constant.

Such a first step toward the kinetic theory of gases has been the
starting point for the building of a consistent bridge, both
theoretical and concrete, between mechanics and thermodynamics.  The
most relevant contribution in such a great project is due to
Boltzmann; we can summarise his grand vision in two points:

\begin{enumerate}
 
\item  the introduction of probabilistic ideas and their
use for the   interpretation of physical observables;

\item the relationship linking the macroscopic world (thermodynamics) to the microscopic one (dynamics).

\end{enumerate}

\begin{figure}
  \begin{center}
    \includegraphics[width=6cm]{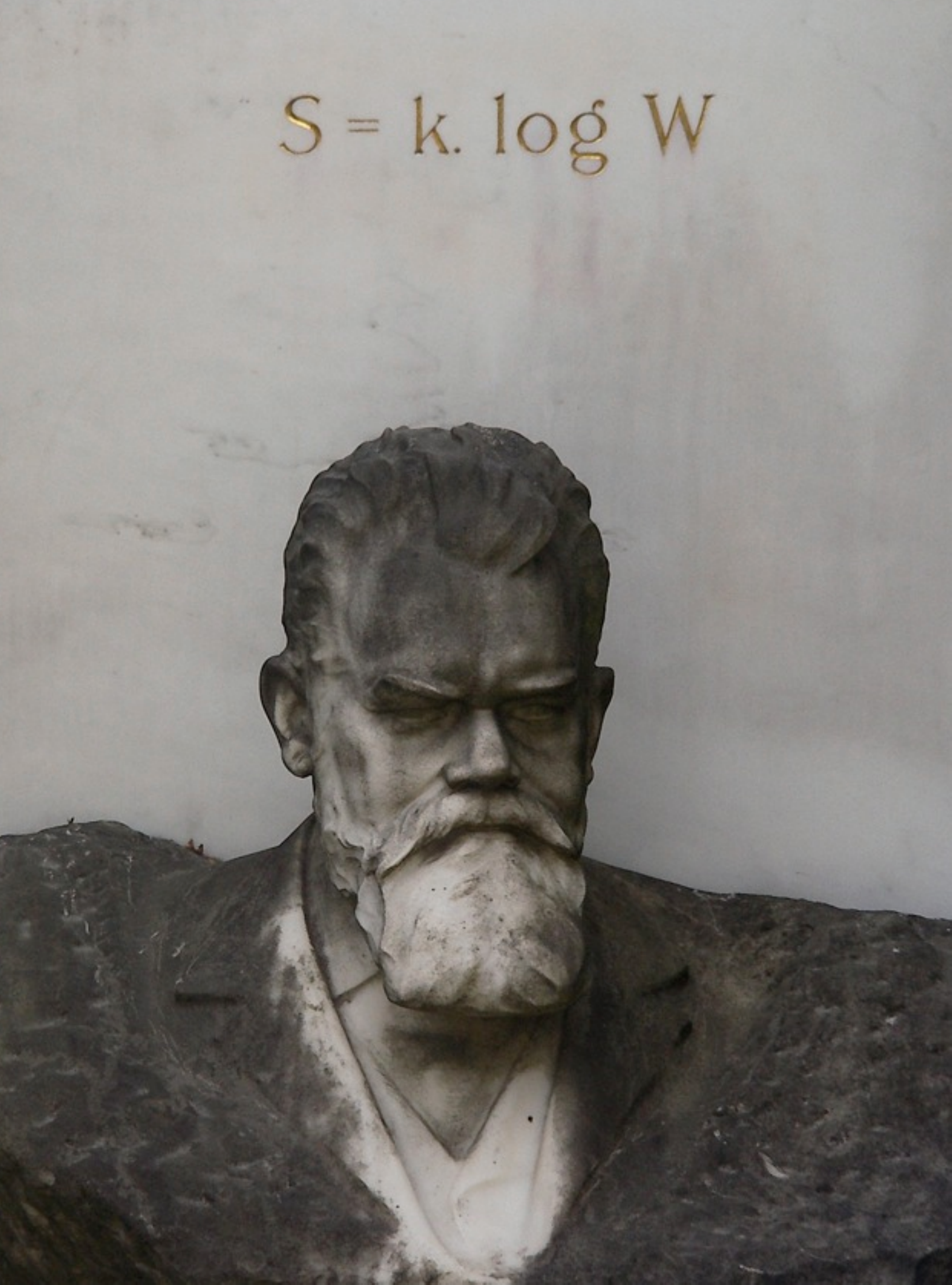}
    \caption{\label{grave}The celebrated Boltzmann's relation engraved on his tomb.}
    \end{center}
  \end{figure}

Point 1 is rather subtle and it is the object of intense study still
today. Boltzmann's idea was to replace time averages with averages
coming from a suitable probability density, which is nothing but the
ergodic hypothesis.  The relation connecting thermodynamics to the
microscopic world (point 2) is given by the celebrated equation
(engraved on Boltzmann's tombstone, see Fig.~\ref{grave}):
\begin{equation}
S= k \log W,
\label{II1}
\end{equation}
where $S$ denotes the entropy of the macroscopic body (a
thermodynamical quantity) and $W$ is the number of microscopic states
(a mechanical-like quantity) realising the macroscopic
configuration. Actually, Eq.~\eqref{II1}, usually called Boltzmann's
law, has been written by Planck~\cite{mehra01}.

If we are able  to express $W$ as function of the energy $E$, from the thermodynamic relation
we have
$$
{1 \over T}= {\partial S \over \partial E},
$$
and therefore we obtain a mechanical definition of temperature.

In the canonical ensemble the temperature appears in the probability distribution
in the phase space:
\begin{equation}
\label{II2}
P({\bf X}) = const. \,  e^{-\beta H({\bf X})}, \,\,\, \beta={1 \over k_B T},
\end{equation}
being ${\bf X}$ the vector describing the microscopic state of the
system, i.e.  the coordinates and momenta of all the particles.  In
the case of systems with short range interactions, the canonical
distribution describes the statistical features of a (small) part of
the system interacting with the remaining (large) part.  Therefore $T$
can be seen as a property of the heat bath with which the system is in
contact.

\subsection{Temperature and fluctuations}

\label{sec:einst}

The bridge law~(\ref{II1}) has important consequences supported by empirical evidence, 
 including, in particular, those derived by Einstein, see for instance~\cite{M15}.
 Denote with
$\alpha_1, \cdots, \alpha_m$  some macroscopic functions
of the microscopic state ${\bf X}: \alpha_j = g_j({\bf X})$, $j = 1, ... ,m$.
Einstein's idea was to use Eq.~(\ref{II1}) with the constraints $ \{ \alpha_k \}$. Since $W( \{ \alpha_k \})$ is proportional to the probability density
of  the variables $\{ \alpha_k \}$, one can invert the formula and obtain
 \begin{equation}
 \label{II3}
 P( \alpha_1, \cdots , \alpha_m) \sim e^{[S(\{ \alpha_k \}) -S_e]/k_B},
 \end{equation}
where $S_e = S(\{ \alpha_k^* \})$ is the equilibrium entropy computed at the equilibrium values $\{ \alpha_k^* \}$, and $S( \{ \alpha_k \})$
 is the entropy of a spontaneous fluctuation of the collective variables 
 $\{ \alpha_k \}$.

Formula (\ref{II3}) is meant to represent the probability
of fluctuations around equilibrium states of macroscopic mechanical quantities.
Since in the limit of large $N$ the fluctuations are small, we can expand
$S(\{ \alpha_k \}) -S_e$ around the equilibrium values $\{ \alpha_k^* \}$:
\begin{equation}
 S(\{ \alpha_k \}) -S_e \simeq { 1 \over 2} \sum_{i, j} 
 {\partial^2 S \over  \partial \alpha_i \partial \alpha_j}\Big|_{\{\alpha_k^* \}}
 \delta \alpha_i \delta \alpha_j,
\end{equation}
where $\delta \alpha_i= \alpha_i - \alpha_i^*$.
With such an approximation,  $ P( \alpha_1, \cdots , \alpha_m)$ is nothing but
 a multivariate
Gaussian distribution
 \begin{equation}
 P( \alpha_1, \cdots , \alpha_m) \sim e^{ -{1 \over 2 k_B} \sum A_{ij}
 \delta \alpha_i \delta \alpha_j },
 \end{equation}
where
$$
A_{ij}= - {\partial S \over \partial \alpha_i \partial \alpha_j}\Big|_{ \{ \alpha_k^* \}},
$$
and therefore one has:
\begin{equation}
\langle \delta \alpha_i \delta \alpha_j \rangle=k_B \Bigl[ {\bf A}^{-1} \Bigr]_{ij}.
\label{II3a}
\end{equation}
Let us comment on the relevance of the above formula: the fluctuations
of observable quantities can be described in terms of macroscopic
equilibrium thermodynamic functions.  For instance, if $\alpha$ is the
energy, one identifies the average $\langle E\rangle$ of the energy with the
internal energy $U$ of the system, and one has the well known formula:
\begin{equation}
\label{II4}
\langle E^2\rangle - \langle E\rangle^2= k_B T^2 C_v,
\end{equation}
which gives us information on the size of the energy fluctuations,
being $C_v$ the heat capacity at constant volume.  Since $C_v$ is
extensive, i.e. proportional to the number $N$ of particles in the
system, the relative size of the energy fluctuations is negligible in
large systems.  Apparently this should lead one to relegate
fluctuations to the set of only marginally interesting phenomena.  For
instance Boltzmann wrote \\ \\ {\it In the molecular theory we assume
  that the laws of the phenomena found in nature do not essentially
  deviate from the limits that they would approach in the case of an
  infinite number of infinitely small molecules.}~\cite{mehra01} \\ \\ Gibbs had
rather similar opinions: \\ \\ {\it [the fluctuations] would be in
  general vanishing quantities, since such experience would not be
  wide enough to embrace the more considerable divergences from the
  mean values.}~\cite{mehra01} \\ \\ On the contrary the fluctuations
have a key role in statistical mechanics.  More than one century ago
Einstein, in his search of an ultimate proof of the existence of
atoms~\cite{einstein04,E05}, realised that Eq.~(\ref{II4}) \\ \\ {\it
  would yield an exact derivation of the universal constant} [$k_B$
  or, equivalently, Avogadro' s number $N_A$] {\it if it were possible
  to determine the average of the square of the energy fluctuations of
  the system.}  \\ \\ Einstein's intuition was correct, and later he
was able to understand how to relate Avogadro's number to a
macroscopic quantity, namely the diffusion coefficient of Brownian
particles; the theoretical work by Einstein and the experiments by
Perrin gave a conclusive evidence of atomism~\cite{perrin13}.  In
addition, Einstein's seminal paper on the Brownian motion contains
another very important result, namely the first example of
Fluctuation-Dissipation Theorem (FDT), see Section 4: a relation
between the fluctuations (given by correlation functions) of an
unperturbed system and the mean response to a
perturbation~\cite{kubo66,kubo86,BPRV08}.  In the specific case of
Brownian motion, FDT appears as a link between the diffusion
coefficient (a property of the unperturbed system) and the mobility,
which measures how the system reacts to a small perturbation: a
fundamental observation is that the temperature of the system appears
in the proportionality factor between these two variables.

Beyond their conceptual relevance and the link with response
functions, fluctuations in macroscopic systems are quantitatively
extremely small and hard to detect (but for the case of second order
phase transitions in equilibrium systems).  On the other hand, in
recent years statistical mechanics of small systems, where
fluctuations are observable and cannot be neglected, is becoming more
and more important due to the theoretical and technological challenges
of micro and nano physics. For instance in~\cite{pekola2016} the
energy fluctuations of a finite free-electron Fermi gas have been
studied; one has quite a non-Gaussian effect at low temperature. 
Such a topic will be treated more in
detail in Section 4.


\section{General aspects of temperature and entropy}
\label{sec:basic}

\subsection{Different equilibrium definitions of temperature and entropy}

Surely it is not possible to underestimate 
the conceptual difficulties around the notion of entropy (and therefore temperature), 
even in equilibrium conditions:
\\
\\
{\it Entropy, like force, is an undefined object, and if you try to define it, you will suffer the same fate as the force definers of the seventeenth 
and eighteenth centuries.
Either you will get something too special or you will run around in a circle.}~\cite{truesdell66}
\\
\\
Let us start noting that there are, at least,  two natural definitions of entropy (and therefore temperature) in equilibrium statistical mechanics.
In the most common cases (i.e. system described by a Hamiltonian with a kinetic contribution and a potential term)
 the two definitions are equivalent in the limit of large systems.
 However it is interesting to discuss in some details   such a topic
 which,  after the publication of experimental measurements of a 
negative absolute temperature~\cite{Braun2013}, has been in the last years  the subject of an intense debate. In Section 3 we discuss 
the issue in more details.

As already mentioned in the Introduction the starting point for a microscopic foundation of temperature is    
the definition of entropy in terms of mechanical quantities.
Even if not historically precise~\cite{mehra01},
 we adopt  the current convention  and call ``Boltzmann entropy'' of a system 
containing $N$ particles, with Hamiltonian 
$H({\bf Q}, {\bf P})$, where ${\bf Q}$ and ${\bf P}$ are vectors in $\mathbb{R}^{dN}$, being $d$ the dimension of the system, the quantity
\begin{equation}
S_B(E,N)=k_B \ln \omega(E,N),
\label{BT}
\end{equation}
where
$$
\omega(E,N)=\int \delta(H({\bf Q}, {\bf P})-E) d^{dN}Q \, d^{dN}P=\frac{\partial \Sigma(E,N)}{\partial E},
$$
and
$$
\Sigma(E,N)=\int_{H({\bf Q}, {\bf P})<E} d^{dN}Q \, d^{dN}P.
$$
Assuming Eq.~(\ref{BT}), one can define the ``Boltzmann temperature'' through
$$
k_B \beta_B=\frac{1}{T_B}=\frac{\partial S_B(E,N)}{\partial E}.
$$

On the other hand, one can adopt a different definition of microcanonical entropy, proposed by Gibbs~\cite{Dunkel2014}.
The Gibbs entropy, which is always monotonically increasing, reads
$$
S_G(E,N)=k_B \ln \Sigma(E,N),
$$ 
and leads to the Gibbs temperature definition, which is always positive:
$$
k_B \beta_G=\frac{1}{T_G}=\frac{\partial S_G(E,N)}{\partial E} \ge 0.
$$
In the rest of this Section we consider systems made of a finite but large number $N \gg 1$
 of particles with local interactions, i.e. we exclude long-range potentials or mean-field models~\cite{R2014}.

\subsubsection{Canonical ensemble and energy fluctuations}

Let us consider a vector ${\bf X}$ in $\mathbb{R}^{2dN_1}$, with $N_1<N$,
 that is a subsystem of the full phase space $({\bf Q}, {\bf P})$
 and let us indicate with $\widetilde{{\bf X}}$ in $\mathbb{R}^{2d(N-N_1)}$ the remaining variables. 
For the Hamiltonian we have
$$
H=H_1({\bf X})+H_2(\widetilde{{\bf X}})+H_I({\bf X}, \widetilde{{\bf X}}),
$$ 
with an obvious meaning of symbols.  Let us consider the case
 $N\gg 1$ and $N_1 \ll N$.  In the microcanonical ensemble with energy
 $E$, the probability density function for the full phase space is
 \begin{equation}
 P({\bf Q}, {\bf P})=\frac{1} {\omega(E,N)} \delta (H({\bf Q}, {\bf P})-E).
\end{equation}
The probability distribution function (PdF) of ${\bf X}$ can be obtained from the latter
with a marginalization procedure, i.e. integrating over $\widetilde{{\bf X}}$.
 Since the Hamiltonian  $H_I({\bf X}, \widetilde{{\bf X}}) $ 
is negligible (a consequence of our assumption of non long- range interaction),
  we have
 \begin{equation}
 P({\bf X})\simeq \frac{\omega(E- H_1({\bf X}), N-N_1)} {\omega(E,N)}.
\label{EA}
\end{equation}
Now  writing $\omega$ in terms of $S_B$:
\begin{equation}
\omega(E,N)= e^{S_B(E,N)/k_B},
\label{EB}
\end{equation}

\begin{equation}
\omega(E-H_1({\bf X}),N-N_1)= e^{S_B(E-H_1({\bf X}),N-N_1)/k_B}, 
\label{EC}
\end{equation}
and  reminding that $N_1 \ll N$ we can assume $H_1 \ll E$, obtaining
 \begin{equation}
  S_B(E-H_1({\bf X}),N-N_1)\simeq S_B(E)- \frac{\partial S_B(E,N)}{\partial E} H_1({\bf X}) + const.
\label{ED} 
\end{equation}
Therefore, using Eqs.~(\ref{EA}, \ref{EB}, \ref{EC}) and~(\ref{ED}) one obtains the PdF in the canonical ensemble:
$$
 P({\bf X})=\frac{1}{Z} \, e^{-\beta_B H_1({\bf X})}.
 $$
 The above derivation is quite standard, and it is  presented in many textbooks. We repeated the reasoning
 with the aim to clarify that, once one assumes the microcanonical distribution,  $T_B$
  is the  ``correct'' temperature ruling the statistics of fluctuations of physical observables in a subsystem. 
  For instance, the PdF of energy $E_1$ in  the subsystem, is
  \begin{equation}
  P(E_1,N_1)\propto \omega(E_1,N_1) e^{-\beta_B E_1}.
  \label{PdFE}
  \end{equation}

\subsection{Large deviation theory and convexity property of entropies}
 
Let us discuss the following relevant property: $S_B(E,N)$  is always convex, i.e. $d^2 S_B(E,N)/dE^2\le0$.
This result is a consequence of the second law of thermodynamics, and, in addition, it can be proved in 
a rigorous way  in the limit of vanishing interaction, and in short-range interacting systems, 
for large $N$~\cite{ellis99,touchette09}.
Actually  $S_B$ is strictly related to the large deviation function associated with the density of states.

 \subsubsection{Convexity of $S_B$ and  the second law of thermodynamics}
 
Let us consider a system ${\cal A}$  of  $N_{{\cal A}}$  particles described by the variables 
$\{ {\bf Q}_{{\cal A}}, {\bf P}_{{\cal A}} \}$
and Hamiltonian $H_{{\cal A}} ({\bf Q}_{{\cal A}}, {\bf P}_{{\cal A}})$, a  
system ${\cal B}$  of  $N_{{\cal B}}$  particles described by the variables $\{ {\bf Q}_{{\cal B}}, {\bf P}_{{\cal B}} \}$
and Hamiltonian $H_{{\cal B}} ({\bf Q}_{{\cal B}}, {\bf P}_{{\cal B}})$, 
and a small coupling among the two, so that the global Hamiltonian is
$$
H=H_{{\cal A}} ({\bf Q}_{{\cal A}}, {\bf P}_{{\cal A}}) + H_{{\cal B}} ({\bf Q}_{{\cal B}}, {\bf P}_{{\cal B}}) +
H_{I} ({\bf Q}_{{\cal A}}, {\bf Q}_{{\cal B}}).
$$
If the two Hamiltonians have the same functional dependencies on the canonical variables 
(i.e. they correspond to systems with same microscopic dynamics, with possibly different sizes $N_{{\cal A}}$  and 
$N_{{\cal B}}$) 
for large $N$, we can introduce the (Boltzmann) entropy per particle $s(e)$
$$
s(e)= \frac{S_B(E,N)}{N}, \,\,\, e=\frac{E}{N},
$$
with $s(e)$ a function which is identical for the two   systems. 

Let us now
suppose that systems ${\cal A}$ and ${\cal B}$ have, respectively, energy 
$E_{{\cal A}}=N_{{\cal A}} e_{{\cal A}}$ and $E_{{\cal B}}=N_{{\cal B}} e_{{\cal B}}$ 
 and the corresponding inverse Boltzmann temperatures are  $\beta_B^{({\cal A})}$ and
$\beta_B^{({\cal B})}$.
When the two systems are put in contact, a new system is realized with 
$N=N_{{\cal A}}+ N_{{\cal B}}$ particles. 
Indicate with   $a=N_{{\cal A}}/N$
the fraction of particles from the system ${\cal A}$, the final energy is 
$E_f=E_{{\cal A}}+E_{{\cal B}}=N e_f$, where 
$e_f= a e_{{\cal A}}+(1-a)e_{{\cal B}}$.
From the second law of the thermodynamics
we have that the final entropy cannot be smaller than the in initial one:
$$
S_{ B}(E_F,N)=N s(e_f)\ge N_{{\cal A}} s(e_{{\cal A}}) +N_{{\cal B}} s(e_{{\cal B}})=
N [a s(e_{{\cal A}}) +(1-a) s(e_{{\cal B}})].
$$
The previous inequality is nothing but a way to express
 the convexity  of  $s(e)$, i.e.
$$
s[a e_{{\cal A}} +(1-a) e_{{\cal B}}]\ge
a s(e_{{\cal A}}) +(1-a) s(e_{{\cal B}}).
$$
The final inverse temperature  $\beta_B^{(f)}$
  is intermediate between   $\beta_B^{({\cal A})}$ and $\beta_B^{({\cal B})}$,
  i.e.  if $e_{{\cal B}} > e_{{\cal A}}$ -  that is 
    $\beta_B^{({\cal A})} > \beta_B^{({\cal B})}$- then
 $$
 \beta_B^{({\cal B})} < \beta_B^{(f)}<\beta_B^{({\cal A})}.
 $$
The energy flux obviously goes from smaller $\beta_B$ (hotter) to larger $\beta_B$ (colder).
 The consequence of convexity is that $\beta_B(E)$  is always decreasing and a negative value does not lead to any ambiguity. 
 
 In the next Section  we'll  discuss a particularly interesting case with different Hamiltonians where   in the system ${\cal A}$ negative temperatures can be present, whereas system ${\cal B}$ has only positive temperatures.
 
 \subsubsection{Entropy and large deviations}
 
If we consider the energy per particle $e=E/N$, Eq.~(\ref{PdFE}), dropping the subscript $1$, can be written in the form
$$
P(e)=\frac{1}{Z}\,  \exp \{ -N\beta_B[e- T_B s(e)]  \},
$$
where
$$
Z \sim  \exp \{ -N \beta_B f(T_B) \},
$$
being $f(T_B)$ the free energy per particle
$$
f(T_B)=\min_e\{ e -T_B s(e) \}.
$$
The value $\overline{e}$ for which $e-T_B s(e)$ reaches its minimum  is given by the condition
$$
\frac{1}{T_B}=\frac{\partial s(e)}{\partial e}\Big|_{\overline{e}},
$$
i.e. it is the value such that the corresponding microcanonical ensemble has temperature $T_B$ .
Therefore we can write
$$
P(e) \sim \exp \{-N{\cal C}(e) \},
$$
with
$$
 {\cal C}(e)=\beta_B [e- T_B s(e) -f(T_B)].
$$
Of course the value of $e$ such that ${\cal C}(e)$ is minimum (zero) is the mean energy:
$\overline{e}=\langle e \rangle$.
With the Gaussian approximation around  $\overline{e}$  one can compute the variance of energy
fluctuations obtaining the (exact) formula~(\ref{II4}).

The  above simple computations are nothing but an example of
large deviations theory (LDT). 
For the sake of self-consistence let us briefly remind the basic aspects of this approach~\cite{touchette09,ld14}.
The general mathematical formulation of LDT has been introduced in the 1930s mainly by Cram\'er for
independent identical distributed  random variables $x_1, x_2, \cdots , x_N$
 with mean value  $\langle x \rangle$, and standard deviation $\sigma$.
 The main aim of the LDT is to go beyond the central limit theorem, which
 is able to describe only the typical fluctuations of the  ``empirical mean'' 
 $y=(x_1+ \cdots + x_N)/N$, i.e. for  $ |y -\langle y \rangle|<O(\sigma /\sqrt{N})$.
Under the rather general assumption of existence of the moment generating function 
$ \langle e^{q x} \rangle$  in some neighbourhood of $q=0$, it is possible to prove that for 
$N \gg 1$ one has
$$
P(y) \sim e^{- N {\cal C}(y)}.
$$
The Cram\'er function ${\cal C}(y)$  depends on the probability distribution of
$x$,  it is positive everywhere but for $y=\langle x \rangle$ where it vanishes. 

From an historical point of view it is interesting  that the first LDT calculation 
has been carried out by Boltzmann~\cite{ellis99}.
He was able to express the asymptotic behavior of the multinomial probabilities in terms of relative entropy.
In the case $x_i$ takes the value $+1$ with probability $p$ and $-1$ with probability $1-p$, using the
Stirling approximation it is possible to
obtain the explicit expression for the Cram\'er function:
\begin{equation}
{\cal C}(y)=\frac{1+y}{2}\ln \frac{1+y}{2 p}+
\frac{1-y}{2}\ln \frac{1-y}{2(1- p)}. \label{cramer}
\end{equation}

\begin{figure}
  \begin{center}
    \includegraphics[width=6cm,angle=-90]{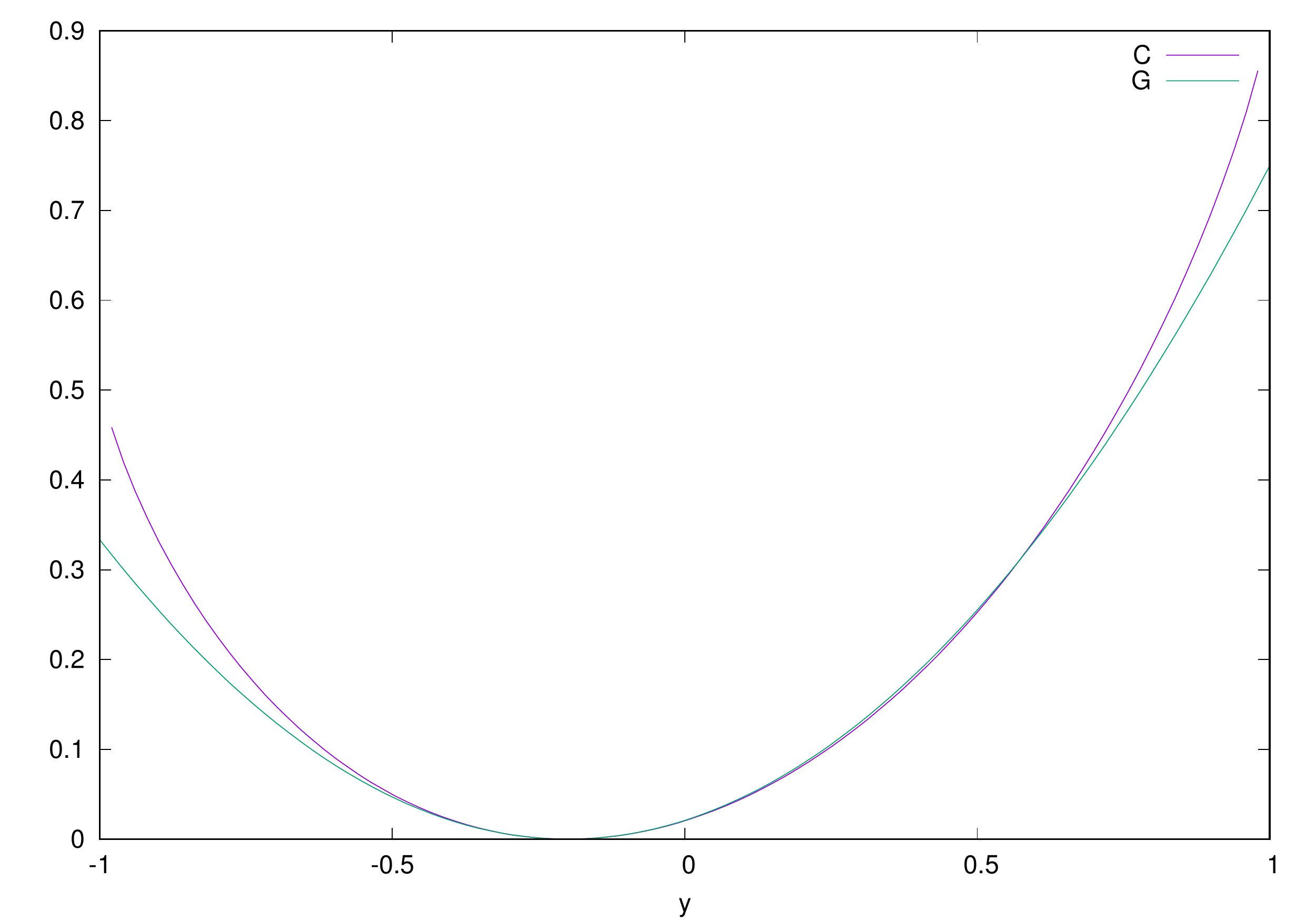}
    \end{center}
  \caption{\label{cramerfig} An example of Cram\'er function as in Eq.~\eqref{cramer} for $p=0.4$. For comparison we show its parabolic approximation (``G'' curve).}
  \end{figure}

Let us stress that the Cram\'er function for independent variables
must obey the following constraints:
\begin{description}
\item (i) ${\cal C}(y) >0$ for $y\ne \langle y \rangle=\langle x \rangle$;

\item (ii) ${\cal C}(y) =0$ for $y= \langle y \rangle$;

\item (iii) ${\cal C}(y)\simeq (y -\langle y \rangle)^2/(2 \sigma^2) $,  where
 $\sigma^2= \langle (y -\langle y \rangle)^2 \rangle$,  if $y$ is close to $\langle y \rangle$;

\item (iv) $d^2 {\cal C}/dy^2 > 0$.
\end{description}
Properties (i) and (ii) are consequences of the law of large numbers, 
and (iii) is nothing but the central limit theorem;
property (iv) has not a simple justification, and in statistical physics it corresponds to a
``mathematical translation'' of the second law of thermodynamics.
Moreover, the Cram\'er function is linked, via a Legendre transform, 
to the cumulant generating function of the variable $x$
$$
{\cal C}(y)=\sup_q \{ qy -L(q) \},
$$
where
$$
L(q)=\ln \langle e^{qx} \rangle.
$$ 
The propetries (i)-(iv) are valid under rather general hypothesis,
in particular it is not necessary that the variables $\{x_n \}$ are
independent, provided that the correlation function
$c(k)=\langle(x_{n+k}- \langle x \rangle) (x_n- \langle x \rangle)
\rangle$ goes to zero faster than $1/k$ as $k \to \infty$~\cite{varadhan84,touchette09,ld14}.

\subsection{Entropy in the $\Gamma$-space and in the $\mu$-space }

Let us now briefly discuss some aspects related to entropy, which,
although simple from a technical point of view, are rather subtle and
can induce confusion.

A popular way  to formulate the second law of thermodynamics is to say 
that the entropy increases in time; 
such a statement deserves careful considerations, in particular on
how to associate an entropy-like quantity to the state of a physical 
system~\cite{lebowitz93,lebowitz93b,cerino16}.
Considering a Hamiltonian system with $N$  weakly interacting particles, its
 microscopic state, in the so called $\Gamma$-space,
is given by the vector ${\bf X}=({\bf q}_1,.., {\bf q}_N, {\bf p}_1,..., {\bf p}_N)$,
where ${\bf q}_n$ and ${\bf p}_n,$ are the position and the momentum of the $n$-th particle respectively.
From the probability distribution  in the $\Gamma$- space $\rho({\bf X}, t)$ at time $t$,
 following Gibbs, we can introduce an entropy
\begin{equation}
S_{\Gamma}(t)=-k_B \int \rho({\bf X}, t) \ln \rho({\bf X}, t) d {\bf X}.
\label{E1}
\end{equation}
If $\rho$ is the stationary  canonical distribution $e^{-\beta H({\bf X})}/Z$, the quantity
$S_{\Gamma}$ is the usual entropy.
More interesting is its dynamical aspect:
using the Liouville theorem it is easy to show that $S_{\Gamma}(t)$ must be  constant.

In order to observe an increase over time for $S_{\Gamma}(t)$-like quantities, 
one can  introduce a coarse-graining of the $\Gamma$-space.
Consider a partition
 of the phase space in cells    $\{ \Lambda_i(\epsilon) \}$ of size $\epsilon$, and
consider  the probability to stay, at time $t$, in the   $i-$th cell:
$$
P_i(\epsilon, t)=\int_{\Lambda_i(\epsilon)} \rho({\bf X}, t) d {\bf X},
$$
the  $\epsilon$ coarse-grained Gibbs entropy is
\begin{equation}
S_{\Gamma}^{(\epsilon)}(t)= -k_B \sum_i P_i(\epsilon, t) \ln P_i(\epsilon, t).
\label{E2}
\end{equation}
Now $S_{\Gamma}^{(\epsilon)}(t)$, for any small $\epsilon \neq 0$  is, typically,
 an increasing function of time: for instance 
in a suitable time interval one has
\begin{equation}
S_{\Gamma}^{(\epsilon)}(t)- S_{\Gamma}^{(\epsilon)}(0) \simeq h_{KS} t + const.,
\label{E3}
\end{equation}
where $h_{KS}$ is the Kolmogorov-Sinai entropy.
The above  behaviour is surely interesting in a dynamical systems context.
Numerical results  show that even such a coarse-grained Gibbs entropy
remains constant up to a crossover time  $t^{*}(\epsilon)\sim \ln (1 / \epsilon)$. 
Only after $t^{*}(\epsilon)$, $S_{\Gamma}^{(\epsilon)}(t)$
 increases, showing the linear growth~(\ref{E3})~\cite{falcioni07}.
 This  $\epsilon$-dependence indicates that 
the increasing of $S_{\Gamma}^{(\epsilon)}(t)$
has no genuine thermodynamic meaning and it is merely  originated by
the coarse-graining procedure;
therefore~(\ref{E3}) cannot be viewed as the second law of thermodynamics~\cite{falcioni07}.

We discuss now another way, due to Boltzmann, to  introduce the concept of entropy in weakly interacting systems.
Instead of dealing with the probability distribution in the $\Gamma$-space, let us define the one
particle distribution function in the $\mu$-space:
\begin{equation}
f({\bf r}, t)={1 \over N} \sum_{n=1}^N \delta ({\bf r}-{\bf r}_n(t)),
\label{E4}
\end{equation}
where ${\bf r}_n=({\bf q}_n, {\bf p}_n)$, and introduce the quantity
\begin{equation}
S_{\mu}(t)=-k_B \int f({\bf r}, t) \ln f({\bf r}, t) d {\bf r}.
\label{E5}
\end{equation}
It is easy to realise that at equilibrium the two  entropies are equivalent: $S_{\Gamma} \simeq N S_{\mu}$.

At a first glance the two definitions of entropy, (\ref{E1}) and
(\ref{E5}), appear rather similar, but a close analysis shows that
their dynamical behaviours are rather different.  We remind that from
the celebrated $H$-theorem there follows that, under suitable
conditions, $S_{\mu}$ increases in
time~\cite{cercignani88,cercignani98}.  It is remarkable the fact that
a coarse-grained procedure for $S_{\mu}$, at variance with the result
for $S_{\Gamma}$, does not change the qualitative behaviour of
$S_{\mu}(t)$ vs $t$~\cite{falcioni07}.

Let us note that the two definitions~(\ref{E1}) and~(\ref{E5}),
somehow, reflect two different approaches to the foundations of the
statistical mechanics and conceptual differences on how to consider
probability.  For the computation of the entropy $S_{\Gamma}$, one
needs $\rho({\bf X}, t)$, namely an ensemble.  It is easy to realise
that $\rho({\bf X}, t)$, and therefore $S_{\Gamma}(t)$, is accessible
only in numerical experiments with systems composed by few degrees of
freedom~\cite{falcioni05}.  On the contrary, if $N\gg 1$, the one
particle distribution function $f({\bf r},t)$, and therefore
$S_{\mu}(t)$, can be seen as an empirical distribution, it is a well
defined macroscopic observable and can be, in principle, measured in a
single system; actually this is the standard procedure in numerical
simulations~\cite{lebowitz93,lebowitz93b,cerino16}.  One may object
that the introduction of $f({\bf r}, t)$ is a form of coarse-graning,
although different from $P_i(\epsilon, t)$. However, as briefly
discussed above, we know that this is the ``correct coarse-graining'',
which is consistent with thermodynamics and in addition it is quite
natural in a numerical approach~\cite{castiglione08,chibbaro14}.

\subsection{Helmholtz's monocycle and  some  Boltzmann's ideas about ergodicity}
\label{monocycle}

As already discussed, usually  Boltzmann's  journey towards a consistent formulation of statistical mechanics 
starting from mechanics,  is  summarised with  two (apparently independent) items:
\\
$\bullet$  the ergodic hypothesis;
\\
$\bullet$ the law relating entropy to mechanics: the celebrated $S=k_B \ln W$.
On the other hand there is an interesting  way to introduce 
 ergodicity  using the second law of thermodynamics and the formula (\ref{II1}).
 Such a link  between ergodicity and entropy,
 had an important role for the development of Boltzmann's ideas,
 and it is based on a result of Helmholtz for one-dimensional Hamiltonian systems.
 Apparently this topic seems to be  almost unknown even to scholars
 interested in the history of physics.
 Luckily it was recently exhumed by  Gallavotti~\cite{gallavotti95} and discussed in a very clear way by  
 Campisi and Kobe~\cite{campisi10}.
 
 Consider a  one dimensional system with   Hamiltonian
 $$
 H(q,p, V)={p^2 \over 2m} + \phi(q,V),
 $$
 where  $V$  is a control parameter, for instance in a pendulum $V$ is the 
  length which can be varied.
 Assume that for each $V$,
 $\phi(q,V)$ has a unique minimun and in addition diverges as  $|q| \to \infty$.
In such a system, for any value of $E$,
the motion is surely periodic;
 denote with $\tau(E,V)$ the  period,  $q_{-}(E,V)$  and $q_{+}(E,V)$ 
 the minimum and maximal value of $q$ respectively.
 Of course the motion is ergodic, i.e. the time averages coincide with the averages 
 computed with the microcanonical
 distribution:
\begin{equation}
d\mu(p,q)=
{ \delta(H(q,p,V)-E) dq dp \over  \int \int \delta(H(q,p,V)-E) dq dp}.
\label{5.9}
\end{equation}
Let us mention Helmholtz's theorem:
\\
Define the temperature $T$ and the pressure $P$ in terms of time averages $ \left<( ....) \right>_t$
computed on the period $\tau(E,V)$:
\begin{equation}
T= {2 \over k_B} \left<{p^2 \over 2m} \right>_t, \,\,
P= -{1 \over k_B} \left<{\partial \phi(q,V) \over \partial V} \right>_t.
\label{5.10}
\end{equation}
Then, the function
\begin{equation}
S(E,V)=k_B \ln 2 \int_{q_{-}{(E,V)}}^{q_{+}(E,V)} \sqrt{2m[E-\phi(q,V)]} dq
\label{5.11}
\end{equation}
satisfies the following relations
\begin{equation}
{\partial S \over \partial E} ={1 \over T}, \,\,
{\partial S \over \partial V} ={P \over T}.
\label{5.12}
\end{equation}
To prove the above formulas it is enough to use
$$
\delta(f(p))=\sum_i {\delta (p-p_i) \over |f'(p_i)|}, 
$$
where  $\{ p_i \}$ are determined by  $f(p_i)=0$, and therefore one has
$$
\left<f \right>_t={2m \over \tau(E,V)} \int_{q_{-}{(E,V)}}^{q_{+}(E,V)} {f(q,p(q)) \over p(q)}  dq
$$
with $p(q)=\sqrt{2m[E- \phi(q,V)]}$,
see~\cite{campisi10} for the details.
Let us note that $S(E,V)$ can be written in the usual form
$$
S(E,V)=k _B\ln \int_{H(q,p,V)<E} dp dq.
$$
The above  results imply a rather interesting consequence, namely
the {\it existence of  a mechanical analogue for the entropy}:
indeed, the following quantity
$$
{dE + PdV \over T},
$$
where $T$ and $P$ are expressed via time averages of mechanical observable, is an exact differential.

Boltzmann's idea was to generalise the above result, which is
surely valid for $1d$ Hamiltonian,   to systems with many particles,
in other words to find a function
$S(E,V)$  such that the relations  (\ref{5.10})  and (\ref{5.12}) are still valid.
In a Hamiltonian system with $N$ particles, assuming ergodicity it is possible to show
a {\it Generalised   Helmholtz's theorem} for the function
$$
S(E,V)=k_B \ln \int_{H({\bf q}, {\bf p},V)<E} d{\bf q} d{\bf p}.
$$
The proof is quite similar to that one for the $1d$ case, see~\cite{campisi10}.

We can summarize Boltzmann's reasoning as follows:
in systems with many particles the periodic trajectory of the $1d$ case is replaced by the hypothesis
that the trajectory will sweep  the whole surface $H=E$ 
(this is a way to say that  the time averages 
can be replaced with the microcanonical averages).
Then from the Generalised   Helmholtz's theorem one has the second law of thermodynamics,
i.e. the existence of a function (the entropy $S$) which can be expressed in mechanical terms,
and, in addition,  $dS/T$  is an exact differential.

In view of the discussion about the different possible definitions of
entropy and temperature (see Section 3.3) we note that all the above
arguments hold for systems with the typical Hamiltonian including a
quadratic kinetic term.

\subsection{Temperature as a time average} 

We now focus on a dynamical approach to temperature, which allows one to measure it as a time average. This result is due to
Rugh~\cite{rugh97,rugh98}, who has been able to show that the (microcanonical) temperature can be written as an average
 of a suitable observable, namely:
\begin{equation}
{1 \over  k_B T}={\partial \ln \omega(E) \over  \partial E}= {1 \over \omega(E)}  {\partial \omega(E) \over \partial E}=
\left<   \nabla \cdot \Big( {\nabla H \over || \nabla H ||^2} \Big) \right>_E.
\label{R1}
\end{equation}
Here the average is  computed with the microcanonical distribution
on the energy constant hypersurface $ \Sigma_E=\{{\bf X}: H({\bf X})=E \}$.

The physical relevance of the above result is that the temperature can be written
in terms of an average of a mechanical observable~\cite{giardina98}, 
and therefore, assuming ergodicity, one can obtain the temperature with a time average along
a trajectory:
$$
{1 \over k_B T}=\lim_{\cal T \to \infty} \frac{1}{\cal T}\int_0^{\cal T} \Phi({\bf X}(t)) \, dt,
$$
where
\begin{equation} \label{rughphi}
\Phi=\nabla \cdot \Big( {\nabla H \over || \nabla H ||^2} \Big).
\end{equation}
If the system is very large, one can write
$$
{1 \over k_B T}=
\left<   {\nabla^2 H \over || \nabla H ||^2}  \right>_E \Big( 1 + O\Big({ 1 \over N}\Big)  \Big)=
{ \left<  \nabla^2 H  \right> _E  \over  \left< || \nabla H ||^2 \right>_E }\Big( 1 + 
O\Big({ 1 \over N}\Big) \Big).
$$
For ideal gases, as well as for harmonic systems, 
it is easy to see that, for $N \gg 1$, the above result coincides with the usual one,
i.e. the temperature is proportional to mean kinetic energy.

Rugh's approach is much more  interesting if the Hamiltonian is 
different from the standard, i.e.  containing a kinetic term
and a potential part:
\begin{equation}
H=\sum_{n=1}^N { {\bf p}_N^2 \over 2 m} + V({\bf q_1}, .... , {\bf q}_N) .
\label{Usual1}
\end{equation}
We discuss such a class of systems in Section 3.3.

\subsubsection{Generalization of Rugh's result}

The result (\ref{R1}) has been generalised, see e.g.~\cite{rugh98,jepps00,powles01}, and
it is possible to show that
\begin{equation}
{1 \over k_B T}=
\left<   \nabla \cdot \Big( { {\bf B} \over  {\bf B} \cdot \nabla H} \Big) \right>_E,
\label{G1}
\end{equation}
where ${\bf B}$ is an arbitrary continuous and differentiable vector 
in phase-space.
The original Rugh's result corresponds to the case ${\bf B}= \nabla H$.

For an Hamiltonian with the shape (\ref{Usual1}), with the choice 
 ${\bf B}=(0, \cdots, 0, {\bf p}_1 , \cdots {\bf p}_N)$, 
Eq.~(\ref{G1}) leads to
  \begin{equation}
{1 \over k_B T}=
\left<  {m \, d \, N \over \sum_n {\bf p}^2_n} \right>_E,
\label{G2}
\end{equation}
where $d$ is the spatial dimension, and in the limit $N \gg 1$ one has the usual relation
$$
k_B T= { 1 \over m \, d \, N} \left<  \sum_n {\bf p}^2_n  \right>_E.
$$
With the choice  ${\bf B}=({\bf q}_1 , \cdots {\bf q}_N, 0, \cdots, 0)$, 
in the limit $N \gg 1$ one obtains
$$
k_B T= { 1 \over \, d \, N} \left<  \sum_n {\bf q}_n\cdot 
{\partial  \over  \partial {\bf q}_n } V
 \right>_E,
$$
which is nothing but the Clausius virial theorem.

Let us now consider the choice ${\bf B}= \nabla V$, for $N \gg 1$, which yields
\begin{equation}
{1 \over k_B T}=
\left<   \nabla \cdot \Big( { {\nabla V} \over  ||\nabla V||^2} \Big) \right>_E \simeq
{ \left<  \nabla^2 V \right>_E  \over
\left<  ||\nabla V||^2 \right>_E }.
\end{equation}
The above formula provides a definition of the temperature only in terms of the coordinates
and not of the momenta. 

The generalization of Rugh's formula, in particular the
configurational temperature, has been successfully used in numerical
simulations, for instance for checking the algorithmic correctness of
Monte Carlo computer programs, and (following the Nos\'e-Hoover
approach) to design new thermostats~\cite{morriss99,jepps00}.

 \subsubsection{A brief mathematical parenthesis: derivation of Rugh's result}
  
For the sake of completeness and selfconsisteny, we report the derivation of
Eq.~(\ref{R1}), which   is a special case of the following result:
 for any  function $F({\bf X})$ one has
 \begin{equation}
{ d \over dE} \int_{H=E} F({\bf X}) \, d \sigma({\bf X})=
\int_{H=E}  \nabla \cdot \Big( F({\bf X}) { \nabla H \over || \nabla H || } \Big)
 { 1 \over || \nabla H || } \, d \sigma({\bf X}),
 \label{R2}
 \end{equation}
where   $d\sigma ({\bf X})$  is the infinitesimal ``area'' on $\Sigma_E$.
Let us introduce the quantity
 $$
 G_F(E,\Delta E)=
{1 \over \Delta E} \Big(  \int_{H=E+\Delta E} F({\bf X}) \, d \sigma({\bf X})-
   \int_{H=E} F({\bf X}) \, d \sigma({\bf X}) \Big),
 $$
and  the unitary inner vector normal to $\Sigma_E$ in ${\bf X}$ is ${\bf n}({\bf X})= \nabla H/ || \nabla H ||$.
Using  the identity $1= \nabla H \cdot \nabla H/ || \nabla H ||^2$, we can write:
 $$
 G_F(E,\Delta E)=
 {1 \over \Delta E} \Big(  
 \int_{H=E+\Delta E} F({\bf X})  { \nabla H \over || \nabla H || } \cdot  {\bf n} \, d \sigma({\bf X})
 - \int_{H=E} F({\bf X})  { \nabla H \over || \nabla H || } \cdot  {\bf n} \, d \sigma({\bf X})
 \Big) \,\, =
 $$
 $$
 {1 \over \Delta E} \Big(  
 \int_{H=E+\Delta E} F({\bf X})  { \nabla H \over || \nabla H || } \cdot  {\bf n}_e \, d \sigma({\bf X})
 + \int_{H=E} F({\bf X})  { \nabla H \over || \nabla H || } \cdot  {\bf n}_e \, d \sigma({\bf X})
 \Big),
 $$
 where  ${\bf n}_e$ is the unitary normal pointing toward the exterior of the region
 $\{ {\bf X}: E < H({\bf X}) < E+\Delta E \}$.
 Now, applying the divergence theorem, we obtain
 $$
 G_F(E,\Delta E)=
 { 1  \over \Delta E} \int_{E<H<E + \Delta E}   \nabla \cdot \Big( F({\bf X}) { \nabla H \over || \nabla H || } \Big)
 \, d {\bf X}.
$$
Since we can write 
\begin{equation}
\label{R3}
d {\bf X}= dE {d \sigma({\bf X}) \over  || \nabla H ||},
\end{equation}
one has
  $$
  G_F(E,\Delta E)=
  { 1  \over \Delta E} 
   \int_E^{E + \Delta E}  d E \int_{H=E }   \nabla \cdot \Big( F({\bf X}) { \nabla H \over || \nabla H || } \Big) \,
{1 \over  || \nabla H ||} d \sigma({\bf X}),
 $$
and,  in the limit $\Delta E \to 0$, one obtains Eq.~(\ref{R2}).
  
For our aim the interesting case is
  $$
  F({\bf X})= {1 \over || \nabla H||}.
  $$
  Reminding that
  $$
  \omega(E)= \lim_{\Delta E \to 0}  {1 \over \Delta E} \int_{E<H<E+\Delta E} d {\bf X},
  $$
  using~({\ref{R3})  and~(\ref{R2}), one has
  $$
\omega(E)  = \int_{H=E} {1 \over || \nabla H||}
   \, d \sigma({\bf X}),  \,\,
 {\partial \omega(E) \over \partial E}=
\int_{H=E}       \nabla \cdot \Big( {\nabla H \over || \nabla H ||^2} \Big)
{1 \over || \nabla H||} \, d \sigma({\bf X}).
$$
Eventually, using the relation
$$
\left<  ( \, \cdot \, ) \right>_E={1 \over \omega(E)}
\int_{H=E} ( \, \cdot \, )
{ d \sigma({\bf X})  \over || \nabla H||},
$$
one has a proof of Eq.~(\ref{R1}).


\section{Temperature beyond the mean kinetic energy}
\label{sec:subtle}

\subsection{Temperature cannot fluctuate}

The energy of a finite system interacting with a thermal reservoir may
fluctuate, while the temperature is a constant representing a
thermodynamic property of the reservoir. However, as discussed in
Sec.~\ref{sec:einst}, Einstein showed that the statistical properties
of macroscopic variables can be determined in terms of quantities
computed in thermodynamic equilibrium~\cite{LandauFisStat,M15}.

Some authors have suggested that Einstein's fluctuation theory can be
extended to give expressions for $\langle (\delta T)^2\rangle$,
leading to apparent relations of complementarity between temperature
and energy, similar to position and momentum in quantum
mechanics~\cite{UL99,S95,F88,MF73}. Discussion and measurements of
temperature fluctuations have been given also in~\cite{CSA92,K85}.  In
contrast, others have stressed the contradictory nature of the concept
of temperature fluctuations~\cite{K88}: in the canonical ensemble,
which describes systems in contact with a thermal reservoir, the temperature is a
parameter, so it cannot fluctuate. Mandelbrot has shown that this problem can receive a satisfactory
answer within the framework of estimation theory~\cite{Mand89}. We
summarize here the main points of this discussion, referring
to~\cite{FVVPS11} for more details.

As seen in Sec.~\ref{sec:einst}, in statistical mechanics the
fluctuations of $n$ macroscopic variables $\{\alpha_1 ... \alpha_n \}$
close to their equilibrium values $\{\alpha_1^* ... \alpha_n^*\}$ obey
a multivariate Gaussian distribution with covariance matrix
\begin{equation} \label{eq:var}
\langle \delta \alpha_i \delta \alpha_j \rangle=k_B \Bigl[ {\bf A}^{-1} \Bigr]_{ij}, \, \, 
A_{ij}= - {\partial S \over \partial \alpha_i \partial \alpha_j}\Big|_{ \{ \alpha_k^* \}},
\end{equation}
where $S$ is the entropy. 

The $A_{ij}$ are functions of quantities evaluated at thermodynamic
equilibrium, so that we can write $\delta
S=S(\{\alpha_k\})-S(\{\alpha_k^*\})$ as a function of different
variables. If we can express $S$ as function of $T$ and
$V$~\cite{LandauFisStat}, then
\begin{equation}
\label{eq-6} \delta S= - {C_V \over 2 T^2} (\delta T)^2 + {1 \over 2T} { \partial P \over \partial V}\Big|_T (\delta V)^2,
\end{equation}
where we assume $\delta T=\delta (\partial E/\partial S)$.
This manipulation of variables is misleading if used inside Eq.~\eqref{eq:var}, as it gives
\begin{equation}
  \label{eq-7} \langle(\delta T)^2\rangle={k_B T^2 \over C_V},
  \end{equation}
whose meaning is not clear, although it appears in many textbooks. Let us note that the fluctuation of kinetic energy per particle
is a well defined quantity, but it is conceptually very different from~(\ref{eq-7}),
although apparently rather similar (just a different numerical constant)~\cite{MH16}.

If we insist in considering Eq.~(\ref{eq-7}) valid, using also Eq.~\eqref{II4}, we get
\begin{equation}
  \label{eq-8} \langle(\delta T)^2\rangle\langle(\delta E)^2\rangle=k_B^2T^4,
\end{equation}
or
\begin{equation}   \label{eq-8B}
  \langle(\delta \beta)^2\rangle\langle(\delta E)^2\rangle=1.
\end{equation}
Equation~\eqref{eq-8} and~\eqref{eq-8B} can be interpreted as
``thermodynamic uncertainty relations'' formally similar to the
Heisenberg principle. Some authors discuss a
``thermodynamic complementarity'' where energy and $\beta$ play the
role of conjugate variables~\cite{UL99}.
Let us briefly explain the origin of the
trouble. Eq.~(\ref{eq:var}) holds if $S$ is function of the
macroscopic variables $\{\alpha_k\}$, which are functions of the
microscopic variables. On the contrary, Eq.~(\ref{eq-6}) gives just
$\delta S$ as function of $\delta V$ and $\delta T$, and it is not
related to a probability distribution.  

Other authors, such as Kittel~\cite{K73,K88}, claim that the concept
of temperature fluctuations is misleading. The argument is simple:
temperature is just a parameter of the canonical ensemble, which
describes the statistics of the system, and therefore it is fixed by
definition. In particular, when a system is in equilibrium with a
thermal reservoir, we can have two situations: either we know the
temperature of the reservoir and can describe the energy distribution
of the system; or we do not know the temperature of the reservoir, and
can determine it from the energy distribution of the system. The
latter situation is called the inverse problem. For such a problem we
can use the tools of estimation theory, which makes it possible to use
the available data (in this case a series of energy values) to
evaluate an unknown parameter (in this case $T$). We will see that
Eq.~(\ref{eq-7}) can be somehow considered valid, but now
$\sqrt{\langle (\delta T)^2\rangle}$ must be interpreted as a measure
of the uncertainty on the temperature.

In order to clarify the discussion, let us recall here a few basic concepts from estimation
theory~\cite{C46,K93}. Consider a probability density function $f (x,
\beta)$ of the variable $x$, which depends on the parameter $\beta$,
together with a sample of $n$ independent events $(x_1, \dots, x_n)$,
governed by the probability density $f$, so that the probability
density of the sample is
\begin{equation}
\label{like} L (x_1, \dots, x_n, \beta) = \prod_{i=1}^n f (x_i, \beta) .
\end{equation}
We would like to estimate the unknown parameter $\beta$ from the
values $\{x_i\}$. For this purpose we have to define a suitable
function of $n$ variables, $\widehat{\beta}(x_1, \dots, x_n)$, to
obtain the estimate of $\beta$ from the available information. The
quantity $\widehat{\beta}$ is, by construction, a random variable. We
call $F (\widehat{\beta}, \beta)$ its probability density, which of
course depends upon the parameter $\beta$. We can calculate, for
instance, its expected value and its variance. When $\langle
\widehat{\beta} \rangle = \beta$, one says that $\widehat{\beta}$ is
an unbiased estimate of $\beta$. It is clear that the usefulness of an
estimating function is tightly linked to its variance.

Given certain general conditions of regularity, the Cram\'er-Rao
inequality~\cite{C46} for unbiased estimators can be established:
\begin{equation}
\label{cramer-rao} \int\!\Big( \widehat{\beta} - \beta \Big)^2 F(\widehat{\beta}) d \widehat{\beta} \geq \left \{ n\!\int\! \Big(\frac{\partial}{\partial \beta} \ln f(x,\beta)\Big)^2 f(x,\beta) dx \right \}^{-1},
\end{equation}
where the quantity in the braces on the right hand side of
Eq.~\eqref{cramer-rao} is known as the Fisher information~\cite{C46}:
it gives a measure of the maximum amount of information we can extract
from the data about the parameter to be estimated. This inequality
puts a limit on the ability of making estimates, and also suggests
that the estimator should be chosen by minimizing the inequality. When
the variance of $\widehat{\beta}$ is the theoretical minimum, the
result $\widehat{\beta}$ is said to be an ``efficient
estimate''~\cite{C46}. Here we follow the convention of distinguishing
between an efficient estimate, which has minimum variance for finite
$n$, and an asymptotically efficient estimate, which has minimum
variance in the limit $n \to \infty$.

It is now useful to relate fluctuations of measurements of energy and fluctuations of estimates of temperature. For this purpose, let us consider a gas of $N$ classical
particles. For simplicity, we begin by measuring the
energy $u$ of a single particle, whose probability distribution we
write as
\begin{equation}
\label{gibbs-2} P (u, \beta) = \frac {g(u) \exp(-\beta u)} {Z (\beta)},
\end{equation}
where the parameter $\beta$ is $1/k_B T$ and the density of single
particle states $g(u)$ is assumed to be known. Suppose that we have
measured $n$ independent values of particle energy $(u_1, \dots,
u_n)$. We can write
\begin{equation}
\label{gibbs-n} P (u_1, \dots, u_n, \beta) = \frac {g(u_1) \exp(-\beta u_1)} {Z (\beta)} \cdots \frac {g(u_n) \exp(-\beta u_n)} {Z (\beta)},
\end{equation}
or \begin{align} P (u_1, \dots, u_n, \beta) & = \frac {g(u_1) \cdots g(u_n) } {G_0(U)} \frac {G_0(U) \exp(-\beta U)} {Z^n (\beta)} \\ & \equiv h (u_1, \dots, u_{n-1} \vert U) P(U, \beta),
\end{align}
where $U = \sum_{i=1}^n u_i$ and $G_0(U)$ is given by
\begin{equation}
\label{GdU} G_0(U) = \!\int\! g(u_1) \cdots g(u_n) \delta\!\left(\sum_{i=1}^n u_i - U \right) du_1 \cdots du_n.
\end{equation}
Because $P(U, \beta)$ is the probability density of measuring a total
energy $U$ in $n$ independent single particle energy measurements, we
see that $h (u_1, \dots, u_{n-1} \vert U)$, which is the conditional
distribution of the energy in the sample given the total measured
energy, does not depend on $\beta$. We conclude that good estimators
of $\beta$ can be constructed as a function of the sum of the measured
energies.

A possible choice of temperature estimator is the maximum likelihood estimator determined by the condition
\begin{equation}
\label{mle-micro} - \frac {\partial }{\partial \beta} \ln Z^n (\beta) \Big \vert_{\widehat{\beta}_{\rm MLE}} = \sum_{i=1}^{n} u_i,
\end{equation}
which - for the case of density of states $g(u) \propto u^{\eta}$ - reads
\begin{equation}
  \widehat{\beta}_{\rm MLE} = \frac{n (\eta +1)}{U},
  \end{equation}
which can be demonstrated not to be an unbiased estimate. 
In general, for large $n$, the values of $\widehat{\beta}$ extracted from
Eq.~(\ref{mle-micro}) are normally distributed around the true value
$\beta$, with variance $1/(n \sigma^2_u)$ where $\sigma^2_u$
is the variance of the single-particle energy calculated with the true
$\beta$. Therefore $\widehat{\beta}$ is asymptotically efficient.

Another estimator for $\beta$ is given by 
\begin{equation}
\label{beta-max} \widehat{\beta}_G = \frac {\partial }{\partial U} \ln G_0(U).
\end{equation}
Unlike the maximum likelihood estimator, $\widehat{\beta}_G$ is an
unbiased estimator of $\beta$ for any $n$, but like the maximum
likelihood estimator, it is not an efficient estimator for finite
$n$. It becomes asymptotically efficient when the density of states
$g(u) \propto u^{\eta}$ is considered~\cite{S73,P77}.

An important point of the preceding discussion is that, due to the
exponential form of the canonical ensemble probability density, all 
the information about $\beta$ is contained in the total energy of an
isolated sample. We gain nothing by knowing the distribution of this
energy among the $n$ elements of the sample. We say that $U= \sum_{i=1}^n
u_i$ is sufficient for estimating $\beta$. Therefore we may also argue
as follows. Instead of $n$ measurements of the molecular energy, we make one
measurement of the energy $E$ on the macroscopic system with density
$P(E, \beta) = G(E) \exp(-\beta E)/Z_N (\beta)$, $G(E)$ being the density
of states of the entire system, which reduces to $G_0(E)$ for systems
made of non-interacting components. The Cram\'er-Rao inequality
becomes
\begin{equation}
\label{CR-macro} \int\! \Big( \widehat{\beta} - \beta \Big)^2 F(\widehat{\beta}) d \widehat{\beta} \geq \frac {1} {\sigma^2_E},
\end{equation}
where $\sigma^2_E$ is the variance of the canonical energy of the macroscopic body.
For an ideal gas of $N$ identical particles, $\sigma^2_E = N \sigma
^2_u$, and Eq.~\eqref{CR-macro} becomes $\sigma^2_{\widehat{\beta}}
\geq 1/N \sigma^2_u$. With regard to the determination of $\beta$, a
single value of the macroscopic energy contains the same information
as $N$ microscopic measurements.

We know that a non-ideal gas of $N$ identical particles with
short-range interparticle interactions behaves (if not at a phase
transition) as if it were composed of a large number, $N_{\rm eff}
\propto N$, of (almost) independent components, and $\sigma^2_E
\approx N_{\rm eff}\,\sigma^2_c$, where $\sigma ^2_c$ is the variance
of one component. For instance, consider a system of $N$ particles in
a volume $V$ with a correlation length $\ell = (c V/N)^{1/3}$, where
$c\gg 1$ indicates strong correlations. We have $N_{\rm eff}\sim
V/\ell^3= c^{-1} N$. Thus, even if $n=1$ in Eq.~(\ref{CR-macro}), that
is, we perform a single measurement of energy, the variance of $E$,
which is the energy of a macroscopic system, is extensive and the
variance of $\widehat{\beta}$ may be small. We have
$\sigma^2_{\widehat{\beta}} \geq 1/N_{\rm eff}\, \sigma^2_c$, with
$N_{\rm eff} \propto N \gg 1$. By looking at $E$ as the result of
$N_{\rm eff}$ elementary energy observations, our preceding
considerations can be applied here with $N_{\rm eff}$ playing the role
of $n$. In particular, the asymptotic properties for large $N_{\rm
  eff}$ of the two estimators are preserved, and the estimates of
$\beta$ obtained by the two
expressions \begin{subequations} \label{mle-E} \begin{align} & - \frac
    {\partial }{\partial \beta} \ln Z_N (\beta) \Big
    \vert_{\widehat{\beta}_{\rm MLE}} = E,\\ \noalign{\noindent and} &
    \widehat{\beta}_G = \frac {\partial }{\partial E} \ln G(E)
\end{align}
\end{subequations} 
approach the same value for $N_{\rm eff} \gg 1$, a condition that is
verified for macroscopic bodies. Therefore, for a macroscopic system,
we can obtain a good estimate of $\beta$ even with a single
measurement of its energy.

In conclusion we have given a justification, in terms of estimation theory, of the
definition of temperature in statistical mechanics either
in the canonical or microcanonical ensemble by means of
Eq.~(\ref{mle-E}). From our discussion we see that the fluctuations of
the random variables $\widehat{\beta}_{\rm MLE}$ and
$\widehat{\beta}_{G}$ when $n \gg 1$ are approximately Gaussian with a
variance $1/(n \sigma^2_u)$. The fluctuations of the total energy of
the sample $U = \sum_{i=1}^n u_i$ also become Gaussian (by the central
limit theorem) with variance $n \sigma^2_u$. Therefore, in this limit,
we have $\sigma^2_{\widehat{\beta}}\,\sigma^2_U = 1$.

Let us notice that, although the
Rao-Cram\'er inequality, Eq.~(\ref{CR-macro}), is formally similar to
Eq.~(\ref{eq-8}), which was obtained by an incorrect use of Einstein's
fluctuation formula, the analogy is inexact and misleading. In
mathematical statistics the quantity $\sigma^2_{\widehat{\beta}}$
measures the uncertainty in the determination of the value of $\beta$
and not the fluctuations of its values.

\subsubsection{About temperature uncertainty in small systems}

It is useful to illustrate the above ideas by means of a mechanical
model for a thermometer~\cite{FVVPS11}.  A box is filled with $N$
non-interacting particles of mass $m$. On the top of the box there is
a piston of mass $M$ which can move without friction in the $\hat{x}$
direction, see Fig.~\ref{pist}. Although the box is three-dimensional, only the motion in
the $\hat{x}$ direction is relevant because we assume that the
particles interact only with the piston. The other directions are
decoupled from $\hat{x}$, independently of their boundary
conditions. The one-dimensional Hamiltonian of the system is
\begin{equation} \label{hpiston}
{\cal H}=\sum_{i=1}^N\frac{p_i^2}{2m}+\frac{P_M^2}{2M}+FX,
\end{equation}
where $X$ is the position along the $\hat{x}$ axis of the piston, and
the positions of the particles $x_i$ along the same axis are
constrained to be between 0 and $X$. A force $F$ acts on the piston,
and in addition there are elastic collisions of the gas particles with
the piston. The particles exchange energy with a thermostat at
temperature $T$ placed on the one side of the box at $x=0$
(see~\cite{FVVPS11} for details of its computational implementation).  In the
following we set $k_B=1$, which is equivalent to measuring the
temperature in units of $1/k_B$.

\begin{figure}
  \begin{center}
    \includegraphics[width=7cm]{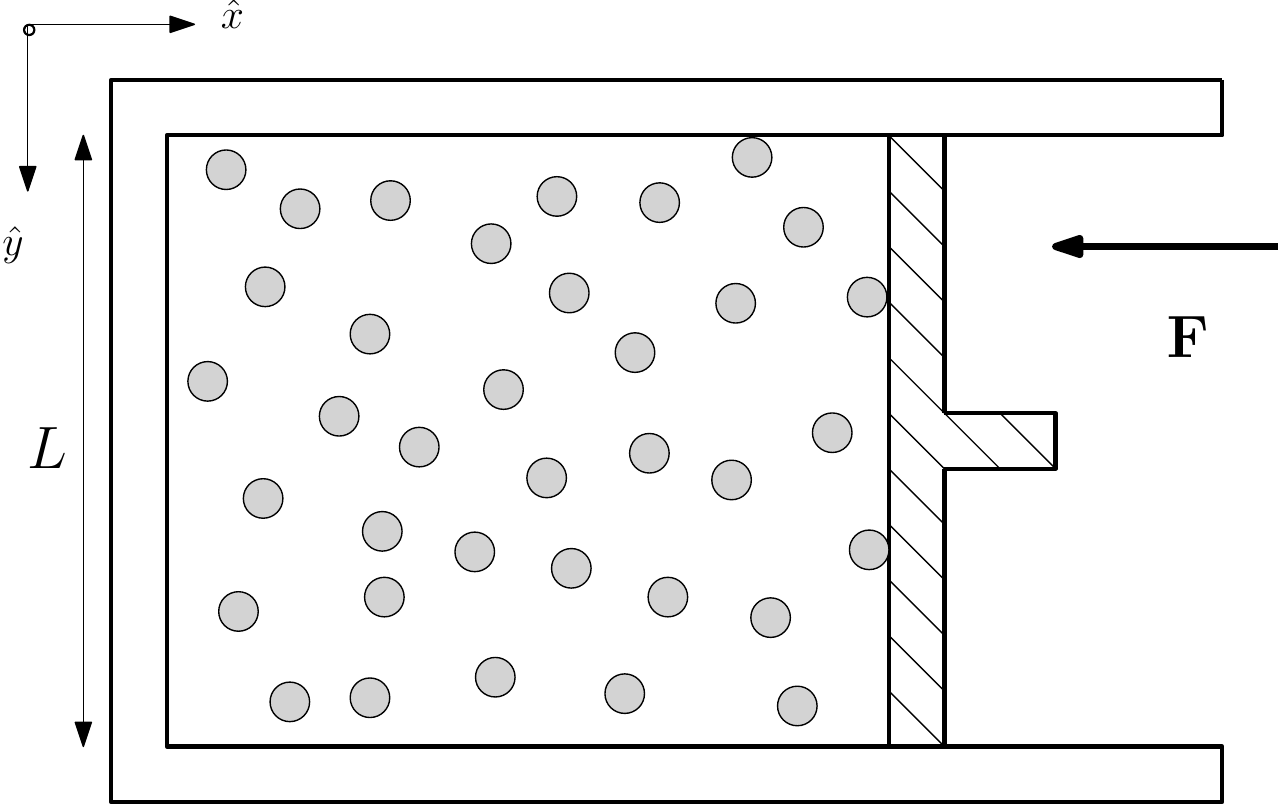}
  \end{center}
  \caption{\label{pist}Sketch of the mechanical model for a thermometer, see Eq.~\eqref{hpiston}}
  \end{figure}

It is straightforward to compute the
mean value of the piston position
\begin{equation}
\langle X\rangle=\frac{(N+1)T}{F}. \label{aveposition}
\end{equation}
As a consequence of this formula, an unbiased estimator for the temperature of the thermostat is
\begin{equation}
\hat{T}=\frac{F\hat{X}}{N+1},
\end{equation}
where $\hat{X}$ is an estimate of the average piston position. We use for it
\begin{equation}
\hat{X}_{\mathcal{N}}= (X^{(1)}+\ldots +X^{(\mathcal{N})})/\mathcal{N}
\end{equation}
where $X^{(1)},\ldots,X^{(\mathcal{N})}$ are $\mathcal{N}$ independent measurements
of the piston position. The probability distribution
function of the piston position is
\begin{equation}
P(X)=\frac{1}{N!}(\beta F)^{N+1} X^N e^{-\beta FX}, \label{pdfX}
\end{equation}
which is an infinitely divisible distribution~\cite{gned-kol}, so that
the variable $\hat{X}_{\mathcal{N}}$ has a probability distribution
function of the same shape, with $N$ replaced by $N\mathcal{N}$. The
variance of $\hat{X}_{\mathcal{N}}$ is
$\sigma^2_{\hat{X}}/\mathcal{N}$, because the values of $X$ are
independent. From Eq.~\eqref{pdfX} we have also that
$\sigma^2_{\hat{X}}=(N+1)/\beta^2 F^2$, and
therefore
\begin{equation}
\sigma^2_{\hat{X}_{\mathcal{N}}}=\frac{1}{\mathcal{N}}\frac{N+1}{\beta^2 F^2}. \label{varposmean}
\end{equation}
This gives the following variance for the temperature estimator $\hat{T}$ obtained by using $\hat{X}_{\mathcal{N}}$ as estimator of $\hat{X}$:
\begin{equation}
\sigma^2_{\hat{T}} = \frac{T^2}{\mathcal{N} (N+1)}, \label{estimator-var} 
\end{equation}
which is exactly the Cram\'er-Rao lower bound, i.e. $\hat{T}$ is
an unbiased and efficient estimator for every $\mathcal{N}$~\cite{FVVPS11}. In
particular, Eq.~\eqref{varposmean} shows that the variance of the
estimator is of order $\sim 1/N$, which can be non-negligible for
single measurements on small systems, but it can be arbitrarily
reduced by increasing the number $\mathcal{N}$ of
measurements.

In a real experiment one cannot be sure that the
$\mathcal{N}$ measurements are independent.  In general, the data are
correlated, and a correlation time $\tau$ must be estimated
numerically. A simple and natural way is to look at the shape of the
correlation functions of the observables of interest. If the distance
in time $\delta t$ between two successive measurements is smaller than
$\tau$, the effective number of independent measurements is
approximately $\mathcal{N}_{\rm eff}=\mathcal{N}\delta t/\tau$. By
plotting $N\sigma^2_{\hat{T}}$ versus $\mathcal{N}_{\rm eff}$ we
expect that the dependence on $N$ disappears, resulting in a collapse
of the curves: this is exactly what happens in Fig.~\ref{figT}, where
we have used for $\tau$ the minimum time such that the autocorrelation
of piston's position $\left\vert C_{X}(t) \right\vert< 0.05$, where
\begin{equation}
C_{X}(t)=\frac{\langle\delta X(t)\delta X(0) \rangle}{\langle\delta X^{2}(0)\rangle}.
\end{equation}

\begin{figure}[h!]
  \begin{center}
    \includegraphics[width=12cm,clip=true]{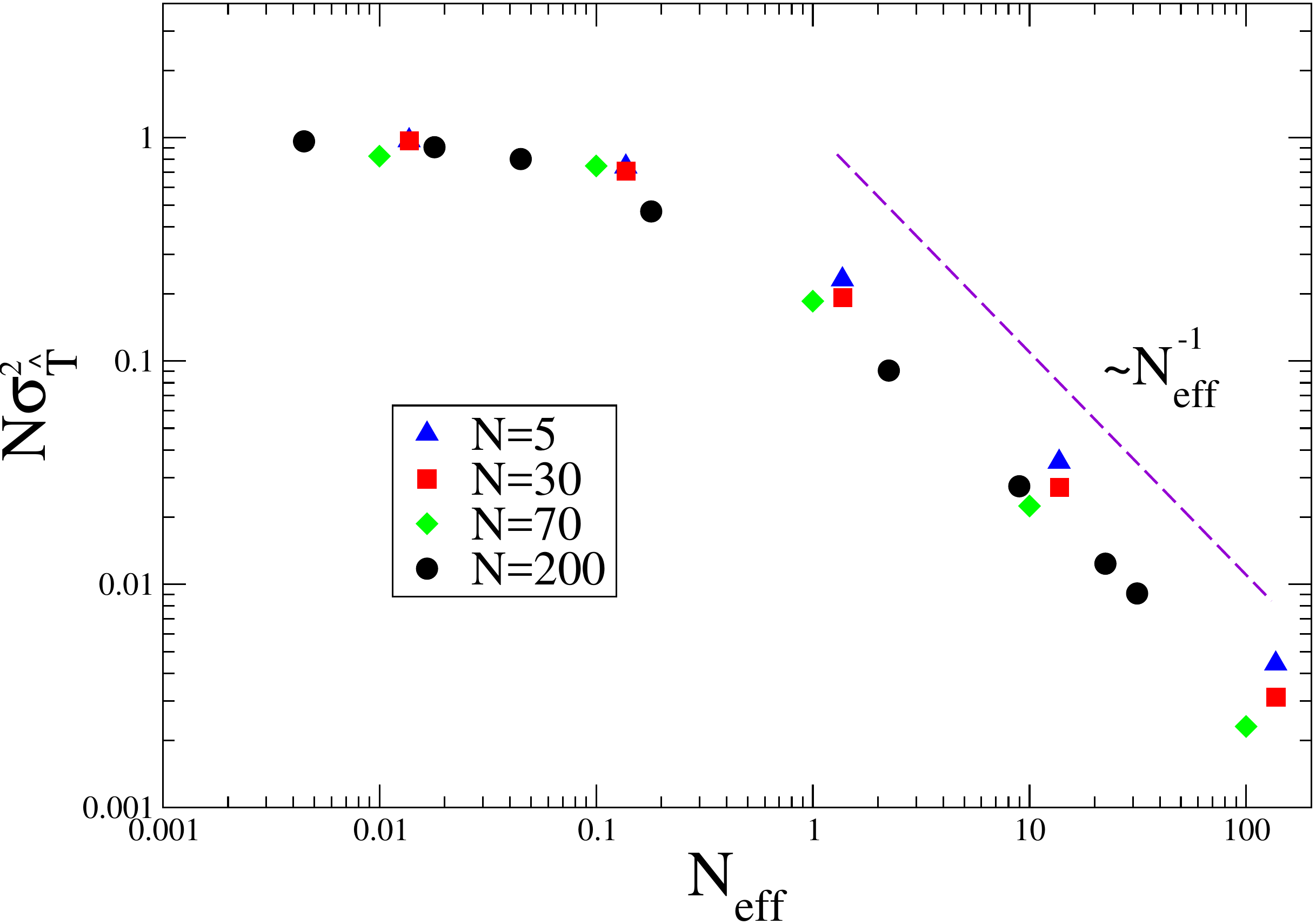}
    \end{center}
\caption{\label{figT}The quantity $N \sigma^{2}_{\hat{T}}$ for different values of $N$ is numerically calculated and plotted as function of $\mathcal{N}_{\rm eff}=\mathcal{N}\delta t/\tau$. For large times, the uncertainty goes to zero as $1/\mathcal{N}_{\rm eff}$. The parameters are $\delta t=0.01$, $M=10$, $m=1$, $F=10$, $T=1$, $N=5$, 30, 70, and 200. It is clear that for the uncertainty of $T$, the relevant quantity is $\mathcal{N}_{\rm eff}$ which depends both on $N$ and $\tau$.}
\end{figure}

\subsection{Small systems: models and stochastic thermodynamics} 

In recent years, the development of experimental techniques allowing
us to observe and even manipulate objects on mesoscopic (nano and
micro) scales, paved the way to the investigation of physical systems
composed by a small number of elementary
constituents~\cite{Haw06}. This ``middle world'' includes several
interesting topics, such as macromolecules (DNA, proteins and molecular
motors)~\cite{R07}, colloidal suspensions~\cite{CRWSSE}, granular
media~\cite{puglio15} and active matter~\cite{marchetti2013}, just to
name a few.  Several results obtained in the last decades proved that the
tools and the concepts introduced in statistical mechanics can predict
with success the behaviour of small systems.

This new range of applicability exceeds the original intent of
statistical physics, i.e. the justification of thermodynamics from the
microscopic level, and introduces new perspectives.
However, extending statistical mechanics to small systems is a very
delicate issue, that can generate misleading conclusions, when treated
without the required care. From a theoretical perspective, a central
issue is the possibility to use the ergodic hypothesis for these systems. Here, we
identify three classes of small systems for which a statistical
approach seems to be still meaningful. In the following, we restrict
our attention, for the sake of consistency and simplicity, to
classical systems: in fact, the description of quantum small systems,
even with some similarities, must be developed independently, since it
presents a very different phenomenology and a huge number of technical
and conceptual subtleties~\cite{talkner2007fluctuation}. \\

\emph{Single (or very few) particle systems.} A Hamiltonian system
composed of a single particle in one dimension that moves in an
external potential is trivially ergodic.  In fact, in the phase
space of such a system (which is two-dimensional) the ``constant
energy surface'', where the system evolves, coincides with
the trajectory of the system.  This is not necessarily true when we
increase the number of particles, since it is possible to exhibit
systems composed of two constituents which are no longer ergodic
(e.g. two coupled harmonic oscillators).  Nevertheless there are many
examples of few-dimensional ergodic systems and several authors claim
that a notion of thermodynamics is possible also for this class of
systems~\cite{Dunkel2014,Hilbert2014}.  It is very interesting to note
that one of the first historical efforts to derive thermodynamics from
 mechanics focused on isolated single particle systems,
 for which Helmholtz~\cite{Helmholtz1884} derived a theorem (on the monocycle) later used by Boltzmann to introduce the ergodic hypothesis: in Sec.~\ref{monocycle} this topic
has been discussed in detail. \\

\emph{Small systems with stochastic dynamics.} This class includes
systems with a non-deterministic dynamics, such as Markov chains or
systems ruled by stochastic differential equations, like the following
Langevin equation
\begin{equation}\label{eq:sde_proto}
\ddot{x}=-\frac{d V(x)}{dx} -\gamma\dot{x} +\sqrt{2D}\eta,
\end{equation}
where $\eta$ represents white noise, $\langle
\eta(t)\eta(t')\rangle=\delta(t-t')$, $V(x)$ is an external potential
and $\gamma$ and $D$ are constants.  The proof of the validity of ergodicity
for Markov stochastic systems is far more accessible than its
counterpart for deterministic systems: in particular it can be proven
that every irreducible and non-periodic Markov process is ergodic
\cite{G90,norris} and that, in addition, it also satisfies the
so-called {\it mixing} condition, i.e. the fact that any (non-pathological) initial
probability distribution $\rho_0({\bf X})$ of the system converges toward an asymptotic
stationary state:
\begin{equation}
\rho({\bf X},t)\overset{t \to \infty}{\longrightarrow} \rho({\bf X}).
\end{equation}
Moreover, for these systems, the typical relaxation times are usually
small, and analytical approaches are possibile. These are some of the
reasons why single particle stochastic systems have been extensively
used in the most recent developments on stochastic
thermodynamics~\cite{seifertrev}.

Another conceptual advantage of small stochastic systems
over deterministic ones is the fact that the aleatory
dynamics can be interpreted as the effect of the interaction of the
system of interest with a large environment. Consider a subsystem of a larger
isolated system, its state is a many-to-one function: $(\bold{X},\bold{Y})\in
\mathbb{R}^{2d(N_1+N_2)}\to \bold{Y}\in \mathbb{R}^{2dN_1}$, with $N_1\ll
N_2$, where $\bold{Y}$ is the variable representing the system of interest,
$N_1$ and $N_2$ are, respectively, the number of particles of the
system and of the environment, and $d$ is the spatial dimension. This
means that many different configurations of the environment correspond
to the same $\bold{Y}$, allowing, at least heuristically, to
interpret the state of the small subsystem as a ``thermodynamical
observable''. Of course, in order to obtain a stochastic description
for the subsystem, one has to assume that fluctuations induced by the
environment are not too large: this condition can be realized, for
instance, when the mass of the system particles is larger than that of
the particles of the environment. \\

\emph{Many-Particle Small Systems (MPSS).} 
By this term we mean systems composed of $\mathcal{O}(10-10^2)$
particles, where the interesting collective typical behaviors of
large systems can be observed and that, at the same time, are far
enough from the infinite-$N$ limit, to exhibit several features
of small systems (large fluctuations, non-standard equilibrium states,
etc...). The question regarding the ergodicity of these systems may
appear more difficult to answer with respect to single-particle or
stochastic systems.
However, from a physical and practical point of view, it is important
to stress that the Khinchin argument for the validity of the ergodicity in systems with $N \gg 1$ for a class of collective observables applies to MPSS.
In fact, even if $N$ is not infinite, many large-$N$ approximations
(e.g. Laplace approximation for the computation of integrals) can be carried on safely,
and discrepancies with the asymptotic behavior are small,
and do not affect the overall phenomenology~\cite{khinchin49}. Regardless,
since $N$ is finite, the fluctuations are visible
and non-negligible, and indeed their study is the main focus of
stochastic thermodynamics.
  
Let us note that MPSS includes the class of systems usually studied in
numerical simulations: in fact, typical computer simulations involve
systems with a number of particles spanning from the few hundreds up
to some tens of thousands. These numbers are clearly much
smaller than the typical size of macroscopic systems
$\mathcal{O}(10^{20})$, nevertheless they are commonly considered
large enough for the study of the properties of macroscopic real systems.

\subsubsection{Stochastic thermodynamics}

As already stressed, the distinguishing feature of small systems is the relevance of
fluctuations, which are negligible only when the number of
constituents is very large, as for macroscopic bodies. The study of
fluctuations of thermodynamics functions, such as energy or entropy,
goes back to Einstein, Onsager and Kubo, but it has recently raised a
renewed interest with the establishing of important results in
response theory~\cite{BPRV08} and in the so-called stochastic
thermodynamics~\cite{VDB85,GC,sekimoto98,LS99,Sekimoto2010,esposito2010,seifertrev}.
The main aim of this theory concerns the statistical
properties of fluctuations in systems which are {\em far} from
thermodynamic equilibrium. Here we briefly give some definitions and
describe the principal conceptual aspects, referring the interested
reader to the review~\cite{seifertrev} and references therein, for an
exaustive presentation of the subject.


The first issue consists in the proper microscopic definition of
work and heat for small systems. In particular, as we will show in the
rest of this Section, the naive definition of infinitesimal work as
$dW=p dV$ (being $p$ the pressure and $V$ the volume), is not
appropriate in a large variety of situations, e.g. when the external
varied parameter is not the volume.  In fact this definition comes
from a basic application of the classical definition of work ($dW={\bf
  F \cdot dx}$) and pressure (${\bf F\cdot \hat n}=p S $, where $S$ is
the surface): in systems beyond the usual applications of
thermodynamics the definition must be reconsidered. We will first
discuss the definition in the context of Hamiltonian systems.

Consider a time-dependent Hamiltonian $\mathcal{H({\bf X}},t)$ and let $
\{{\bf \widetilde{X}}(t)\}_{t=0}^{\mathcal{T}}$ be a solution of Hamilton equations of
motion; then, the time derivative of a generic function $A(t)=A({\bf
  \widetilde{X}}(t),t)$ reads
\begin{equation}\label{eq:liouville}
\frac{d A(t)}{d t}=\frac{\partial A({\bf X},t)}{\partial {\bf X}}\dot{{\bf X}}+\frac{\partial A({\bf X},t)}{\partial t}=\left\{A,\mathcal{H}\right\}+\frac{\partial A}{\partial t}.
\end{equation}
It is natural to identify the internal energy of the system
with the value of the Hamiltonian $E(t)=\mathcal{H}({\bf \widetilde{X}} 
(t),t)$ and compute its time derivative via Eq. \eqref{eq:liouville}
\begin{equation}
\frac{d E(t)}{dt}=\left\{\mathcal{H},\mathcal{H}\right\}+\frac{\partial \mathcal{H}}{\partial t}=\left.\frac{\partial \mathcal{H}}{\partial t}\right|_{\widetilde{\bf X}(t)}.
\end{equation}
Therefore, if the Hamiltonian depends on an external varying parameter, the total energy of the system changes and it
is not difficult to see that, from a thermodynamic point of view, this
change corresponds to the time derivative of the work $\dot{W}(t)$,
leading to
\begin{equation}\label{def1}
\dot{W}=\frac{\partial \mathcal{H}}{\partial t}.
\end{equation}
It is important to remark that this convention implies
that, whenever the energy of the system increases, the work 
is positive, whereas, in the opposite case, the work is
negative: this is in contrast with
the usual thermodynamic convention, but has a far more transparent
energetic interpretation.

Suppose now that the explicit time dependence of the Hamiltonian is
given by an additional term, namely $\mathcal{H}=\mathcal{H}_0+h(t)$,
with $h(t)=\lambda(t)X_i$, $X_i$ being one of the component of the
phase space variable ${\bf X}$ and $\lambda(t)$ a time-dependent
parameter; in this case one has
\begin{equation}
\dot{W}(t)=\widetilde{X}_i(t) \dot{\lambda}(t),
\end{equation} 
which is equivalent to $dW=Vdp$, but differs from the standard
definition $dW=-p dV$. The explanation of this result was reported in
Ref.~\cite{jarzynski2007comparison}: assume that, 
generalizing the above situation, the Hamiltonian depends
upon a certain number of external parameters $\lambda_k(t)$,
with ${k=1,\ldots,M}$: 
\begin{equation}
\mathcal{H}=\mathcal{H}_0({\bf X})+\sum_k\lambda_k(t)g_k({\bf X}),
\end{equation}
where $g_k({\bf X})$ are functions of ${\bf X}$, which define the
macroscopic states.  In this case one can apply an alternative
definition of internal energy, $E(t)=\mathcal{H}_0(t)$, whose time
derivative reads
\begin{equation}
\frac{d \mathcal{H}_0}{dt}=\{\mathcal{H}_0,\mathcal{H}\}=-\sum_k \lambda_k(t)\{g_k,\mathcal{H}\}=-\sum_k\lambda_k(t)\dot{g}_k({\bf X}(t)).
\end{equation}
Therefore, using this definition where all the time-dependent terms
are considered {\it external}, the work
takes the form
\begin{equation}\label{def2}
\dot{W}=-\sum_k\lambda_k(t)\dot{g}_k({\bf X}(t)).
\end{equation}
In particular, when $k=1$ and $g({\bf X})=X_i$, we
recover the usual thermodynamic definition of work
\begin{equation}
\dot{W}(t)=-\lambda(t) \dot{X}_i,
\end{equation}
that is equivalent to $dW=-p dV$. It is important to remark that this
definition gives a nonzero work also when the external parameters are
fixed, $\dot{\lambda}=0$. This occurs because some energy may be
exchanged between the internal energy $\mathcal{H}_0$ and the external
terms appearing in the Hamiltonian, despite the fact that the value of
the Hamiltonian $\mathcal{H}({\bf X}(t))$ does not change over
time. In the rest of this section we will use the first definition of
work, Eq.~\eqref{def1}.

It is not possible to define the heat exchange in the context 
of purely Hamiltonian systems, since these systems do not
transfer energy with the external environment. Moreover, 
in general, there are several different ways to define the 
interaction of a system with an external thermostat at temperature
$T$ (stochastic, deterministic, etc\ldots). For this reason,
to keep the discussion on general terms, it is important
to give a definition of heat that does not rely on the
specific model. In the following we will discuss
this topic in the context of coarse-grained stochastic 
differential equations.

When a time-dependent Hamiltonian system is coupled to an external
thermostat, the variation of the total energy $E(t)=\mathcal{H}({\bf
  X}(t),t)$ is due to different causes: indeed, a part of energy is
funneled through the external parameter that varies in time, whereas
another part is exchanged with the thermostat to which the system is
attached. It is customary to denote by the term heat all the energy
that is not exchanged in ``Hamiltonian manner'', i.e.
\begin{equation}\label{eq:heat}
\dot{Q}(t)=\dot{E}(t) -\dot{W}(t),
\end{equation}
where $\dot{W}(t)$ is the quantity defined in Eq.~\eqref{def1}. This
last equation is the microscopic equivalent of the first
principle of thermodynamics, when the appropriate sign convention
is chosen. From a practical point of view, there may be some
ambiguity in the definition of $\dot{Q}$ in the context of
stochastic differential equations: in particular, as we will see
later, it is very important to specify whether the derivative 
is taken according to the It\^o or the Stratonovich convention.
To avoid such ambiguities it can be useful to report the
integrated version of Eq. \eqref{eq:heat}:
\begin{equation}
Q(\mathcal{T})=\Delta E - \int_0^\mathcal{T} \frac{\partial \mathcal{H}(t)}{\partial t} dt,
\end{equation}
where $\mathcal{T}$ is the total time of the measurement, and
$\Delta E=E(\mathcal{T})-E(0)$ is the total energy variation in such
an interval. This last equation does not present any ambiguity, since
the integration variable of the integral appearing on the r.h.s.  is
the time $t$ (and not the phase space position ${\bf X}$), and
therefore $E(t)$ is a well-defined function of time.


Analogous definitions of heat and work can be given in the framework
of stochastic differential equations. In particular, following
Ref.~\cite{Sekimoto2010}, we will examine the simple case of a
unidimensional Brownian particle, with mass $m$, position $x$ and velocity $v$,
in contact with a thermal bath at temperature $T$, and moving in an
external time-dependent potential $V(x,t)$.  The stochastic
differential equation describing this system in the underdamped regime
reads
\begin{eqnarray}\label{eq:underdampedsde}
  \dot{x}&=&v\nonumber\\
  \dot{v}&=&-\gamma v - \frac{\partial_xV(x,t)}{m} + \sqrt{\frac{2k_B T \gamma}{m}}\eta,
\end{eqnarray}
where $\eta$ is white noise with $\langle 
\eta(t)\eta(t')\rangle=\delta(t-t')$ and $\gamma$ a constant parameter.


From the natural choice for the energy of the particle $E(t)$
 \begin{equation}
E(t)=\frac{1}{2} m v(t)^2 + V(x(t),t),
\end{equation}
there follows that the work
performed on the system is
\begin{equation}\label{eq:def_work_stoc}
\dot{W}(t)=\left.\frac{\partial V(x,t)}{\partial t}\right|_{x=x(t)}.
\end{equation}
Those definitions, despite their apparent coherence,
conceal some inconsistencies due to the fact that 
Eq.~\eqref{eq:underdampedsde} is a coarse grained equation.
It is very easy to show that, in the simple case $\partial_t V=0$,
the (equilibrium) invariant distribution of the stochastic system is 
\begin{equation}\label{eq:dist}
\rho(x,v)\propto\exp\left\{-\beta\left(\frac{m}{2} v^2 +V(x)\right)\right\}.
\end{equation}
On the other hand, when considering $x$ as a single
component of a much larger Hamiltonian system
${\bf X}=(x,v,x_1,v_1,x_2,v_2\ldots,x_N,v_N)$
with Hamiltonian $\mathcal{H}({\bf X})$, we have, at equilibrium,
\begin{equation}\label{eq:dist_single}
\rho^{eq}(x,v)=\int \frac{dx_1dv_1\ldots dx_N dv_N}{Z} \,\,e^{-\beta \mathcal{H}({\bf X})},
\end{equation}
where $Z$ is the partition function. Whenever the
Hamiltonian can be split into the sum of a kinetic term of the
particle and another part involving all the remaining degrees of
freedom of the system,
\begin{equation}
\mathcal{H}({\bf X})=\frac{mv^2}{2}+\mathcal{H}_{int}(x,x_1,v_1,\ldots,x_N,v_N),
\end{equation}
it is possible to recast Eq. \eqref{eq:dist_single} in the following form
\begin{equation}
\rho^{eq}(x,v)= \frac{Z(x)}{Z_0}e^{-\beta m \frac{v^2}{2}},
\end{equation}
where $Z(x)=\int \exp\{-\beta \mathcal{H}_{int}\} dx_1 dv_1\ldots dx_N dv_N$, 
and $Z_0$ is the normalizing constant. Therefore, by comparing the last
expression with Eq. \eqref{eq:dist}, it is easy to see that
\begin{equation}
V(x)= -\frac{1}{\beta}\ln Z(x),
\end{equation}
which is interpreted by many authors (see~\cite{Sekimoto2010}) as a
conditional free energy function, rather than a proper potential
energy. The simplest case is when
$\mathcal{H}_{int}(x,x_1,v_1,\ldots,x_N,v_N)$ can be split in an external potential $V(x)$ depending only on $x$, and another
contribution with all the remaining variables. 

The above considerations show that the microscopic definition,
Eq.~\eqref{def1}, and the coarse-grained one
Eq.~\eqref{eq:def_work_stoc}, are, in general, not
equivalent. Therefore one should take some care in doing energetic
considerations when starting from a Langevin equation, without having
an underlying microscopic description of the system~\cite{CP15}.


When Eq.~\eqref{eq:def_work_stoc} is adopted, the heat $Q$ is the 
difference between energy and work (first principle of thermodynamics).
An explicit formula for heat can be readily obtained:
\begin{eqnarray}\label{strato_no_av}
\dot{Q}&=&\frac{d E}{dt} - \dot{W}=(\partial_v{E})\dot{v}+(\partial_xE)\dot{x}\nonumber \\
&=&mv\left(-\gamma v -\frac{\partial_x V}{m} +\sqrt{\frac{2\gamma k_BT}{m}}\eta\right)+ v\partial_x V \nonumber\\
&=&m\left(-\gamma v +\sqrt{\frac{2\gamma k_BT}{m}}\eta\right)v,
\end{eqnarray}
where, since we applied the usual calculus rules, the differential
equation must be interpreted with the Stratonovich
convention~\cite{R89,G90}. Nevertheless, in order to compute 
$\langle \dot{Q} \rangle$, it is useful to derive the equivalent result with the It\^o
convention:
\begin{equation}\label{ito_no_av}
\dot{Q}=m\left(-\gamma v +\sqrt{\frac{2\gamma k_BT}{m}}\eta\right)v+\gamma k_BT.
\end{equation}
Of course, the average of Eq.~\eqref{ito_no_av} and~\eqref{strato_no_av} must give the same result, the It\^o
expression being more explicit:
\begin{equation}\label{ito}
  \langle \dot{Q}\rangle = -\gamma m \langle v^2 \rangle +k_BT \gamma=
  -2\gamma\left(\frac{m\langle v^2\rangle}{2}-\frac{k_BT}{2}\right).
\end{equation}
It is clear that the quantities $\dot{Q}$ and $\dot{W}$, and thus
the integrated heat and work, are fluctuating quantities 
because they depend upon
the single trajectories $\{x(t),v(t)\}_{t=0}^{\mathcal{T}}$.


The stochastic equation \eqref{eq:underdampedsde} gives a description
of the system at the level of single trajectories: naturally, an
equivalent description can be given in terms of the probability
distribution function $\rho(x,v,t)$, i.e. the probability density of
finding the particle at time $t$ with position $x$ and velocity $v$.
The time evolution of this quantity is given by a partial differential
equation, the Fokker-Planck equation \cite{G90}:
\begin{equation}
\partial_t \rho(x,v,t)= -{\bf \nabla \cdot J}(x,v,t),
\end{equation}
where the two components of the current ${\bf J}=(J_x,J_v)$ are
\begin{eqnarray}
J_x&=&v \rho(x,v,t),\\
J_v&=&-\gamma v \rho(x,v,t) -\frac{\partial_x V}{m} \rho(x,v,t) -\frac{k_BT \gamma}{m} \partial_v \rho(x,v,t).
\end{eqnarray}
The average values of the thermodynamic quantities introduced above
can be obtained by taking the time derivative of the average energy of
the system, i.e.
 \begin{equation}
\langle E(t) \rangle = \int dxdv\, \left(\frac{1}{2} m v^2 +V(x,t)\right)\rho(x,v,t),
\end{equation}
that yields 
\begin{eqnarray}
  \langle \dot{E} \rangle &=& \int dxdv\, \partial_t V \rho(x,v,t) + \int dxdv\, E(x,v,t) \partial_t \rho(x,v,t)\nonumber\\
                          &=& \langle \dot{W}\rangle +\langle \dot{Q} \rangle.
\end{eqnarray}
The identification of the first term of the r.h.s. in the first line
with the average mechanical power is immediate, since this quantity
trivially coincides with Eq. \eqref{eq:def_work_stoc}.  The other term
of the sum is associated with the average heat rate, because, in
agreement with the first principle of thermodynamics, the sum of the
two terms must represent the total (internal) energy variation of the
system. In order to show that this second definition, i.e.
\begin{equation}
\langle \dot{Q} \rangle =\int dx dv  E(x,v,t) \partial_t \rho(x,v,t),
\end{equation}
is consistent with Eq.~\eqref{ito}, it is necessary to perform some algebra:
\begin{eqnarray}
  \langle \dot{Q} \rangle &=&- \int dxdv\, E(x,v,t){\bf \nabla \cdot J}(x,v,t)=\int dxdv\, {\bf J}(x,v,t)\cdot\nabla E(x,v,t)\nonumber \\
&=&\int dxdv\,  (-\gamma v)\left(mv +k_BT\partial_v \ln \rho\right)\rho(x,v,t).
\end{eqnarray}
By performing one last integration by parts, one  gets
Eq. \eqref{ito} from the above expression.  It is important to
remark that it is very useful to have two different formalisms
describing the stochastic thermodynamics of one-particle systems: in
fact, on the one hand, Eqs.~\eqref{eq:def_work_stoc} and~\eqref{ito_no_av} give a
recipe on how to measure the work and heat exchanged on a single
trajectory. Therefore, e.g. in numerical simulations where the state
of the system is accessible at every time $t$, one can integrate the
two expressions and obtain the (fluctuating) heat and work exchanged
in a single realization of the experiment.  On the other hand, the
expressions involving the distribution $\rho(x,v,t)$ and its derivatives
are more useful in an analytic context, and, as we will show in the
following subsection, are necessary to obtain a connection between
these quantities and the entropy production.

\subsubsection{Connection with standard thermodynamics and entropy
  production}

Let us briefly recall that in thermodynamics the entropy $S(A)$ of a state $A$ is well
defined only if $A$ represents a set of thermodynamic variables
identifying an equilibrium state. In particular, the entropy is a
state function obtained through the formula
\begin{equation}
\Delta S=S(B)-S(A)=\int_A^B \frac{dQ}{T},
\end{equation}
where $dQ$ is the heat exchanged with the exterior, and $T$ the
temperature of the system on any quasi-static transformation, i.e. a
generic path consisting in a succession of equilibrium states that
connects $A$ to $B$.
The second principle of thermodynamics, in the form of Clausius
inequality, reads
\begin{equation}
\oint \frac{dQ}{T} \leq 0,
\end{equation}
i.e. $\Delta S \geq \int_B^A \frac{dQ}{T}$, and the equality only
holds in the case of quasi-static transformations. Assuming that the
system interacts with a thermostat at constant temperature $T$, when
considering the system of interest and the thermostat attached to it,
one simply gets
\begin{equation}
\Delta S_{tot}=\Delta S_{sys}+ \Delta S_{therm}\geq 0
\end{equation}
or, equivalently,
\begin{equation}
\Delta S_{sys}=- \Delta S_{therm} + \Sigma.
\end{equation}
Here $\Sigma$ is a positive quantity which, with an evident meaning,
is called total entropy production (it is the increase of entropy in
an isolated system due only to the irreversibility of the dynamics),
and the other term of the sum is
\begin{equation}
\Delta S_{therm}= -\frac{Q}{T},
\end{equation}
where $Q$ is the heat {\em absorbed} by the system, i.e. $Q=\int_A^B
dQ$.  This last quantity has been called in recent years {\it entropy
  production of the medium}~\cite{seifert05}, since it is the
variation of the entropy of the thermostat, as explained below.

These quantities can be reinterpreted in the framework of stochastic
thermodynamics as follows. The entropy function in
this the context is provided by the Gibbs
entropy, previously defined in Eq.~(\ref{E1}),
\begin{equation}\label{entropia}
S_{\Gamma}(t)=-k_B\int dx dv \,\,\rho(x,v,t) \ln \rho(x,v,t).
\end{equation}
It is possible to compute the derivative of
the Gibbs entropy and express it as a sum of different 
terms: 
\begin{eqnarray}
  \frac{d S_{\Gamma}(t)}{dt}&=& -k_B\int dxdv\, \left[\partial_t \rho(x,v,t) \ln \rho(x,v,t)+ \partial_t \rho\right]
  =-k_B\int dxdv\, \frac{{\bf J}\cdot\nabla \rho}{\rho} \nonumber \\
&=&-k_B\int dxdv\, \left[v\partial_x\rho-\gamma v \partial_v\rho -\frac{\partial_x V}{m}\partial_v\rho -\frac{k_BT \gamma}{m}\frac{(\partial_v \rho)^2}{m}\right] \nonumber \\
  &=&k_B\int dxdv\, \left[\gamma v \partial_v \rho + \frac{k_BT
      \gamma}{m}\frac{(\partial_v \rho)^2}{\rho}\right],
\end{eqnarray}
where, in the last line, we used an integration by parts. Carrying on
the calculations further, one can identify two contributions in the
entropy variation
\begin{eqnarray}
  \frac{dS_{\Gamma}(t)}{dt}&=&k_B\int dxdv\,\left(2\gamma v \partial_v \rho +\frac{k_BT
      \gamma}{m}\frac{(\partial_v \rho)^2}{\rho}\right)-k_B\int dxdv\,\gamma v \partial_v \rho \nonumber \\
&=&k_B\int dxdv\,\frac{1}{\rho}\left(\sqrt{\frac{k_BT \gamma}{m}}\partial_v \rho+\sqrt{\frac{\gamma m}{k_BT}}v \rho\right)^2 \nonumber \\
&-&\frac{1}{T}
\int dxdv\,\left(\gamma mv^2 \rho + \gamma v k_BT\partial_v \rho \right)
\equiv \dot{A}(t) + \frac{\langle \dot{Q}\rangle}{T},
\end{eqnarray}
where $\dot{A}(t)$ is a positive quantity. The identification of the
two contributions with their thermodynamic counterparts is immediate:
$\dot{S}_{\Gamma}(t)$ is the entropy production of the system,
$\dot{A}(t)=\dot{\Sigma}$ is the total entropy production (i.e. of the system and the thermostat), and the
last term is the entropy production of the medium $\dot{S}_{therm}$.

This calculation also allows us to obtain a single-trajectory,
fluctuating entropy variation of the system. The state-dependent
entropy of the system is defined as
\begin{equation}
s(x,v,t)=-k_B\ln \rho(x,v,t),
\end{equation}
so that its average gives the Gibbs entropy in
Eq.~\eqref{entropia}. This quantity can be measured once the state of
the system $x$ and the probability distribution function $\rho(x,v,t)$,
i.e. the solution of the Fokker-Planck equation, are known. For
instance, in every trajectory that originates at time $t_0$ in
$(x_0,v_0)$ and finishes at time $t_f$ in $(x_f,v_f)$ one can measure
the entropy difference:
\begin{equation}
\Delta s= -k_B\ln \frac{\rho(x_f,v_f,t_f)}{\rho(x_0,v_0,t_0)}.
\end{equation} 
Averaging on all the possible trajectories one gets
$\langle \Delta s\rangle =\Delta S_{\Gamma}$. Moreover, 
the single-trajectory equivalent of the total entropy production $\dot{\Sigma}$ can be obtained 
as the sum of the two terms
\begin{equation}
\dot{\Sigma}(x,v,t)=\dot{s}(x,v,t)-\frac{\dot{Q}}{T}(x,v,t).
\end{equation}
The value of this observable depends on the single realization of the
process, while $\langle \dot{\Sigma} \rangle$ is positive on
average. This observation constitutes a probabilistic interpretation
of the second principle of the thermodynamics: from the point of view
of the single realization, there may happen ``violations'' of the
second principle, i.e. trajectories on which the entropy production is
negative, but on average this occurrences must been compensated in
order to give a positive average. Fluctuation Relations are obeyed by the above quantities, as discussed below in Section 4.

Let us recall that, whenever the dynamics is conservative, i.e. a
Liouville theorem holds, the value of the Gibbs entropy is always
constant, for every initial distribution $\rho(x,v,0)$, as already
discussed: for this reason this entropy is not appropriate to describe
some irreversible processes like the free expansion of a gas in a
container, as discussed in detail in~\cite{cerino16}. In the present context, on the
other side, the system has a stochastic dynamics and - in addition -
because of the very few degrees of freedom, the criticisms to the use
of $S_\Gamma$, discussed in Section 2.3, are not relevant.

\subsection{Negative temperature}
\label{sec:temp_neg}

In Section 2 we already mentioned that in equilibrium statistical
mechanics one can introduce two different definitions of temperature.
Such a topic has been recently the subject of an intense
debate~\cite{Vilar2014,Schneider2014,Wang2014,Frenkel2014,Dunkel2014,Hilbert2014,Campisi2015,buonsante2015,buonsante17}.
This new interest is due to the publication of experimental
measurements of a negative absolute
temperature~\cite{Braun2013,Carr2013}: it was demonstrated the
possibility to prepare a state where the observed distribution of the
modified kinetic energy {\em per atom} appeared to be inverted,
i.e. with the largest population in the high energy states, yielding a
{\em de facto} negative absolute temperature.

The possibility of a negative absolute temperature is well known since
the work by Onsager on the statistical hydrodynamics of point
vortices~\cite{Onsager1949} and the results on nuclear spin systems by
Pound, Ramsey and Purcell (see~\cite{Ramsey1956,landsberg77,landsberg}
for a review and discussion).  In those investigations, it was clear
that an inverse temperature parameter $\beta_B$ ranging in the full
infinite real line $(-\infty,\infty)$ did not lead to any
inconsistency or paradox.

It is interesting that $T_G$ appears in the theory of Helmholtz
monocycles, which had an important role in the development of 
Boltzmann's ideas for the ergodic theory, for one-dimensional
systems~\cite{Helmholtz1884,campisi10}, see Section 2.4. Therefore one
could conclude that the ``correct'' temperature is $T_G$.
However the approach based on the Helmholtz monocycle can be used only in
systems whose Hamiltonian contains a quadratic kinetic term and a potential
part.  For such systems in the limit $N \gg 1$ one has $T_G= T_B +O(1/N)$.  Since in the reasoning in terms of monocycles the
temperature is defined via the average of the kinetic energy, in a
generic system it is not possible to follow the argument discussed in
Section 2.4.

In the following we present a line of reasoning where Boltzmann
temperature $T_B$ (positive or negative), in systems with many degrees
of freedom and short range interaction, is the (unique) proper
parameter which is relevant for the statistical properties of the
energy fluctuations, as well as in determining the flux of energy
between two systems at different temperatures, in addition it is
measurable, without the appearance of any evident inconsistency. Let
us remark that the systems discussed in~\cite{Hilbert2014}, from which
the authors try to show that only $T_G$ is the ``correct'' temperature,
are small ($N=\mathcal{O}(1)$) and/or with long-range interactions.

In our discussion, we assume that $S_B(E,N)$ is always convex,
i.e. $d^2 S_B(E,N) /dE^2 \le 0$. This is certainly true in the limit
of vanishing interaction and in short-range-interacting systems for
large $N$, since $S_B$ is strictly related to the large deviation
function associated with the density of states, see Section 2.2.2.

\subsubsection{Point-Vortex systems}

One of the first, and very important, systems showing  negative temperature has been
studied by Onsager in a seminal paper at the origin of the modern
statistical hydrodynamics~\cite{onsager49}.
Because of its historical and technical relevance we briefly summarize the main results.

Consider a two-dimensional incompressible ideal flow ruled by Euler equation
 \begin{equation}
 \partial_t {\bf u}+({\bf u} \cdot \nabla) {\bf u}= -\frac{ \nabla p}{\rho_0}, \,\, \nabla \cdot {\bf u}=0,
 \label{EU0}
 \end{equation}
 where $\rho_0$ is the constant density and $p$ the pressure.
 The vorticity  can be written as $\nabla \times {\bf u}=\omega \hat{\bf z}$,
 where $\hat{\bf z}$ is the unitary vector perpendicular to the plane of the flow, 
 and $\omega$ evolves according to
\begin{equation}
\partial_t \omega +({\bf u} \cdot \nabla) \omega=0.
\label{EU1}
\end{equation}
The previous equation expresses the conservation of vorticity along
fluid- element paths~\cite{kraichnan80}.  From the incompressibility
we can write the velocity in terms of the stream function, ${\bf
  u}=\nabla^{\perp}\psi=(\partial_y, -\partial_x)\psi$, while the
vorticity is given by $\omega=- \Delta \psi$.  Therefore, the velocity
can be expressed in terms of $\omega$
$$
{\bf u}({\bf x}, t)=-\nabla^{\perp}\int d {\bf x}' {\cal G}({\bf x}, {\bf x}')
\omega ({\bf x}', t), 
$$
where  ${\cal G}({\bf r}, {\bf r}')$  is the Green function of the Laplacian operator 
 $\Delta$, e.g. in the infinite plane
 ${\cal G}({\bf r}, {\bf r}')=-1/(2 \pi) \ln |{\bf r} -{\bf r}'|$.
Consider now an initial condition at $t=0$ such that the vorticity is localized on $N$
point-vortices 
$$  
 \omega({\bf r}, 0)=\sum_{i=1}^N \Gamma_i \delta({\bf r} -{\bf r}_i(0)), 
$$
where $\Gamma_i$  is the circulation of the $i-$th vortex.
The Kelvin theorem ensures that the vorticity remains localized at any time, and therefore
$$  
 \omega({\bf r}, t)=\sum_{i=1}^N \Gamma_i \delta({\bf r} -{\bf r}_i(t)), 
$$
which, plugged in Eq.~(\ref{EU1}), implies that the vortex positions 
${\bf r}_i=(x_i, y_i)$  evolve according to
$$
\frac{ d x_i}{dt}=\frac{1} {\Gamma_i} \frac{\partial H} {\partial y_i}, \,\,\,
\frac{ d y_i}{dt}=-\frac{1} {\Gamma_i} \frac{\partial H} {\partial x_i}, 
$$
with
$$
H=\sum_{i \neq j}  \Gamma_i \Gamma_j {\cal G}({\bf r_i}, {\bf r}_j).
$$
So the  $N$ point-vortices constitute a $N$ degrees of freedom
Hamiltonian system~\cite{newton01} with canonical coordinates 
$$
q_i=\sqrt{|\Gamma_i|} x_i, \,\,\,
p_i=\sqrt{|\Gamma_i|} sign(\Gamma_i)  y_i.
$$

Consider now  $N$ point-vortices confined in a bounded domain $\Omega$ of area $A$.
 Since for each point-vortex ${\bf r}_i \in \Omega$  one has
  $$
 \Sigma(E)=\int_{H<E} d q_1 \cdots d q_N d p_1 \cdots d p_N\le C_N A^N, \,\,\, C_N=\prod_{i=1}^N |\Gamma_i|.
 $$ Therefore $\omega(E)=d \Sigma(E)/d E$ must approach to zero for $E
 \to \pm \infty$, and must attain its maximum at a certain value
 $E_M$, so that for $E > E_M$ the entropy $S_B(E)=k_B \ln \omega(E)$
 is a decreasing function and hence $T_B(E)$ is negative.

 \begin{figure}
\begin{center}
\includegraphics[width=7cm,clip=ture]{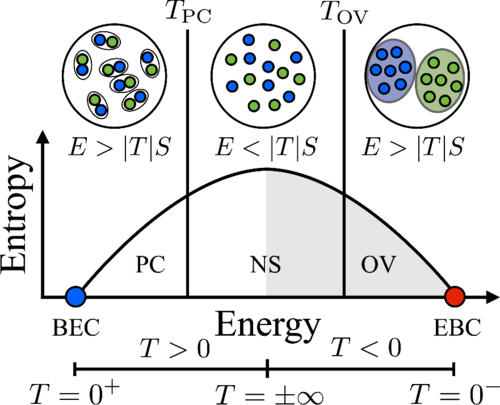}
\caption{\label{tapio} A schematic plot of entropy versus energy for
  the point-vortex model of~\cite{sdh14}.  Blue and green colors correspond
  to vortices and anti-vortices respectively. The vortex
  binding-unbinding phase transition separates the normal state (NS)
  from the pair-collapse (PC) state at positive temperature, whereas
  there is a transition to the coherent Onsager vortex (OV) state at a
  vortex-number-dependent negative temperature. Reprinted Figure 2
  with permission from [T Simula, M J Davis and K Helmerson,
    Phys. Rev. Lett., 113, 165302 (2014)]. Copyright 2014 by the
  American Physical Society.}
\end{center}
   \end{figure}
 
The states at large energy, $E \gg E_M$, are those in which the
vortices are crowded in special way.  Since in any domain $\Omega$ for
${\bf r} \sim {\bf r}'$ we have ${\cal G}({\bf r}, {\bf
  r}')\simeq-1/(2 \pi) \ln |{\bf r} -{\bf r}'|$, the configurations
with very large energy are those where point vortices, whose
$\Gamma_i$ have the same sign, are close.  Therefore in a system
with positive and negative $\Gamma_i$, negative temperature states
correspond to the presence of well separated clusters of vortices with
the same vorticity sign.  On the contrary for $E < E_M$ (positive
temperature) the vortices of opposite $\Gamma_i$ tend to remain
close~\cite{onsager49,kraichnan80,newton01}. Such a clustering
phenomenon has been observed also in simulations of quasi-2d superfluid Bose-Einstein
condensates~\cite{sdh14}, see Fig.~\ref{tapio}.

\subsubsection{Systems with negative temperature are not pathologic}

The Hamiltonian of the point-vortex system contains a long range
interaction, however the presence of negative $T_B$ does not depend on
such a peculiarity.  Replacing ${\cal G}({\bf r}_i, {\bf r}_j)$ with a
bounded function with a maximum for ${\bf r}_i= {\bf r}_j$ and fastly
decreasing to zero for large values of $|{\bf r}_i- {\bf r}_j|$, we
have a short range system and it is straightforward to show the
presence of negative temperature.

In Section 2 we already discussed the second law and energy flux
between two systems in contact.  The energy flux obviously goes from
smaller $\beta_B$ (hotter) to larger $\beta_B$ (colder), and a
negative $T_B$ does not lead to any ambiguity. Confusion may arise
from the fact that $T_B<0$ is, for the purpose of establishing the
energy flux, hotter than $T_B>0$. However using the variable $\beta_B$, the
confusion is totally removed~\cite{Ramsey1956}.  Let us now briefly
discuss a particularly interesting case of two systems with different
Hamiltonians.  Suppose that for the system $\mathcal{A}$ negative
temperatures can be present, whereas system $\mathcal{B}$ has only
positive temperatures; it is quite easy to see that the coupling of
the system $\mathcal{A}$ at negative temperature with the system
$\mathcal{B}$ at positive temperature always produces a system with
final positive temperature.  Indeed, at the initial time the total
entropy is
\begin{equation}
S_{I}=S^\mathcal{A}(E_\mathcal{A})+S^\mathcal{B}(E_\mathcal{B}),
\end{equation}
while, after the coupling, it will be 
\begin{equation}
S_{F}=S^\mathcal{A}(E'_\mathcal{A})+S^\mathcal{B}(E'_\mathcal{B}),
\end{equation}
where $E'_\mathcal{A}+E'_\mathcal{B}=E_\mathcal{A}+E_\mathcal{B}$ and, within our assumptions, $E'_\mathcal{A}$ is
determined by the equilibrium condition~\cite{huang} that $S_{F}$ takes the maximum possible
value, i.e.
\begin{equation}
\beta_\mathcal{A}={\partial S^\mathcal{A}(E'_\mathcal{A}) \over \partial E'_\mathcal{A}}=
\beta_\mathcal{B}={\partial S^\mathcal{B}(E'_\mathcal{B}) \over \partial E'_\mathcal{B}}.
\end{equation}
Since $\beta_\mathcal{B}$ is positive for every value of
$E_\mathcal{B}'$, the final common temperature must also be
positive. The above result, which can also be found, without a
detailed reasoning, in some textbooks \cite{KTH91,callen}, helps to
understand why it is not easy to observe negative temperature for a
long time. Experiments showing negative temperatures have been
recently realized, by means of an efficient isolation of the system of
interest~\cite{Braun2013}.

\subsubsection{The generalised Maxwell-Boltzmann distribution}

In systems whose Hamiltonian contains the  kinetic term $\sum_n p_n^2/(2 m)$,
we know the probability distribution density for the momentum of a single particle:
\begin{equation} 
P(p) \propto  e^{-\beta {p^2 \over 2m} },
\end{equation}
in such a case, for $N \gg 1$, $\beta=\beta_B=\beta_G$.
Let us  wonder about the same problem  in the case with
\begin{equation}
H=\sum_{n=1}^N g(p_n) + \sum_{n,k}^N V(q_n,q_k),
\end{equation}
where the variables $\{ p_n \}$, as well as the function $g(p)$, are
limited. 
With standard arguments one may compute the probability density for
the distribution of a single momentum $p$, obtaining the generalised
Maxwell-Boltzmann distribution:
\begin{equation} \label{maxbol}
P(p) \propto  e^{-\beta_B g(p)},
\end{equation}
which is valid for both positive and negative $\beta_B$.  We
mention that in the experiment in~\cite{Braun2013}, the
above recipe has been applied to measure both positive and negative system's
temperatures.

From (\ref{maxbol}) and the PdF of the energy in the canonical ensemble,
 the deep meaning of the
(Boltzmann) temperature is quite transparent: it is a quantity which
rules the statistical features of energy of a subsystem (as well as
 the ``momentum'' of a single particle). 
Let us stress again that since $T_B$ is associated with the large
microcanonical system (in physical terms, the reservoir) it is a
non-fluctuating quantity~\cite{FVVPS11} also for each sub-system and, 
in general, for non-isolated systems, see Section 3.1. 

It is remarkable that (a quantum version of) Eq.~\eqref{maxbol} has
been observed in~\cite{Braun2013} in an experiment with a bosonic
system of cold atoms in $2d$. The generalised kinetic energy as a function of quasi-momenta $(p_x,p_y)$ - in the
first Brillouin zone - is $g(p_x,p_y)=-2J[\cos(p_x \lambda)+\cos(p_y
  \lambda)]$ where $\lambda$ is the lattice constant and $J$ a
coupling strength. The experimental result is shown in Fig.~\ref{figbraun}.

\begin{figure}
  \begin{center}
    \includegraphics[width=8cm]{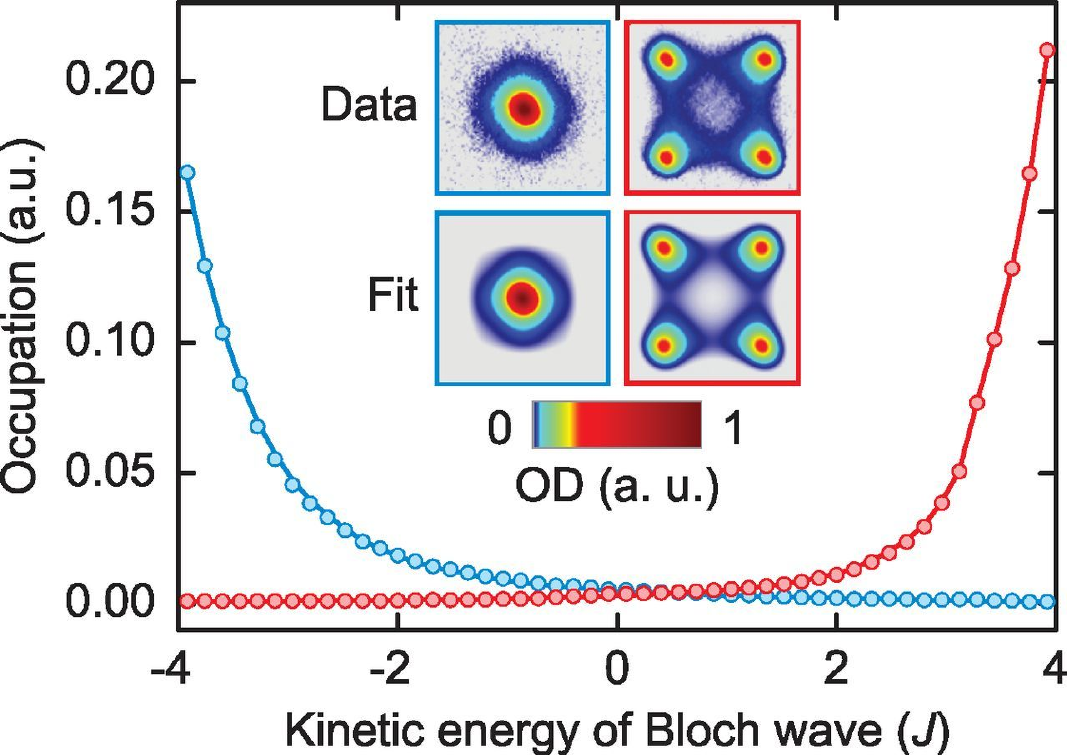}
    \end{center}
\caption{\label{figbraun} Probability distribution of the ``kinetic
  energies'' in the cold atoms experiment of~\cite{Braun2013}. The
  blue and the red circles are data at two
  different energies (red is with higher energy). The blue line is a
  fit through a Bose-Einstein distribution with positive temperature,
  while the red one is with a negative temperature. The inset shows
  the distributions of $(p_x,p_y)$ in the two situations (left for the
  low energy, or positive temperature). From S. Braun et al., Science 339, 52 (2013)~\cite{Braun2013}. Reprinted with permission from AAAS. }
  \end{figure}

\subsubsection{About the measurement of  $T_B$ and $T_G$}

The definitions of $\beta_B$ and $\beta_G$  discussed in Section 2
are based on the functional dependence of $\omega(E)$ and $\Sigma(E)$ upon the energy. 
In a real or numerical experiment it is pretty impossible
to make use of such an approach.
On the other hand, after the work of Rugh \cite{rugh97},
we know that, assuming the ergodicity, 
$\beta_B$ can be computed with a molecular dynamics simulation, and,
at least in principle, by a long-time series from an experiment.

Coming to $\beta_G$, a way, even discussed in textbooks and
considered sometimes rather important~\cite{Hilbert2014}, to determine
 its measurement is via the equipartition theorem,
which states
\begin{equation} \label{eqpart}
\left\langle x_i \frac{\partial H}{\partial x_j} \right\rangle = \delta_{ij} T_G.
\end{equation}
However the usual derivation of Eq. (\ref{eqpart}) implies the
possibility to neglect boundary terms in an integration by parts. Such
a possibility is challenged in the class of systems with bounded energy and phase space 
that we are considering.
In particular it is easy to show that (\ref{eqpart}) does not hold under the simultaneous realization of the following conditions:

$\bullet$ bounded space of the canonical variables;

$\bullet$  bounded derivatives of the Hamiltonian $ \frac{\partial H}{\partial x_j} $;

$\bullet$  bounded energy from above and below: $E_m\le E \le E_M$;

$\bullet$ vanishing density of states at the boundaries, i.e. $\omega(E_M)=0$.
\\
Given such conditions, one has that, on the one side,
$T_G(E)={\Sigma(E)}/{\omega(E)}$
diverges when $E \to E_M$. On the other side,
$\left\langle x_i \frac{\partial H}{\partial x_j} \right\rangle$ is limited,
resulting in a contradiction.

We note that a failure of (\ref{eqpart}) is  possible even  in the
absence of  negative temperatures, i.e.
 $T_G \simeq T_B >0$ for all $E$. Consider, for instance, the following
Hamiltonian
\begin{equation}
H=\sum_{n=1}^N {p_n^2 \over 2} +\epsilon \sum_{n=1}^N [ 1- \cos (\phi_n-\phi_{n-1}) ]
\end{equation}
where $\phi_n \in [-\pi , \pi)$.  For large $E$, i.e.
$E \gg \epsilon N$, the contribution to $\Sigma(E)$ of the variables
$\{ \phi_n \}$ does not depend too much on the value of $E$, so that
$\Sigma_{\epsilon}(E)\simeq \Sigma_{0}(E)\propto \, E^{N/2}$,
and  $T_G\simeq 2E/N$ and, for large $N$,  $T_B=T_G+O(1/N)$.
On the other hand, it is easy to see that
\begin{equation}
\left|  \phi_n {\partial H \over \partial \phi_n} \right| \le 2 \pi \epsilon,
\end{equation}
and, therefore, the  formula (\ref{eqpart})  cannot hold for large value of $E$ and $N$.

\subsubsection{A case study for negative temperatures}
\label{model}

In order to discuss  how systems with negative $T_B$ 
are not atypical at all, 
we  treat a system composed of
$N$ ``rotators'' with canonical variables
$\phi_1,...,\phi_N,p_1,...,p_N$ with all $\phi_i$ and $p_i$ defined in
$[-\pi,\pi)$, and with Hamiltonian
\begin{equation} \label{modham}
  H(\phi_1,\ldots,\phi_N,p_1,\ldots,p_N)=\sum_{n=1}^N [1-
  \cos(p_n)] + \epsilon\sum_{n=1}^N [1 -\cos (\phi_n -\phi_{n-1})].
\end{equation}
Choosing, as boundary condition, $\phi_0=0$ guarantees that the only
conserved quantity by the dynamics is the total energy $E$. Let us
note that the shape of the kinetic part in the previous Hamiltonian is
a one-dimensional version of the model used in~\cite{Braun2013}.

It is immediate to verify that the energy has a maximum value
$E_M=2N(1+\epsilon)$ which is realised when $p_n=\pi$ and
$\phi_n-\phi_{n-1}=\pi$ for every $n$.
When $\epsilon=0$ it is  easy  to see that Hamiltonian in
Eq. (\ref{modham}) implies negative Boltzmann temperatures. Indeed at small
energy one has $1-\cos(p_n)\simeq p_n^2/2$ so that
$$
\Sigma(E) \simeq C_N E^{N/2}, \,\, \omega(E) \simeq  \frac{N}{2} C_N E^{N/2-1},
$$
with $C_N=(2\pi)^N\frac{\pi^{N/2}}{\Gamma(N/2+1)}$. Close to $E_M=2N$ one has
\begin{equation}
\omega(E) \simeq  \frac{N}{2} C_N (E_M-E)^{N/2-1} \,\, .
\end{equation}
In conclusion we have that $\omega(E)=0$ if $E=0$ and $E=E_M$, which
implies a maximum in between and a region (at high energies) with
negative $\beta_B$.  The previous scenario is expected to hold also in the
presence of a small interaction among the rotators, this   can be
numerically confirmed with a sampling of the phase-space (see~\cite{cerino15}):
the density of
states $\omega(E)$ has a maximum in $\tilde{E}\approx E_M /2$; it is
an increasing function for $E<\tilde{E}$ whence $T_B>0$; it decreases
for $E>\tilde{E}$ whence $T_B<0$. 


Let us now  discuss some  statistical features of systems with negative temperature,
stressing in particular  the differences, and the analogies, with
the more common cases with $T_B >0$.

A rather natural way to determine $T_B$  is via Eq. (\ref{maxbol}):
computing the following average (over a single long trajectory of the
system)
\begin{equation}
\rho(p)=\lim_{\tau \to \infty} \frac{1}{N\tau} \int_0^\tau dt\,\, 
\sum_{i=1}^N \delta \left[ p_i(t) -p\right],
\end{equation}
for different values of $p$, and assuming that the system is ergodic,
it is possible to recover the single-particle-momentum probability density function
$P(p)$. 
The result of such a measure
is reported in Fig. \ref{fig:boltz_temp}: for two different values of
energy $E_+<\tilde{E}$ and $E_->\tilde{E}$ the measured $\rho(p)$ is
plotted as a function of the ``kinetic energy'' of the individual rotator
$g(p)=1-\cos(p)$. The presence of a negative temperature at $E=E_-$
can be readily indentified.
Of course the clear positive slope of the function at $E=E_-$ is due to   the fact that
$T_B(E_-)<0$: the opposite situation is encountered at $E=E_+$, where
the decreasing behavior of $\rho(p)$ indicates a temperature
$T_B(E_+)>0$.  These conclusions can also be drawn by measuring the
time average of the function $\phi(X)$ (see Section 2); in the inset of
Fig. \ref{fig:boltz_temp} we report the temperature obtained 
with the Rugh's method.

\begin{figure}[!hbtp]
  \begin{center}
    \includegraphics[width=8cm]{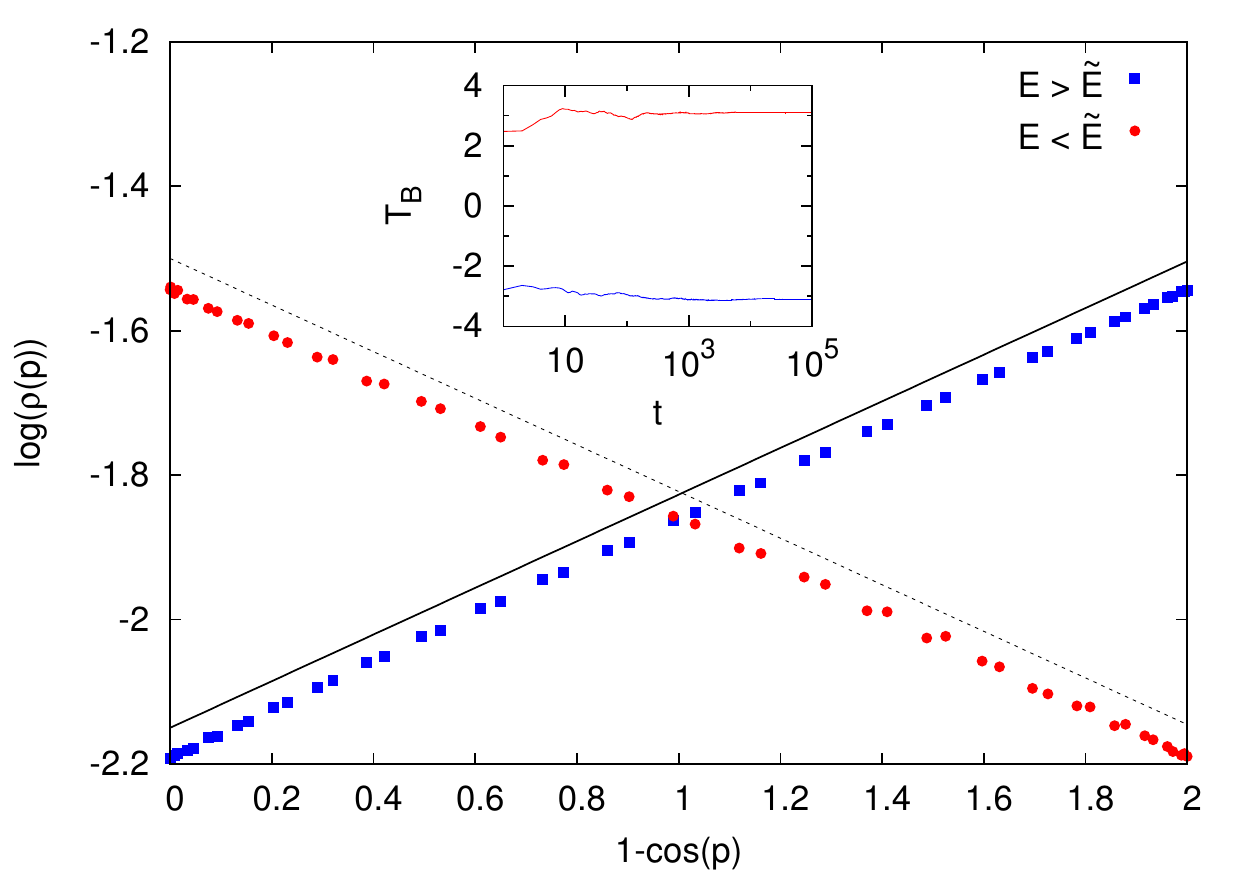}
    \end{center}
\caption{Measure of the Boltzmann temperature in the rotators chain
  for $N=100$ and $\epsilon=0.5$. Probability distribution function of
  the momentum of the rotators as a function of their ``kinetic energy''
  $g(p)=1-\cos(p)$ at energy $E=E_-=170$ (blu squares) and
  $E=E_+=130$. The slopes of the two black straight lines are
  $-1/T_B^\infty(E)$, where $T_B^\infty(E)$ is the asymptotic value of
  the corresponding curve in the inset. Inset: The $T_B$ obtained from
  the cumulated average of the observable $\phi({\bf X}(t))$ (see Eq.~\eqref{rughphi}) over a
  trajectory up to time $t$ at $E=170$ (blue line) and $E=130$ (red
  line). }\label{fig:boltz_temp}
\end{figure}

\subsection{Equivalence of ensembles and the equipartition formula}

Following the usual treatment of textbooks, assuming
that $S_B(E,N)=N s(e)$, where $e=E/N$ and $s(e)$
is convex and performing a steepest descent analysis, for
large $N$, one obtains the canonical functions from the (Boltzmann) microcanonical ones, e.g.:
\begin{equation}
T_B(e) s(e) = e-f(T_B(e)),
\end{equation}
where $f(T)$ is the free energy per particle in the canonical ensemble.
In such a derivation,
the relevant point is  the convexity of $S(e)$ and nothing about
its first derivative is required. Therefore, the equivalence of ensembles
naturally holds under our hypothesis even for negative $T_B$. Since
$T_B$ and $T_G$ can be different even for large $N$, as in our model
defined with Eq. (35), it is evident that $T_G$ is not relevant for
the ensemble equivalence.

A proposal \cite{Hilbert2014} to measure the Gibbs
temperature is by means of the equipartition formula,
Eq. (\ref{eqpart}): for the Hamiltonian in Eq. (\ref{modham}) one should
get
\begin{equation} \label{micro}
\langle p_n \sin p_n \rangle_E =T_G(E),
\end{equation}
for every $1\le n \le N$. 
Here   we use the
notation $\langle \, \rangle_E$ to denote the average in the
microcanonical ensemble, in order to distinguish it from a canonical
average $\langle \, \rangle_\beta$ which is useful to get some analytic 
expressions and better investigate 
the validity of Eq.~(\ref{micro}).

An explicit expression (see details of analogous calculations in
Ref.~\cite{Livi1987}) can be derived for the mean energy
\begin{equation}\label{eq:av_energy}
  U(\beta)=\langle H\rangle_\beta= - \frac{\partial}{\partial \beta} \ln Z (\beta)=
  N \left(1+\epsilon  - \frac{I_1(\beta)}{I_0(\beta)}-\frac{\epsilon I_1(\beta \epsilon)}{I_0(\beta \epsilon)}\right),
\end{equation}
where $I_0(x)$ and $I_1(x)$ are, respectively, the zeroth and the
first modified Bessel function of the first kind. Analogously, one can
get an analytic formula for the equipartition function
\begin{equation}
\label{eq:av_gibbs}
\langle p \sin(p)\rangle_\beta = \frac{1}{\beta}-\frac{e^{-\beta}}{\beta I_0(\beta)}.
\end{equation}
Let us remark that Eqs.~(\ref{eq:av_energy}) and~(\ref{eq:av_gibbs})
hold for both positive and negative $\beta$.

In Fig.~\ref{fig:gibbs} we report the plot of the parametric curve
$(U(\beta),\langle p \sin(p)\rangle_\beta)$ obtained by varying $\beta$ both in the
positive and in the negative
region of the real axis.
This curve can be  compared with measures of $\langle p \sin(p)\rangle_E$ computed from
molecular dynamics simulations in the microcanonical ensemble at different
values of the energy $E$ (Fig. \ref{fig:gibbs}). Such a comparison
clearly shows that the results obtained in the two different ensembles
are identical, a transparent evidence that the equivalence of ensemble
already exists for this system quite far from the thermodynamic limit
($N=100$).

In addition  Fig. \ref{fig:gibbs}  shows that the equipartition
formula cannot be used to measure the Gibbs temperature: indeed, as
already pointed out, the equipartition
theorem can fail if the density of states $\omega(E)$ vanishes. This
is the case of our system  where
$T_G=\Sigma(E)/\omega(E)$ should diverge for $E\to2 N(1+\epsilon)$: on
the other hand the results obtained in the canonical and in the
microcanonical ensemble clearly indicate that $\langle p \sin(p)\rangle_E\to 0$ as
$E\to 2N(1+\epsilon)$.

\begin{figure}[!hbtp]
  \begin{center}
    \includegraphics[width=8cm]{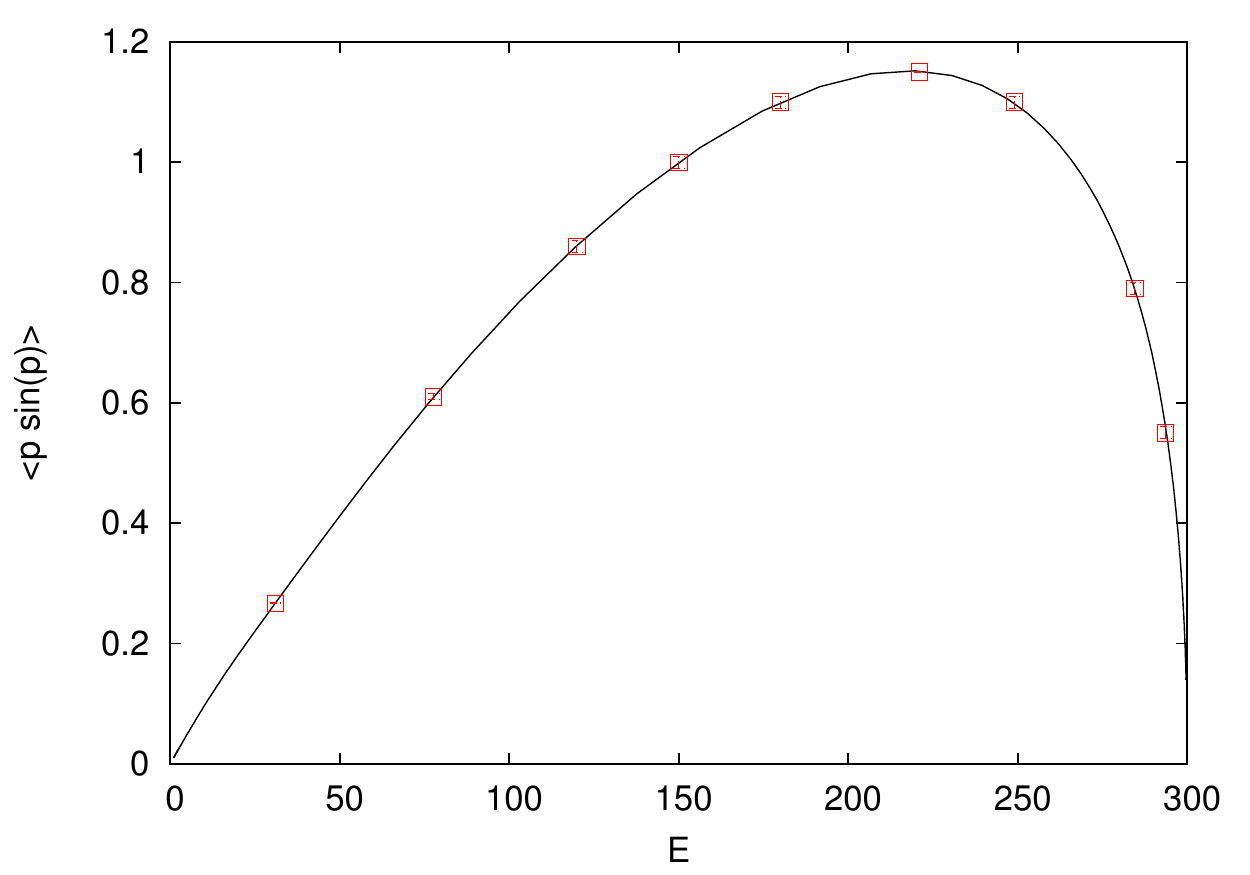}
    \end{center}
\caption{ 
Black line: $\langle p\sin(p)\rangle_\beta$ vs $U(\beta)$ in
the canonical ensemble
 and
as parametric functions of  $\beta\in(-\infty,\infty)$. 
Red squares: time averages of the
equipartition function in molecular dynamics simulations at fixed
 energy $E$ (microcanonical ensemble). The values for the parameters
   of the model are $N=100$ and $\epsilon=0.5$.}
\label{fig:gibbs}
\end{figure}

\subsubsection{Spatial coherence}

In analogy with systems of point vortices  previosly discussed, as well as in other systems e.g. discrete non-linear Schr\"odinger equation~\cite{flop11},
the rotators model in Eq. (\ref{modham})
possesses a spatially ordered phase at large values of $E$: this can
be easily understood by noting that the density of states $\omega(E)$
vanishes in $E=E_M$, i.e. that there is a small number of microscopic
configurations corresponding to large values of $E$. In particular,
the maximum of the energy $E_M=2 N(1+\epsilon)$ is attained by the
unique microscopic state where, for every $n$, $p_n=\pi$ and
$\phi_n-\phi_{n-1}=\pi$; that is, where all the rotators are fixed
($\dot{\phi}=\sin \pi =0$) and the distance among two consecutive
rotators is $\Delta \phi=\pi$. As a consequence, since $\phi_0=0$, all
the particles with even index ($n=0,2,4\ldots$) must be at $\phi=0$
and the others ($n=1,3,\ldots)$ in $\phi=\pi$. At smaller values of
$E\lesssim E_M$, such
considerations can be extended, yielding a very similar situation:
even and odd rotators must be close, respectively, to $\phi=0$ or
$\phi=\pi$.

\subsection{Temperature in small isolated systems?}

Let us briefly discuss about the  possibility 
to introduce, or not,  the concept of temperature in small systems. 
As already mentioned, such a topic, beyond its  interest
in the general context of the statistical physics, can be relevant
in the treatment of nanosystems which can involve just 
few particles.
\\
As discussed in Section 2,  for Hamiltonian systems,
following the elegant approach introduced by Rugh,
we can  define (and compute)
 the temperature $T_B$ in terms of
time average $\left<( ....) \right>_t$
of a suitable  observable $\phi({\bf X})$:
\begin{equation}
{1 \over T_B}=  \left< \phi({\bf X}) \right>_t \,\, .
\end{equation}
In a similar way one can compute (at least for a certain class
of systems) the $T_G$ using the relation
\begin{equation}
k_B T_G=  \left< x_i {\partial H \over \partial x_i} \right>_t \,\, .
\end{equation}
\\
A quite natural, and general, question is the possible dependence of the time
averages on the initial condition ${\bf X}(0)$.
If the number of degrees of freedom is very large, one can invoke
a well know important result due to Khinchin~\cite{khinchin49}
which can be summarized
as follows: the physically relevant observables (i.e. that ones involving all the 
degrees of freedom) are selfaveraging; 
they are practically constant (except in a region of small measure) on a constant- energy
surface. 
More precisely   in system with $N \gg 1$ the ergodicity is not a real problem,
at least at a physical practical level:
 the  time average of ``physical observables'' (e.g. the kinetic
energy) is very close to the microcanonical average, with the exception of initial
conditions in a (small) region whose measure is  $O(N^{-1/4})$, and therefore  goes to zero a $N \to \infty$.
\\
Unfortunately such a result does not hold in small systems.
Some authors computed numerically $T_B$ and $T_G$ in small Hamiltonian systems,
e.g. the celebrated H\'enon- Heiles model, or systems with quartic nonlinearity,
usually with $N=2$ or $N=4$~\cite{bannur97,bannur98,berdi91}.
They observed that $T_G \neq T_B$ and in addition, as expected from very general
results of the Hamiltonian systems, the results depend on the initial conditions.
Let us note that the investigated
systems do not show negative temperature, the difference 
between $T_G$ and $T_B$  are just a consequence of the small value of N.
\\
In our opinion the  results in~\cite{bannur97,bannur98,berdi91} 
show, in a rather clear way, 
that in  small isolated systems the concept
of temperature (both $T_B$ and $T_G$) is rather misleading: 
its value can depend on the initial condition,
and more important, does not have a clear physical relevance.
\\
In Section 3.2 we discussed  small systems interacting with a thermal bath.
We saw how, in spite of the limited value of $N$, the temperature 
is well defined and rules the  statistical feautures,
e.g. it appears in the probability distribution of the energy.
In an analogous way also  for a class of  Langevin equations with detailed balance (see Section 4.1)
e.g. modelling the evolution of a  
colloidal particle interacting  with a thermal bath,
 it  is not difficult to  introduce the temperature.
\\
The mathematical reason of such a difference is quite transparent:
at variance with the deterministic case previously discussed,
because of  the presence of a bath (noise)
 we are dealing with  stochastic
systems: the problem of the validity of the ergodicidy
is solved.
So we have that for small, but non isolated, systems
we can introduce in a coherent way the concept
of temperature which  is a property of the reservoir,
and rules the statistical features of the system.


\section{Temperature and Response Theory}
\label{sec:fluct}

In the present Section we discuss the role played by temperature in
the behavior of a system which is driven out of equilibrium. A weak
perturbation is treated by means of the close-to-equilibrium linear
response theory, summarized in the Equilibrium Fluctuation-Dissipation
Relation (EFDR), where temperature has a clear status of proportionaly
factor between response function and correlation of fluctuations (see
Subsection~\ref{sub41}). Far from equilibrium, things are much more
complex and - in general - temperature loses its preeminent
role. Nevertheless, several theoretical results in
far-from-equilibrium statistical systems have been obtained, in the
last twenty years, chiefly ascribable to two lines of research: linear
response of systems which are already out of equilibrium (see
Subsection~\ref{sub42}), and relations between asymmetric fluctuations
of entropy production (partially discussed in both
Subsections~\ref{sub41} and~\ref{sub42}). In the close-to-equilibrium
limit all those results coincide with the EFDR. In particular cases,
as in systems with well-separated timescales, it is possible to extend
the notion of temperature-like quantities to the non-equilibrium realm
(see Subsection~\ref{sub43}).

\subsection{Close to equilibrium: Temperature from Kubo relation in Hamiltonian systems and Langevin equations} 
\label{sub41}

In Section 1.4 we have seen the Einstein theory for the fluctuations of macrostates, repeated here for convenience:
\begin{equation} \label{einst0}
P(\alpha_1,...,\alpha_m) \sim e^{[S\{\alpha_k\}-S_e]/k_B}.
\end{equation}
Eq.~\eqref{einst0} expresses the probability of a spontaneous fluctuation
as the exponential of the entropy difference with respect to
equilibrium. Such a formula leads, by means of a simple perturbative
expansion, to Eq.~\eqref{II3a} which relates the amplitude of
macro-fluctuations, quantified by the covariances $\langle\delta
\alpha_i \delta \alpha_j \rangle$, to the second derivatives of the
entropy which involve temperature. The latter represent, in general, the susceptibilities,
i.e. the long time responses of the system to external perturbations. 

Summarizing, Eq.~\eqref{einst0} teaches us
that in statistical mechanics - under equilibrium conditions - three
fundamental concepts, i.e. spontaneous fluctuations, responses to a
perturbation and temperature, are connected in a unique relation. One
of the simplest examples of such a threefold connection is the
textbook formula
\begin{equation} \label{enflu}
\langle E^2 \rangle-\langle E\rangle^2=k_B T^2 C_v,
\end{equation}
where the variance of the energy distribution is proportional to the
heat capacity (a susceptibility, i.e. a response function) and to the square of the
temperature. An analogous formula, due to Einstein, relates the
diffusivity $D$ to the mobility $\mu$ for a Brownian particle
disperesed in a solvent fluid:
\begin{equation} \label{einst}
D= k_B T \mu,
\end{equation}
where again the temperature appears as the constant of proportionality
relating a measure of fluctuations ($D$) and a measure of response
($\mu$). 

The above examples involve quantities which do
not depend upon time and, for this reason, are often referred to as {\em
  static} EFDR. In the first half of the
$20$th century a series of important results, with different physical
systems and observables, showed that many other similar relations
exist, tying in the same way those three ingredients: spontaneous
fluctuations, response and
temperature~\cite{KTH91,BPRV08}. The generalisation to a
time-dependent - or dynamic - relation with the same form was stated in universal terms for the first time
by Onsager in 1931~\cite{O31,O31b}. For instance, by recalling the
general relation between diffusivity and the velocity autocorrelation,
i.e. that
\begin{equation}
D=\int_0^{\infty} dt \langle v(t)v(0)\rangle,
\end{equation}
it is seen that  Eq.~\eqref{einst} is equivalent to
\begin{equation} \label{einst2}
\langle v(t)v(0)\rangle= k_B T  R_{v F}(t),
\end{equation}
with the identification
\begin{equation} \label{mob}
\mu = \int_0^\infty dt R_{vF}(t).
\end{equation}
In the r.h.s. of Eq.~\eqref{einst2} it appears the so-called response
function, $R_{vF}(t)$, which is the mean variation at time $t$ to an
impulsive perturbation at time $0$: in our examples $R_{vF}(t)$ relates
the rate of variation of colloidal (tracer) particle's velocity and a perturbation of the
external force applied at time $0$. In order to discuss in full
generality the Fluctuation-Dissipation Relation (FDR), it is useful to recall the
definition of response functions which is the central object of linear
response theory. For simplicity we restrict the discussion to the
linear perturbation of stationary states, i.e. states which are
invariant under translations of time. For this reason time-dependent
correlation functions and response functions only depend on
differences of times. The general case is treated in Section 4.2.

The response function $R_{\mathcal{O}\mathcal{F}}(t)$ for an observable $\mathcal{O}(t)$
to a time-dependent perturbation of a parameter or degree of freedom
$\delta \mathcal{F}(t)$ is implicitly defined in the following relation
\begin{equation}
\overline{\Delta \mathcal{O}(t)} = \int_{-\infty}^t dt' R_{\mathcal{O}\mathcal{F}}(t-t') \delta\mathcal{F}(t'),
\end{equation}
where $\overline{\Delta \mathcal{O}(t)} = \overline{\mathcal{O}(t)}-\langle\mathcal{O}(t)
\rangle_0$ represents the average deviation, at time $t$, of the
observable $\mathcal{O}$ with respect to its average value in the unperturbed
stationary system. Here we use $\overline{f(t)}$ to denote a
time-dependent (perturbed) average of the observable $f$, and $\langle f \rangle_0$ to denote
its stationary unperturbed average. It is clear that, taking an
impulsive shape for the external perturbation i.e. $\delta\mathcal{F}(t)=\Delta \mathcal{F}\delta(t)$ (with $\delta(t)$ the Dirac delta distribution), one has
\begin{equation}
\frac{\overline{\Delta \mathcal{O}(t)}|_{imp}}{\Delta \mathcal{F}} = R_{\mathcal{O}\mathcal{F}}(t),
\label{imp}
\end{equation}
which gives a more operational definition to the response function~\footnote{Be careful that $\Delta \mathcal{F}$ has the dimensions of a time-integral of $\mathcal{F}(t)$.}. It
is also useful to see what happens when the perturbation takes the
shape of a Heaviside unit step function, i.e. $\delta\mathcal{F}(t)=\delta
\mathcal{F}_0 \Theta(t)$:
\begin{equation}
\frac{\overline{\Delta \mathcal{O}(t)}|_{step}}{\delta \mathcal{F}_0} = \int_0^t dt' R_{\mathcal{O}\mathcal{F}}(t'),
\end{equation}
If $\mathcal{O}(t)$ is the tracer velocity along one axis and
$\mathcal{F}(t)$ is the external force applied from time $0$ to time
$\infty$ to the tracer (parallel to that axis), the final velocity
reached by the tracer is exactly $\delta \mathcal{F}_0\int_0^\infty
dt' R_{vF}(t')$, which explains the connection with the identification made
in Eq.~\eqref{mob}.

Equipped with the response function definition, it is possible to
 enunciate the FDR for  Hamiltonian $\mathcal{H}$ systems at equilibrium with a thermostat at temperature
$T$~\cite{K57}:
\begin{equation} \label{fdt}
R_{\mathcal{O}\mathcal{F}}(t) = \frac{1}{k_B T}\langle \mathcal{O}(t) \dot{A}(0)\rangle_0 = -\frac{1}{k_B T}\langle \dot{\mathcal{O}}(t) A(0)\rangle_0,
\end{equation}
where $A$ is the observable (or degree of freedom) which is coupled to
$\mathcal{F}(t)$ in the Hamiltonian to produce the perturbation,
i.e. $\mathcal{H}(t)=\mathcal{H}_0 - \mathcal{F}(t)A$. We immediately
see that if $\mathcal{O}$ is the tracer's velocity and $\mathcal{F}(t)$ is an
external force applied to its $x$ coordinate, Eq.~\eqref{fdt} is nothing but 
 Eq.~\eqref{einst2}. In conclusion the Einstein relation is a
particular case of the EFDR.

From Eq.~\eqref{fdt} one may get several possible variants, which are
useful in different physical situations. A large amount of remarkable
results concern, for instance, the time-Fourier transform of
Eq.~\eqref{fdt}, as well as the relation connecting currents and
conductivities (the so-called Green-Kubo relations, see below)~\cite{KTH91,BPRV08}.

An interesting generalisation of the EFDR
concerns the realm of stochastic processes. For instance, the
so-called Klein-Kramers model, for a single particle in $1$
dimension~\cite{G90}, well describes the dynamics of a system at
thermal equilibrium:
\begin{subequations} \label{KK}
\begin{align} 
\frac{d x(t)}{dt} &= v(t) \\
m\frac{d v(t)}{dt} &= -\frac{d U(x)}{dx} -\gamma v(t) + \sqrt{2 \gamma k_B T}\xi(t),
\end{align}
\end{subequations}
where $\xi(t)$ is a white Gaussian noise with $\langle \xi(t)
\rangle =0$ and $\langle \xi(t)\xi(t') \rangle = \delta(t-t')$,
$\gamma$ is the viscosity, $U(x)$ is an external potential. The model
can be easily generalised to $N>1$ interacting particles in any
dimensions. In the absence of the external potential, Eq.~\eqref{KK}
coincides with the original Langevin equation proposed a few years
after the theories of Einstein~\cite{E05} and Smoluchovski~\cite{s06} to explain diffusion
in Brownian motion~\cite{L08}. Eq.~\eqref{KK} actually describes more than spatial diffusion
(which is the {\em long time} behavior of Eq.~\eqref{KK}), as it
includes short times ``ballistic'' effects. From a mathematical point
of view, it describes a Markovian time-continuous process.

Most importantly, the distribution of its stationary states (achieved
with the condition $\gamma >0$ and confining potential) is given by
the Gibbs measure $P(x,v) \propto e^{-\mathcal{H}(x,v)/(k_B T)}$ with
$\mathcal{H}(x,v)=mv^2/2 + U(x)$: such a stationary state satisfies
detailed balance, which is equivalent to say that the unconditioned
probability of observing a trajectory is equal to the unconditioned
probability of the time-reversed trajectory~\cite{R89}. Linear
response theory, when applied to the Klein-Kramers model in its
stationary state, gives exactly the same result as
Eq.~\eqref{fdt}~\cite{R89,BPRV08}.

The Klein-Kramers process is Markovian with respect
to the variables $(x,v)$, usually such a property is just a rough
approximation for the dynamics of a tracer which interacts
with other particles in a fluid. For instance the Langevin equation -
i.e. the Kramers equation with $U(x)=0$ - must be  generalised to
take into account retarded hydrodynamic effects, by the introduction of linear memory
terms, e.g. by writing a generalized Langevin Equation
(GLE)~\cite{KTH91}:
\begin{equation} \label{gle}
m \frac{d v(t)}{dt} = -\int_{-\infty}^t dt' \Gamma(t-t') v(t') + \eta(t),
\end{equation}
where $\Gamma(t)$ is a memory kernel representing retarderd damping,
and $\eta(t)$ is a stationary stochastic process with zero average
$\langle \eta(t) \rangle =0$ and time-correlation which - at equilibrium
- satisfies the so-called FDR of the second kind:
\begin{equation} \label{fdt2}
\Gamma(t) = \frac{1}{k_B T} \langle \eta(t) \eta(0) \rangle.
\end{equation}
It is clear that Eq.~\eqref{fdt2} has the same structure of
Eq.~\eqref{fdt} and this motivates the name of the relation. The
Markovian case (damping with zero memory) is obtained when
$\Gamma(t)=2\gamma \delta(t)$ (recalling that $\int_{-\infty}^t dt' 2\gamma \delta(t') v(t')=\gamma v(t)$).  

In the discussion of the Fluctuation-Dissipation theorem of the second
kind, Kubo put in evidence that such a result is necessary to
guarantee the equipartition theorem and this is the reason why it is a
signature of thermodynamic equilibrium. A condition for (Markovian)
microscopic dynamics which is necessary and sufficient for
thermodynamic equilibrium is detailed balance~\cite{T03}, commonly
assumed in stochastic thermodynamics~\cite{LS99,db05}. It is
interesting to discuss a simple example which connects the concept of
detailed balance to the FDR of the second
kind.

Let us consider the following bivariate Ornstein-Uhlenbeck process
\begin{subequations} \label{twovar}
\begin{align} 
m \frac{dv(t)}{dt} &=-\gamma v(t) + \sqrt{M\Gamma_0} z(t) + \sqrt{2 \gamma k_B T_0} \phi_1(t) \label{twovar1}\\
M\frac{dz(t)}{dt} &= -\sqrt{M \Gamma_0}v(t)-\frac{M}{\tau}z(t)+\sqrt{2 \frac{M}{\tau} k_B T_\tau}\phi_2(t), \label{twovar2}
\end{align}
\end{subequations}
where $v(t)$ is the velocity of a Brownian particle and
$\sqrt{M\Gamma_0}z(t)$ represents an effective fluctuating
force\footnote{Therefore we assume that $z(t)$ is a variable with even
  parity under time-reversal. The parity of coarse-grained variables
  is, in general, a subtle problem, see for
  instance~\cite{CPV12,CP15}.}  induced by the surrounding solvent
upon the particle: we assume that it evolves according to a separate
Ornstein-Uhlenbeck process with characteristic relaxation time $\tau$,
effective mass $M$, effective ``temperature'' $T_\tau$ and feedback
from the particle itself (with $\phi_1$ and $\phi_2$ Gaussian white
noises with unitary variances).  Recently such a model has been used to study a massive tracer diffusing through a moderately dense
granular material~\cite{sarra10b} (we discuss this case below, in
Subsection~\ref{sub51}). A similar model has also been proposed to
represent a kind of active Brownian particles, where $z(t)$ represents
self-propulsion~\cite{MBGL15} (this model is briefly described in Subsection~\ref{sub53}).  The system in Eqs.~\eqref{twovar} is
linear and therefore fully solvable. The stationary state is
characterized by a bivariate Gaussian distribution
\begin{equation}
p_{st}(\mathbf{X}) \propto \exp \left(-\frac{1}{2}\mathbf{X}\Sigma^{-1}\mathbf{X}^T \right),
\label{pst}
\end{equation}
where $\mathbf{X}=(v,z)$
with covariance matrix $\Sigma$ given by 
\begin{align}
\Sigma_{vv}&=\langle v^2 \rangle=k_B T_0/m+ \langle v z \rangle \sqrt{\Gamma_0 M}/\gamma \\
\Sigma_{zz}&=\langle z^2 \rangle=k_B T_\tau/M- \langle v z\rangle \sqrt{\Gamma_0 M} \tau/M \\
\Sigma_{vz}&=\Sigma_{zv}=\langle v z \rangle=\frac{\Gamma_0 \tau \gamma}{(\gamma+\Gamma_0 \tau)(m+\gamma   \tau)\sqrt{\Gamma_0 M}} k_B(T_\tau-T_0).
\end{align}

The probability of a trajectory going from time $0$ to time $t$,
denoted as $\{v(s),z(s)\}_{s=0}^t$, conditioned to the initial point
$(v(0),z(0))$ is given by the Onsager-Machlup expression~\cite{OM53}
\begin{multline} \label{fw}
\ln P(\{v(s),z(s)\}_{s=0}^t) = \\ \mathcal{C} -\int_0^t ds \left\{ \frac{[m \dot{v}(s) +\gamma v(s)-\sqrt{M \Gamma_0}z(s)]^2}{\gamma k_B T_0}+\frac{[M\dot{z}(s)+\sqrt{M\Gamma_0}v(s)+z(s) M/\tau]^2}{k_B T_\tau M/\tau}\right\},
\end{multline}
where $\mathcal{C}$ is a constant.
According to the above expression, the probability (with the condition of starting at $(-v(t),z(t))$) of the time-reversed trajectory reads
\begin{multline} \label{bw}
\ln P(\{-v(t-s),z(t-s)\}_{s=0}^t) = \\ \mathcal{C} - \int_0^t ds \left\{ \frac{[m \dot{v}(s) -\gamma v(s)-\sqrt{M \Gamma_0}z(s)]^2}{\gamma k_B T_0}+ \frac{[-M\dot{z}(s)-\sqrt{M\Gamma_0}v(s)+z(s) M/\tau]^2}{k_B T_\tau M/\tau}\right\}.
\end{multline}
The ratio between the probabilities of the forward (Eq.~\eqref{fw}) and
backward (Eq.~\eqref{bw}) trajectories is obtained straightforwardly, by
noting that many terms cancel out and some of the remaining ones are
integral of exact time-differentials. In few passages it is verified
that
\begin{equation} \label{detbal}
\frac{P(\{v(s),z(s)\}_{s=0}^t)}{P(\{-v(t-s),z(t-s)\}_{s=0}^t)} = \frac{p_{st}[-v(t),z(t)]}{p_{st}[v(0),z(0)]} ,
\end{equation}
if and only if $T_\tau=T_0$. Eq.~\eqref{detbal} shows, in this case,
the equivalence between thermodynamic equilibrium and detailed
balance.

The linear response of variable $v(t)$ in Eqs.~\eqref{twovar} to an
infinitesimal perturbation of $v$ of intensity $\delta v(0)$ (which is
equivalent to an impulsive force $m \delta v(0) \delta(t)$), can be
obtained by recalling that - if the average dynamics is linear, i.e.
\begin{equation}
\dot{\overline{\mathbf{X}}} = A \overline{\mathbf{X}},
\end{equation}
as it is in this case, then the linear response is obtained straightforwardly as
\begin{equation} \label{lr}
\overline{\delta \mathbf{X}(t)} = e^{A t} \delta \mathbf{X}(0),
\end{equation}
while the covariance matrix (at positive time-delay $t$) in the steady state reads
\begin{equation} \label{cm}
\langle \mathbf{X}(t) \mathbf{X}(0)\rangle = \Sigma e^{A t}.
\end{equation}
Putting together results~\eqref{lr} and~\eqref{cm} we get
\begin{equation}
m R_{vF}(t)=\frac{\overline{\delta v(t)}}{\delta v(0)} = \Sigma^{-1}_{vv} \langle v(t) v(0) \rangle + \Sigma^{-1}_{vz} \langle v(t) z(0) \rangle.
\label{respz}
\end{equation}
Again it is immediate to see that the Einstein relation,
Eq.~\eqref{einst2}, is recovered when $T_\tau=T_0$ which coincides
with $\Sigma^{-1}_{vz} = 0$ and $\Sigma^{-1}_{vv}=m/(k_B T_0)$.

It is interesting to note~\cite{VBPV09,PV09,CPV12} that a formal solution of $z(t)$ from
Eq.~\eqref{twovar2}, put into Eq.~\eqref{twovar1}, leads to a GLE of
the kind in Eq.~\eqref{gle} with memory kernel given by
\begin{equation}
\Gamma(t)=2\gamma \delta(t) + \Gamma_0 \exp(-t/\tau) \;\;\;(t>0),
\end{equation} 
and noise made of two uncorrelated terms: $\eta(t)=\eta_0(t) +
\eta_\tau(t)$, with both terms having zero average and
time-correlations
\begin{align}
\langle \eta_0(t)\eta_0(t') \rangle&= 2 k_B \gamma T_0 \delta(t-t') \\
\langle \eta_\tau(t)\eta_\tau(t') \rangle&= k_B T_\tau \Gamma_0 \exp(-|t-t'|/\tau).
\end{align}
Again we see that Eq.~\eqref{fdt2} is satisfied only when
$T_0=T_\tau$. This clarifies the deep connection between the symmetry
under time-reversal, thermodynamic equilibrium and the
Fluctuation-Dissipation theorem of the second kind.

To conclude this succinct summary of a wide topic of non-equilibrium
statistical mechanics (close to equilibrium), it is interesting to
discuss the case of currents, which has been mentioned a few pages
above. Currents are physically relevant 
{\em responses} to external perturbations: examples include the
electrical current in a conductor under an external electrical
potential, the heat flow in a material under a temperature gradient,
or the transverse momentum current in a fluid under shear. Most
importantly, a measurement in a finite sample, observed for a finite
time, will result in a {\em fluctuating} current: this has become a real
possibility in the last decades, with the advent of experiments at
micro or nano scale, as well as with the increase in the time-resolution of
instruments~\cite{liphardt,seifertrev}. The typical quantity which is measured is
a finite-time (over a duration $\tau$) averaged current, i.e.
\begin{equation} \label{curr}
J_{\tau}(t) = \frac{1}{\tau} \int_t^{t+\tau} ds j(s),
\end{equation}
where $j(s)$ is an instantaneous current at time $s$,
defined over a finite volume of the system: for instance, in the case
of a simple current of mass of a certain species of molecules
(e.g. tracers in a mixture), it is proportional to the empirical
average (in the volume) of the instantaneous velocity of such
molecules. The situation, and the following discussion, is greatly
simplified under the assumption of stationarity.

Physical currents have the property of changing sign under the
action of the operation of time-reversal. Indeed if a given trajectory
of the system (in the full phase space) corresponds to observing a
value $x$ for $J_{\tau}(t)$, the time-reversed trajectory corresponds
to observing $-x$ for $J_{\tau}(t)$. In the mentioned example, where
the current is related to a flow of massive tracers, a time-reversed
microscopic trajectory corresponds to observe - in a chronologically
inverted order - all tracers' velocities with a change of sign, i.e. $j(s) \to
-j(s)$ and therefore $J_\tau \to -J_\tau$~\footnote{the change of time-ordering does not affect the
  integral in Eq.~\eqref{curr} as $ds$ changes sign together with an
  inversion of the extremes of integration.}

A system in equilibrium does not contain net physical
currents, i.e. the stationary average of $J_{\tau}(t)$ is zero. A
stationary state with a small non-zero average current $\langle j
\rangle$ is, therefore,
understood as the $t \to \infty$ response to a small external
perturbation $\delta \mathcal{F}_0 \Theta(t)$, applied since time $0$, in an equilibrium system. For this
reason it has to satisfy the Fluctuation-Dissipation theorem near
equilibrium, Eq.~\eqref{fdt}, that in this particular case takes the
form
\begin{equation}
\langle j \rangle = \delta \mathcal{F}_0 \int_0^\infty dt' R_{J\mathcal{F}}(t') = \frac{\delta \mathcal{F}_0}{k_B T} \int_0^\infty dt' \langle j(t') \dot{A}(0)\rangle_0,
\end{equation}
with $A(t)$ the observable conjugated to the perturbation
$\mathcal{F}$.  An interesting
simplification comes by observing that the observable conjugate to the
perturbation needed to produce a current $J$ is the time-integral of the current itself~\cite{KTH91},
which leads to the classical Green-Kubo relation
\begin{equation} \label{ave}
\langle j \rangle = \langle J_{\tau} \rangle = \frac{\delta \mathcal{F}_0}{k_B T} \int_0^\infty dt' \langle j(t') j(0)\rangle_0.
\end{equation}
What happens to the fluctuations, e.g. the variance, of the $\tau$-averaged current $J_\tau$?
At equilibrium (when $\langle j \rangle =0$), and for large $\tau$, it is always possible to write
\begin{equation} \label{var}
\tau^2 \langle J_{\tau}^2 \rangle_0 = \left\langle\left[\int_0^\tau dt' j(t') \right]^2 \right\rangle_0 \approx 2 \tau \int_0^\tau dt' \langle j(t')j(0) \rangle_0.
\end{equation}
In the framework of linear response theory, where $O(\delta
\mathcal{F}^2)$ and higher order terms are neglected, the variance of
a current is not changed too much by the perturbation, that is $\langle
J_{\tau}^2 \rangle - \langle J_\tau \rangle^2 \approx \langle
J_{\tau}^2 \rangle_0 $. In conclusion, putting together
Eq.~\eqref{ave} and~\eqref{var}, it is found for large $\tau$
\begin{equation}
\langle J_\tau^2 \rangle - \langle J_\tau \rangle^2 \approx 2 \frac{k_B T \langle J_\tau \rangle}{\tau \delta \mathcal{F}_0}.
\end{equation}
At equilibrium, and close to it (at first order in the perturbation),
the statistics of $J_\tau$ is well described by a Gaussian probability
density function~\cite{db05}, which - according to the last result - reads at large
$\tau$ 
\begin{equation} \label{gausgc}
p(J_\tau) \propto \exp\left[ -\tau\frac{(J_\tau-\mu \delta \mathcal F_0)^2}{4 \mu k_B T}\right],
\end{equation}
where we have defined the generalised mobility $\mu=\langle J_\tau \rangle/\delta \mathcal{F}_0$.

Remarkably, Eq.~\eqref{gausgc} - which is a mere consequence of the
Fluctuation-Dissipation relation (and the limit $\tau \to \infty$)
appears to be the particular form, when close to equilibrium, of a
more general ``Fluctuation Relation'' which establishes a symmetry
between time-reversed values of the current, i.e. $J_\tau$ and $-J_\tau$. Indeed it is immediate to
verify that Eq.~\eqref{gausgc} satisfies
\begin{equation} \label{fr}
p(J_\tau)/p(-J_\tau) = \exp\left(\tau \frac{\dot{S}_\tau}{k_B}\right),
\end{equation}
where we have introduced the ``fluctuating entropy production rate'' $\dot{S}_\tau$ averaged over the time $\tau$, defined as
\begin{equation} \label{clausius}
\dot {S}_\tau = \frac{J_\tau \delta \mathcal{F}_0}{T}.
\end{equation}
In the following sections, together with some relevant
non-equilibrium generalisation of the Fluctuation-Dissipation theorem,
we discuss the universality of Eq.~\eqref{fr}, which also applies to
cases with a large perturbation $\delta \mathcal{F}_0$, where
$p(J_\tau)$ is no more guaranteed to be Gaussian, and even to cases
without a single thermostat, where Eq.~\eqref{clausius} has to be
replaced by a more general definition of fluctuating entropy
production. 

In conclusion it is rewarding, after a journey through linear response
theory, stochastic processes, time-reversal symmetry and current
fluctuations, to arrive to relation Eq.~\eqref{fr} which is
reminiscent of the starting point, Eq.~\eqref{einst0}. It is important
to underline that the coherence among the many concepts touched in the
last pages is a {\em gift} of thermodynamic equilibrium, where
temperature is a well defined concept.~\footnote{All along the present
  Subsection we have always considered stochastic processes whose
  invariant probability distribution is the canonical one. For this
  reason there is no ambiguity in the definition of temperature.}

\subsection{Extension to far-from-equilibrium systems: a generalized fluctuation-dissipation relation}
\label{sub42}

The significance of an equation is somehow determined by the difference
of definition between the concepts underlying the left and right hand
sides of the equality sign~\cite{donth}.  The FDT discussed in the
previous Section represents a clear instance of such a deep connection
between different quantities: spontaneous fluctuations and response to
external stimuli.  On the one hand, the theoretical interest in
deriving such kind of relations relies on the possibility to go deep
into the mechanisms governing a given phenomenon, uncovering hidden
links between its facets and shedding further light on its nature. On
the other hand, from the experimental perspective, such relations
allow us to obtain information about a certain quantity by measuring a
related one, which could be more easily accessible in the lab.

Unfortunately, the particularly elegant and simple forms of the FDT,
Eqs.~(\ref{einst2}) and~(\ref{fdt}), do not hold when the considered system
is in nonequilibrium conditions, for instance when it is
coupled to thermostats at different temperatures, or during the
relaxation from an initial to a final state. In these cases, due to the presence of
currents (of energy, entropy or particles) crossing the
system, the dynamics is not invariant under time reversal. Nevertheless, fluctuation-dissipation relations (FDRs) between
the response function to an external perturbation in a nonequilibrium
state and the correlation functions computed in the unperturbed
dynamics, can still be derived for a large class of systems.  The
price to pay is that such relations do not share the same generality
of the EFDR, and their explicit forms depend on the system
under investigation.  Several efforts have been devoted to give an intelligible
physical meaning to the extra (nonequilibrium) contributions appearing
in the FDRs, but a clear and shared understanding in terms of
macroscopic thermodynamic quantities is still lacking.

The literature on the generalized FDRs has rapidly increased in the
last decades and we refer the interested reader to the very accurate
reviews recently appeared on this wide
topic~\cite{BPRV08,cuglirev,seifertrev,BM13} to have a complete
overview. In this section we present a concise perspective,
identifying two main classes of FDRs:
\begin{itemize}
\item A) FDRs involving the explicit shape of the invariant probability distribution of the process;
\item B) FDRs involving quantities defined in terms of the microscopic
  dynamical rules (transition rates) of the model.
\end{itemize}
These two classes may have applications in complementary cases,
depending on the problem at hand.

\subsubsection{Class A}

The first kind of nonequilibrium FDR we consider expresses the average
response of the observable $\mathcal{O}$, in terms of correlation
functions, which involve (derivatives of) the stationary distribution
of the system. As shown in~\cite{A72,FIV90}, for a system with states
${\bf X}$, the response of $\mathcal{O}(t)$ to an impulsive
perturbation applied at time $s$, defined in Eq.~(\ref{imp}), takes the
form
\begin{equation}
R_{ \mathcal{O} \mathcal{F}}(t-s)=-\sum_j\left\langle
\mathcal{O}(t)\frac{\delta \log p_{st}({\bf X})}{\delta
  X_j}\Bigg|_s\right\rangle,
\label{vulp}
\end{equation}
where $p_{st}({\bf X})$ is the invariant probability distribution of the system. This
result can be obtained both for deterministic and stochastic dynamics.
Similar formulae derived following different approaches, and
generalized even to non-stationary conditions, have been proposed
recently~\cite{prost,ss10,verley}.

Form Eq.~(\ref{vulp}), the EFDR is immediately recovered replacing
$p_{st}$ with the equilibrium distribution. In general conditions,
however, if the explicit form of $p_{st}$ is known or if some physical
assumptions can be made on it, then Eq.~\eqref{vulp} gives information
on the coupling among degrees of freedom that have to be taken into
account out of equilibrium to reconstruct the response function from
unperturbed correlators. In the following, we will discuss an explicit
example where this formalism can be applied, in the context of
granular systems (see Section 5).

\subsubsection{Class B}

Let us consider a stochastic continuous variable $\alpha$ evolving
according to the following stochastic differential equation
\begin{equation}
\dot{\alpha}(t)=\mathcal{A}[\alpha(t)]+\eta(t),
\end{equation}
where $\mathcal{A}[\alpha]$ is a known function of the variable
$\alpha$ and $\eta$ is a Gaussian noise, with zero mean and variance
$\langle \eta(t)\eta(t') \rangle=2D\delta(t-t')$.  Then, expressing
the response function as a correlation with the
noise~\cite{n65,cugli1},  one obtains the following FDR
\begin{equation}
R_{\mathcal{O} \mathcal{F}}(t,s)
=\frac{1}{2D}\left\{\langle
\mathcal{O}(t)\dot{\alpha}(s)\rangle- \langle
\mathcal{O}(t)\mathcal{A}[\alpha(s)]\rangle\right\},
\label{resp}
\end{equation}
where $\langle\cdots\rangle$ denotes an average over initial
conditions and noise realizations.  The above result is valid in
general non-equilibrium conditions, in stationary or transient
regimes, where an explicit dependence on the two times $(t,s)$ appears
in the response function, and can be easily generalized to vectorial
variables or field theories~\cite{aron}.

From the general expression Eq.~(\ref{resp}), one obtains the FDR for
the Klein-Kramers model~(\ref{KK}), with the following substitutions:
$\alpha=v$, $\mathcal{A}=-\gamma v-dU(x)/dx$ and $D=\gamma T$, which
yields
\begin{equation}
R_{\mathcal{O} \mathcal{F}}(t,s)=\frac{1}{2\gamma T}\Big\{\langle
\mathcal{O}(t)\dot{v}(s)\rangle- \left\langle
\mathcal{O}(t)\left[-\gamma
  v(s)-\frac{dU(x)}{dx}\right]\right\rangle\Big\}.
\label{resp2}
\end{equation}
The equilibrium form of the FDT, Eq.~(\ref{fdt}),
can be recovered from Eq.~(\ref{resp2}) exploiting time translation
invariance and causality. Indeed, the first term in the r.h.s. of
Eq.~(\ref{resp2}) can be rewritten as
\begin{eqnarray}
\langle \mathcal{O}(t)\dot{v}(s)\rangle&=&-\langle \dot{v}(t)\mathcal{O}(s)\rangle
=-\left\langle[-\gamma v(t)-\frac{dU(x)}{dx}+\eta(t)]\mathcal{O}(s)\right\rangle   \nonumber \\
&=&\gamma\langle v(t)\mathcal{O}(s)\rangle + \left\langle\frac{dU(x)}{dx}\mathcal{O}(s)\right\rangle \nonumber \\
&=&\gamma\langle \mathcal{O}(t)v(s)\rangle-\left\langle \mathcal{O}(t)\frac{dU(x)}{dx}\right\rangle, 
\label{qed}
\end{eqnarray}
where we used the Onsager relations and we have assumed that
$\mathcal{O}$ is odd under time inversion.  Finally, substituting
Eq.~(\ref{qed}) into Eq.~(\ref{resp2}) one gets
Eq.~(\ref{fdt}).

Analogously, the FDR for an overdamped Langevin system is obtained
from the relation~(\ref{resp}) by taking $\alpha=x$,
$\mathcal{A}=F(x)/\gamma$ and $D=T/\gamma$, which yields
\begin{equation}
R_{\mathcal{O} \mathcal{F}}(t,s)=\frac{\gamma}{2T}\left\{\langle \mathcal{O}(t)\dot{x}(s)\rangle-
\langle \mathcal{O}(t)\mathcal{A}[x(s)]\rangle\right\}.
\label{resp1}
\end{equation}
These kinds of relations (and their generalizations) have been derived
in different contexts and with different
approaches~\cite{cugli1,ss06,BMW09,BMW10,seifertrev}, providing
several physical interpretations for the nonequilibrium contributions,
in terms of stochastic entropy, entropy production, dynamical
activity, etc. We will briefly discuss some of these concepts below.

The generalization of Eq.~(\ref{resp}) to systems with a
discrete state space, such as the Ising model or spin glasses, is
non-trivial and requires the solution of some technical problems. For
instance, the FDR for discrete variables $\sigma=\pm1$ evolving
according to a Master Equation with transition rates
$w(\sigma\to\sigma')$, in contact with a reservoir at temperature $T$,
takes the following form~\cite{LCZ05}
\begin{equation}
R_{\mathcal{O}\mathcal{F}}(t,s)=\frac{1}{2T}\left\{\frac{\partial}{\partial
  s}\langle \mathcal{O}(t)\sigma(s)\rangle-\left\langle
\mathcal{O}(t)B(s)\right\rangle\right\},
\label{discreto}
\end{equation}
where the quantity $B[\sigma]$ is defined by
\begin{equation}
B[\sigma(s)]=\sum_{\sigma'}[\sigma'-\sigma(s)]w[\sigma(s)\to\sigma'].
\label{B}
\end{equation}
From the relation~(\ref{discreto}), the equilibrium FDT~(\ref{fdt}) is
immediately obtained exploiting the property
\begin{equation}
\left\langle
\mathcal{O}(t)\sum_{\sigma''}[\sigma''-\sigma(s)]w[\sigma(s)\to\sigma'']\right\rangle_{eq}=-\frac{\partial}{\partial
  s}\langle \mathcal{O}(t)\sigma(s)\rangle_{eq},
\label{eqprop}
\end{equation}
valid when the average is taken in the equilibrium
state~\cite{LCZ05}. A unified formalism for the derivation of FDRs
including both the cases of continuous and discrete variables is
presented in~\cite{LCSZ08b}.

In order to illustrate with an explicit example the form assumed by
the FDRs in a specific case, let us consider the two-variable model
introduced previously in Eq.~(\ref{twovar}).  For this system one can
apply both the FDR of class A, Eq.~(\ref{vulp}), and the FDR of class
B, Eq.~(\ref{resp}). Using the known stationary distribution
Eq.~(\ref{pst}), from Eq.~(\ref{vulp}) one immediately gets the
expression (\ref{respz}). Alternatively, applying Eq.~(\ref{resp}), one has
\begin{equation}
R_{v\mathcal{F}}(t)=\frac{1}{2\gamma T_0}\left[-\frac{d}{dt}\langle
  v(t)v(0)\rangle + \frac{\gamma}{m}\langle v(t)v(0)\rangle
  -\frac{\sqrt{M\Gamma_0}}{m}\langle v(t)z(0)\rangle\right],
\end{equation}
where the variable $z(t)$ is an effective field velocity, defined in the
model~\eqref{twovar}.  Obviously the two formulae are equivalent, as it
can be checked.

\subsubsection{Nonlinear FDRs}

The FDRs in the form (\ref{resp}) or (\ref{discreto}) can be
generalized to nonlinear orders, providing a relation between
nonlinear response functions and multi-point correlation functions.
These cases are relevant for instance in the context of disordered
systems, where usually growing characteristic lengths near critical
points cannot be measured through linear response functions or
two-point correlators, which remain always short-ranged.  The problem
of deriving nonlinear FDRs in this context has been addressed
initially in~\cite{BB05}, and the general formalism valid for
arbitrary order has been set in~\cite{LCSZ08a,LCSZ08b}. As an example
we report here the form of the second order response function for a
system of discrete variables $\sigma$ perturbed by two fields
$\mathcal{F}_1$ and $\mathcal{F}_2$ at times $t_1$ and
$t_2$~\cite{LCSZ08b}
\begin{eqnarray}
R_{\mathcal{O}\mathcal{F}}^{(2)}(t,t_1,t_2)&\equiv&\left . \frac{\delta
  \langle \mathcal{O}(t)\rangle_\mathcal{F}}{\delta
  \mathcal{F}_1(t_1)\delta \mathcal{F}_2(t_2)}\right|_{h=0}\nonumber
\\ &=&\frac{1}{4T^2}\left\{\frac{\partial}{\partial
  t_1}\frac{\partial}{\partial t_2}\langle \mathcal{O}(t)
\sigma(t_1)\sigma(t_2)\rangle -\frac{\partial}{\partial t_1}\langle
\mathcal{O}(t) \sigma(t_1)B(t_2)\rangle \right.\nonumber \\ &-&\left
. \frac{\partial}{\partial t_2}\langle \mathcal{O}(t)
B(t_1)\sigma(t_2)\rangle+\langle \mathcal{O}(t)
B(t_1)B(t_2)\rangle\right\},
\label{second}
\end{eqnarray}
that in equilibrium, exploiting the property~(\ref{eqprop}),
simplifies to
\begin{eqnarray}
R_{\mathcal{O}\mathcal{F}}^{(2)}(t,t_1,t_2)=\frac{1}{2T^2}\Big\{\frac{\partial}{\partial
  t_1}\frac{\partial}{\partial t_2}\langle \mathcal{O}(t)
\sigma(t_1)\sigma(t_2)\rangle -\frac{\partial}{\partial t_2}\langle
\mathcal{O}(t) B(t_1)\sigma(t_2)\rangle,
\end{eqnarray}
with $t>t_1>t_2$.  Note that even at equilibrium there remains a dependence
on the quantity $B$, which explicitly involves the transition rates of
the model, making the second order FDR less general than the linear
one. This result has several consequences, implying that kinetic
effects (combined in the quantity $B$), irrelevant at linear order,
have to be taken into account for the nonlinear behavior of the
system, even in equilibrium conditions.  This can be exploited to
obtain information on the dynamical rules governing the system from
the study of the equilibrium second order response, as proposed
in~\cite{basu}. Experiments on colloidal particles in anharmonic
potentials confirm the validity of this theoretical
approach~\cite{helden}.

Another interesting result recently derived in~\cite{FB16}, shows that the
second order response function to an external force, in a form similar
to Eq.~(\ref{second}), can be related to the thermal response of the
system. More specifically, in the case of models described by an
overdamped Langevin dynamics, the linear response to perturbations of
the noise intensity can be expressed as a nonlinear response to a
constant force, leading to the study of nonequilibrium heat capacities
and thermal expansions coefficients.

Finally, let us mention that nonlinear response functions also play a
crucial role in nonlinear optics and quantum spectroscopy.  In
particular, in this context, the lack of a unique nonlinear FDR is
responsible for the differences between classical and quantum
response~\cite{mukamel1,mukamel2,foini17}.

\subsubsection{Applications}

The exact relations derived above find many applications in different
contexts. Here we briefly discuss some of them, referring the
interested reader to the abundant literature on the
subject~\cite{BPRV08,cuglirev,seifertrev,BM13}. \\
 
\emph{Zero-field algorithms.} In general, in
statistical systems, the numerical measure of the linear response
function is a very time-demanding task, due to a poor signal to noise
ratio. Moreover, the linear regime has always to be
checked with repeated measures in a range of different values of
the external field. This problem is overcome by FDRs, which allow one to obtain information on the
response by the measure of suitable correlation functions, without
applying any external perturbation. Moreover, the linear regime is
already ``built-in'' and does not need any further
check. These kinds of algorithm have been used in several
nonequilibrium contexts, from spin systems showing aging
dynamics~\cite{CR03,chatelain,ricci,LCZ05,CLSZ10}, to
glasses~\cite{berthier} and active matter~\cite{szamel}. Note that
some of the relations derived for the purpose of a numerical
implementation~\cite{chatelain,ricci,berthier} cannot be applied to
real systems because the terms appearing in the unperturbed
correlators are only accessible in numerical protocols
(e.g. measurement of attempted, but rejected, moves in Monte Carlo simulations) and
do not correspond to physical quantities experimentally observable. On
the contrary, the form of FDR reported in Eq.~(\ref{discreto})
involves the quantity $B$ defined in~(\ref{B}), which only depends on
the state of the system at a give time, and therefore is an observable
quantity (see~\cite{CLSZ10} for an accurate discussion of this issue
in the context of spin models). \\

\emph{Harada-Sasa relation.} Analogously to the numerical studies, in
real experiments, the use of FDRs can simplify the measurement of some
quantities which are otherwise not directly accessible.  A major
example is the Harada-Sasa relation~\cite{harada1,harada2,deutsch},
where the stationary heat dissipated in the system $J_x$, is expressed
in terms of the violations of the equilibrium FDT in the frequency
domain. For the Langevin equation~(\ref{KK}), this relation reads
\begin{equation}
J_x=\gamma\left\{\langle
\dot{x}\rangle^2+\int_{-\infty}^{\infty}\frac{d\omega}{2\pi}[\tilde{C}(\omega)-2T\tilde{R}'(\omega)]\right\},
\end{equation}
where $\tilde{C}(\omega)$ and $\tilde{R}'(\omega)$ are the Fourier
transforms of the velocity autocorrelation and the real part of the
response function, respectively.  This result has been recently
generalized to systems evolving according to a Master Equation and in
non-stationary states~\cite{LBS14}, and with separate time
scales~\cite{wangsasa}.  These relations turned out to be very useful
in different experimental contexts; for instance to evaluate the
single-molecule energetics of a rotary molecular motor
F1-ATPase~\cite{toyabe}, or to measure the heat dissipated by an
optically confined colloidal particle in an aging gelatin droplet
after a fast quench~\cite{gomez,berut15}. \\

\emph{Nonequilibrium extra-terms.} More generally, from a theoretical
perspective, the off equilibrium FDRs highlight the coupling of the
system variables with new quantities, which play a central role in the
nonequilibrium dynamics. 

A recently proposed line of research focuses on the FDR of class B,
pointing out the symmetry properties under time reversal of the
different terms~\cite{BMW09,BMW10,BM13,basu,basu2}. In particular,
using a path integral formalism, Maes and coworkers have shown that
the part of the action functional which is symmetric under time
reversal, called dynamical activity (traffic, or ``frenesy''), enters
the nonequilibrium FDR and plays a role complementary to the entropy
production (that is antisymmetric for time reversal). The dynamical
activity, which for instance for discrete spin systems takes the
explicit form given in Eq.~(\ref{B}), is responsible for the
nonequilibrium behaviors~\cite{BMW09}. In particular, it is shown that
kinetic factors, such as details of the coupling between the system
and the environments or symmetric prefactors appearing in the
transition rates, play a central role in the characterization of
nonequilibrium response.

On the other hand, following the approach of the FDRs of class A, the
appearance of extra terms singles out the coupling arising out of
equilibrium among degrees of freedom and can be exploited to validate
phenomenological models introduced to describe the system under study.
An enlightening example of such an approach in the context of granular
fluids will be discussed in Section 5.

\subsection{Effective temperature: a subtle concept} 
\label{sub43}

Within the context of non equilibrium phenomena, the first natural
attempts to formulate a general theory start from the concepts well
defined in the consolidate equilibrium theory.  In particular, the
possibility that also out of equilibrium a characterization of the
statistical features of the system in terms of a few variables is still feasible, leads to
extend familiar concepts as temperature to the non equilibrium
realm. As a matter of fact, the question of whether and in what way
such a quantity can be properly defined in this context is a
fundamental issue in the construction of a general theory of
nonequilibrium systems~\cite{cugli2,casas2003}.  This topic is really vast and we refer the
interested reader to the reviews~\cite{leuzzi,cuglirev} for a wider
discussion.

In the next subsections we recall how the effective temperature can be
introduced via the linear FDR and discuss some specific
issues. Explicit applications in the framework of granular systems,
turbulence and active matter, will be discussed in Section 5.

\subsubsection{Definition of effective temperature}

One of the first attempts to define the concept of an effective
temperature $T_{eff}$ in systems out of equilibrium is based on the
linear FDR~\cite{cugli2}. Considering the case of an Ising spin
system, let us start by writing the linear susceptibility, using the FDR~(\ref{discreto})
\begin{eqnarray}
\chi(t,t_w) \equiv \int_{t_w}^t ds R_{\sigma \mathcal{F}}(t,s) =
\frac{\beta}{2} \int_{t_w}^t ds \left [\frac{\partial}{\partial s} C(t,s)
-  \langle \sigma (t) B(s) \rangle \right ],
\label{eff1}
\end{eqnarray}
where $C(t,s)=\langle \sigma(t)\sigma(s)\rangle$ and $t_w$ is a
reference waiting time.  We then introduce the quantity
\begin{equation}
\psi(t,t_w) =  \int_{t_w}^t ds \frac{\partial}{\partial s}C(t,s)  = 1- C(t,t_w),
\label{eff3}
\end{equation}
which, for fixed $t_w$, is a monotonously increasing function of
time. Then, it allows us to reparametrize $t$ in terms of $\psi$ and to
write $\chi(t,t_w)$ in the form $ \chi(\psi,t_w)$, with a little abuse of notation.

In equilibrium, where time translation invariance holds, the
dependence on $t_w$ disappears and the parametric representation
becomes linear, yielding
\begin{equation} \chi(\psi) = \beta \psi,
\label{eff6}
\end{equation}
with the obvious consequence
\begin{equation}
\beta = \frac{d\chi(\psi)}{d\psi}.
\label{eff7}
\end{equation}
Out of equilibrium, the parametric representation is not linear and
an effective temperature can be defined by the generalization of the
above relation
\begin{equation} 
\beta_{eff}(\psi,t_w) =\frac{ \partial \chi(\psi,t_w)}{\partial\psi},
\label{eff8}
\end{equation}
with $\beta_{eff} = 1/T_{eff}$.

The underlying idea for the introduction of an effective temperature
can be precisely illustrated within the picture of coarsening systems,
which after a sudden quench to a low temperature $T$, evolve from the
disordered initial condition at high temperature, via the growth of
domains of positive and negative magnetic order~\cite{zannetti}.
These systems are characterized by a slow relaxation due to the time
dependent correlation length which typically increases with a
power-law behavior.  In order to give a full description of the non
equilibrium dynamics, two-time observables have to be considered. In
particular, the spin-spin autocorrelation $C(t,t_w)$ and the linear
response $R(t,t_w)$ to an external field, in the off-equilibrium aging
regime show a separation of time scales.  This means that, for $t_w$
sufficiently large, such quantities display quite different behaviors
in the short and in the long time regimes, defined by the conditions
$t-t_w \ll t_w$ and $t-t_w\gg t_w$, respectively. More specifically,
$C(t,t_w)$ first approaches a plateau at $M_{eq}^2$ in a stationary
manner, $M_{eq}$ being the equilibrium magntization, and then it
decays below this value with an explicit dependence on $t_w$. At fixed
$t_w$, it decays to zero for sufficiently large $t$.

Such a phenomenology reflects the existence of \emph{fast} and
\emph{slow} degrees of freedom in the system. For ordering kinetics,
the fast degrees of freedom represent the thermal fluctuations within
ordered domains, while the slow ones (e.g. the spontaneous magnetization)
represent domains. In this context
the off-equilibrium behavior during the slow dynamics can be
described by the separation of the time scales for different
groups of degrees of freedom. Then, each of these is considered as in
equilibrium with a different thermostat at some 
effective temperature, which depends on the time scale and
differs from the physical temperature of the real external thermostat. 
The value of $T_{eff}$ can be therefore obtained by forcing
the out of equilibrium linear FDR in the equilibrium form of FDT, as
in Eq.~(\ref{eff8}).  Since in coarsening systems the full response
function obeys the asymptotic form
\begin{equation}
\lim_{t_w \to \infty}\chi(\psi,t_w) = \left \{ \begin{array}{ll}
        \beta \psi, \;\;$for$ \;\; 0 \leq  \psi \leq 1-M_{eq}^2  \\
        \beta (1-M_{eq}^2),  \;\; $for$ \;\; 1-M_{eq}^2 < \psi \leq 1,
        \end{array}
        \right .
        \label{eff10}
\end{equation}
from Eq.~(\ref{eff8}) one obtains
\begin{equation}
\lim_{t_w \to \infty}\beta_{eff}(\psi,t_w)    = \left \{ \begin{array}{ll}
        \beta, \;\;$for$ \;\; 0 \leq  \psi \leq 1-M_{eq}^2  \\
        0,  \;\; $for$ \;\; 1-M_{eq}^2 < \psi \leq 1.
        \end{array}
        \right.
        \label{eff12}
\end{equation}
Namely, the effective temperature coincides with the temperature of
the thermostat in the short time regime, where the system seems to be
equilibrated, while it is drastically different from it in the out of
equilibrium regime, at long times, where it takes the value
$T_{eff}=\infty$.  This result suggests that in the quench below the
critical point, the fast degrees of freedom thermalize, whereas the slow
ones do not interact with the thermal bath and keep on
remaining to the temperature of the initial condition.

\begin{figure}
  \begin{center}
\includegraphics[width=7cm,clip=true]{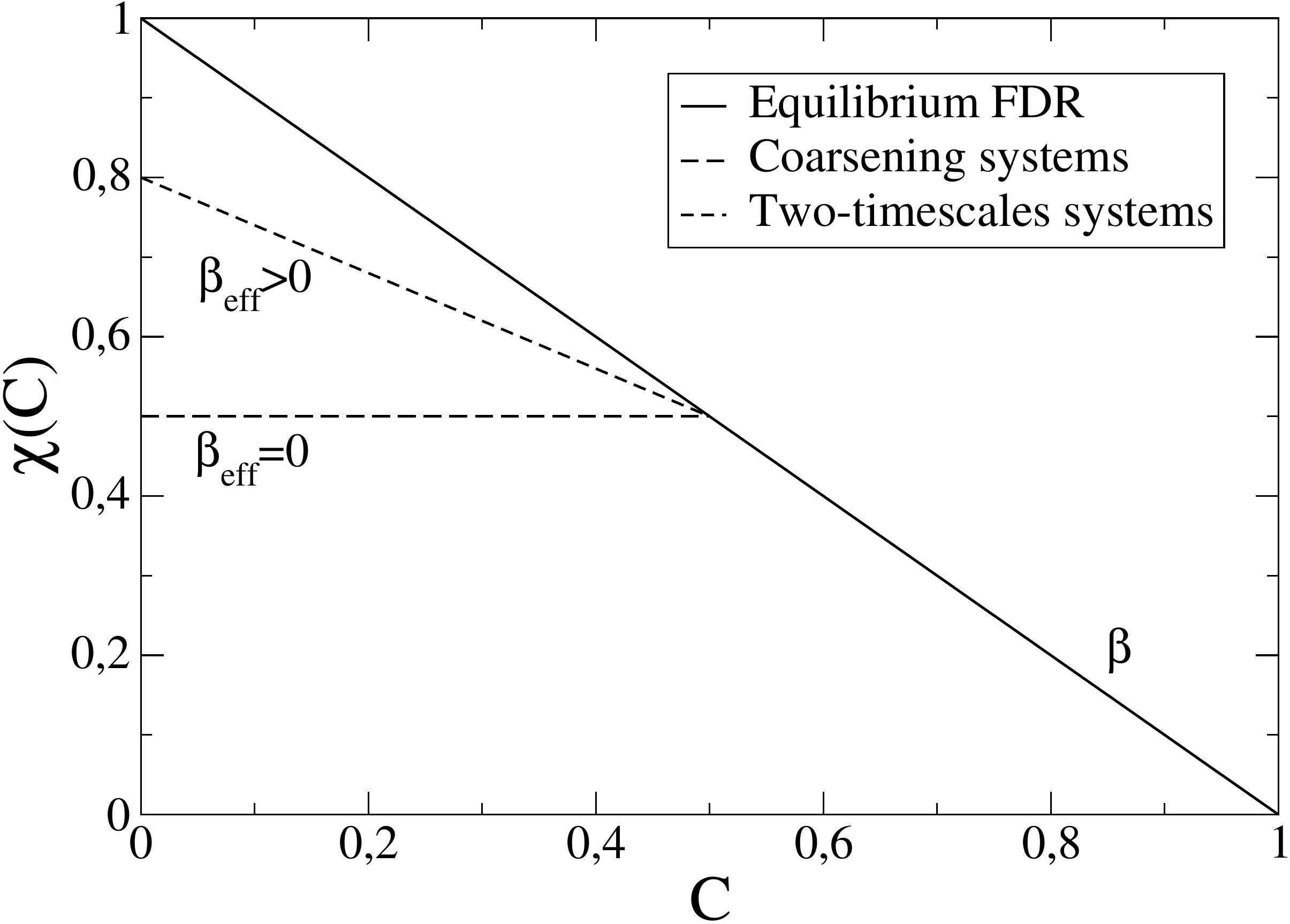}
\caption{\label{efftemp} Schematic example of a parametric plot $\chi$ vs. $C$ showing effective temperatures in different systems.}
\end{center}
\end{figure}

In disordered glassy systems the above scenario can show different,
more complex behaviors~\cite{CR03}. For instance, in models of fragile
structural glasses the effective temperature takes two values: $T$ in
the stationary regime, but $T<T_{eff}<\infty$ in the aging
one~\cite{cugli2,kob2000,dileonardo2000,berthier,gnan2010}, see
Fig.~\ref{efftemp} showing a typical parametric plot $\chi$
vs. $C$. Models of spin glasses are instead characterized by the
presence of a continuous range of temperatures in the aging regime,
extending from a lower bound $T^* > T$ up to infinity~\cite{CR03}.

\subsubsection{Effective temperature and Fluctuation Relations}

Fluctuation Relations include recent theoretical results on general
symmetry properties of the density probability of quantities such as
entropy production, heat and work in non-equilibrium
systems. Initially discovered in numerical simulations of molecular
fluids under shear~\cite{ECM}, these relations have been proven for a
large class of models~\cite{GC,CJ,K98,LS99,GK}, and represent an important
step forward in the construction of a general theory for off
equilibrium processes (see~\cite{seifertrev} for a recent review).

We have already encountered an example of these relations, when
studying some general properties of non-equilibrium currents in
stationary states, see Eq.~(\ref{fr}) above. For a system
in contact with a single thermostat at temperature $T$, a small
external perturbation $\delta \mathcal{F}_0$ produces a current
$J_\tau$, which is proportional to the entropy production associated
with the driving via the temperature $T$, see Eq.~(\ref{clausius}).
In this simple case, the asymmetry function, defined as
\begin{equation}
R(J_\tau) =\frac{1}{\tau}\log\frac{p(J_\tau)}{p(-J_\tau)},   
\end{equation}
shows a linear behavior $R(J_\tau)=J_\tau\delta\mathcal{F}_0/k_BT$,
with slope $\delta \mathcal{F}_0/k_BT$.  

In more general situations, defining the fluctuating entropy production $\Sigma_\tau$ as 
\begin{equation}
\Sigma_\tau=\log\frac{\mathcal{P}(\{\bold{X}\}_0^\tau)}{\mathcal{P}(\mathcal{I}\{\bold{X}\}_0^\tau)},
\label{ep}
\end{equation}
where $\mathcal{P}(\{\bold{X}\}_0^\tau)$ represents the probabiltiy to
observe the trajectory $\{\bold{X}\}_0^\tau$ in the time interval
$[0,\tau]$, and $\mathcal{I}$ denotes the time-inversion operator, one
can prove, under quite general assumptions, that the density
distribution of $\Sigma_\tau$, $p(\Sigma_\tau)$, satisfies the following relation
\begin{equation}
\log\frac{p(\Sigma_\tau)}{p(-\Sigma_\tau)} =\Sigma_\tau.
\end{equation}
The properties of $\Sigma_\tau$ have been investigated in great detail
in several cases. In particular, the study of the distribution
$p(\Sigma_\tau)$, in many experiments and numerical works, revealed a
non-universal form and the thermodynamic interpretation in terms of
macroscopic heat, work and entropy is, in general, an open issue. For
practical purposes, a central point relies on the fact that the
definition in Eq.~(\ref{ep}) is not always useful, due to the
difficulty to get access directly to $\Sigma_\tau$ (for instance in
experiments, where usually measures are constrained to somehow
coarse-grained quantities), and one has to resort to an expression
based on phenomenological grounds. In these cases, a possible way out
is to assume that the entropy production is the heat exchanged by
the system with the environment (or the work applied on the
system), divided by a parameter playing the role of an effective
temperature.  However, extreme care must be used, because
``reasonable'' candidates for this parameter, such as the bath
temperature or the (rescaled) kinetic energy of some degrees of
freedom, could be misleading. Therefore, an accurate analysis of the
model is necessary, as highlighted in different non-equilibrium
situations, where, for instance, the system is coupled with themostats
at different temperatures, or where the action of the thermal bath is
irrelevant, as in athermal
systems~\cite{FM04,PVBTW05,sarra10b,GPSM12,EWS11}. An explicit example
in the context of granular fluids will discussed in
Subsection~\ref{sub51}.

In the context of non-stationary, slow relaxing systems, which cannot
equilibrate even in the absence of external driving, the issue of the
relation between the effective temperature defined through the FDR and
the Fluctuation Relations for the entropy production has been addressed
in~\cite{sellitto} for kinetically constraint models, and
in~\cite{ZBCK05} for mean-field glassy models and coupled Langevin
equations. Other studies focused on the symmetry properties of the
distribution of the heat exchanged by the system with the bath during
the relaxation after a quench to the low temperature phase, in spin
galss models and for Brownian oscillators~\cite{CR04,picco,CSZ}.

\subsubsection{Some caveats on effective temperature}

The real usefulness of the concept of effective temperature and its
range of applicability is still under debate~\cite{cuglirev}.  The
first issue is concerned with the properties satisfied by the standard
equilibrium temperature.  In fact, in order to get the status of
temperature, the new quantity ought to satisfy the properties imposed
by thermodynamics.  For non-equilibrium stationary systems,
characterized by the presence of currents due to the simultaneous
action of driving and dissipation, forces of non-thermal origin are
also present, and one wonders whether the observed
fluctuations of the relevant quantities under investigation can be
described through a single parameter playing the role of an effective
temperature. For instance, granular media - as discussed in Section 5 - are usually subjected to
vibration and dissipate energy in the particle-particle collisions,
whereas systems of active particles present an internal conversion of
energy into directed motion.  In these cases, an equilibrium-like
description of the system is then reliable if this parameter satisfies
some requirements, dictated by the equilibrium picture for Brownian
fluctuations. In particular, it should appear i) in the amplitude of
the effective noise in the system, ii) in the proportionality constant
between the linear response of an observable and the corresponding
correlation function, iii) in the variance of the distribution
probability of relevant quantities (such as energy), if Gaussian. In
Section 5 we will discuss in more details these issues.

Another important requirement is that the effective temperature ought
to be independent of the observable used to construct the correlations
and responses functions appearing in the FDR. This condition seems
quite strong and turns out to be not satisfied in different
systems~\cite{fielding,calabrese,martens}.  However, within the
mean-field theory, the effective temperature has been shown to display
many of the required properties~\cite{cugli2}. In particular, within a
time sector, it is the reading which would be shown by a thermometer
tuned to respond on that timescale.  Moreover, it is independent of
the observable used to construct the FDR.  Conversely, in non-mean
field models, such properties are not yet well established.
Nonetheless, the concept remains quite useful and suggestive.

The presence of a separation of time scales in the system is certainly
a necessary (but not sufficient) condition for the application of the concept of effective
temperature. For instance, it is possible to chose the parameters in the model introduced
in Eqs.~(\ref{twovar}) in order to have a scenario with time-scale separations which is quite different from that observed in aging glassy systems,
as accurately discussed in~\cite{VBPV09}. 
In the general case, the response-correlation
parametric plot may be more difficult to read,
showing intermediate ``effective temperatures'', as well as a more
complex nonlinear shape.

A preliminary exploration of another open end in the important
question of how general the effective temperature can be, focused on
whether it is possible to extend this definition to the nonlinear FDR,
consistently with what it is done in the linear case. In fact, if such
a concept actually provides important information in the
characterization of the off-equilibrium behavior, one expects it to be
independent of the order of the FDR used to construct it.
In~\cite{LCSZ08b}, it has been shown that the concept of effective
temperature can be extended to the second order, drawing a consistent
picture with that obtained in the linear case, within the context of
non disordered coarsening systems.  The results obtained provide a
further support to the idea that the effective temperature can play an
interesting role in the description of the off equilibrium dynamics of
such systems. It would be very interesting to apply the same line of
reasoning presented in~\cite{LCSZ08b} to the case of disordered
systems, such as structural or spin glasses.

Finally, let us note that the concept of effective temperature has
also been revisited within the theoretical framework for linear
response proposed by Maes and coworkers~\cite{BMW09b}. From this
perspective, the effective temperature is given by the ratio between
the dynamical activity and the entropic terms, measuring the weight of
``active states'' in the dynamics.

Explicit examples of the application of an effective temperature,
together with a critical analysis of its practical relevance, will be
discussed in Section~\ref{sec:nonham} in different contexts, from
granular systems to turbulence and active matter.


\section{Towards a temperature for non-Hamiltonian systems?}
\label{sec:nonham}

\subsection{Thermodynamics  and Chaos}

About three decades ago, large deviations theory inspired a sort of
``thermodynamic formalism'' for chaotic systems and fully developed
turbulence~\cite{benzi84,benzi85,paladin87,bohr98,beck97}.  For sake
of completeness we briefly discuss, avoiding too technical details,
such a topic stressing its relation with statistical mechanics~\cite{chaos}.

\subsubsection{Multifractal measures for chaotic attractors}

Let us condider a smooth measure  $\mu({\bf y})$. For any ${\bf x}$, one has
$$
\int_{|{\bf x} -{\bf y}|<  \ell}\ d \mu({\bf y})
 \sim  \ell^{d},
$$
where $d$  is the (integer) dimension of the phase space.
In the case of a chaotic attractor one has a
singular behavior: in the simplest case (homogeneous fractal measure) one has
$$
\int_{|{\bf x} -{\bf y}|<  \ell}\ d \mu({\bf y})
 \sim  \ell^{D_F},
$$
where ${\bf x}$ belongs to the attractor, and $D_F$ is the fractal dimension.
On the other hand,  usually the measure on a chaotic attractor
(as well as in fully developed turbulence) is not homogeneous, 
and  a single fractal dimension
is not enough to give a complete description of the singularities.

Consider a partition of  the attractor in cells (boxes)    $\{ \Lambda_i(\ell) \}$ 
of size $\ell$, and introduce the probability to stay in the   $i-$th cell:
\begin{equation}
P_i(\ell)=\int_{\Lambda_i(\ell)} d\mu({\bf x}).
\label{C1}
\end{equation}
Let us now  introduce  Renyi's dimensions
 $d_q$ (where $q$ is a real number)~\cite{paladin87,beck97}:
\begin{equation}
{\cal M}_q(\ell)=
\sum_i P_i(\ell)^q \sim  \ell^{(q-1)d_q},
\label{C2}
\end{equation}
or, in a more formal way
$$
d_q=\lim_{\ell \to 0} \, {1 \over q-1} \,  { \ln  {\cal M}_q (\ell) \over \ln \ell}.
$$ 
It is possible to show that  $d_q$ must be a non increasing function, and, in addition, 
for $q=0$ one has $d_0=D_F$.
Of course if the measure is homogeneous one has
$d_q=D_F$, and more   $d_q$  is far from the horizontal line $D_F$, more the disomogeneity of the measure is strong.

The information dimension $d_1$ is the most
relevant one:  for almost any
 ${\bf x}$ on the attractor  one has
 $$
\int_{|{\bf x} -{\bf y}|<  \ell}\ d \mu({\bf y})
 \sim  \ell^{d_1}.
$$
There exists a nice way to describe the  singularities of the measure
 using a procedure which follows an approach in terms of  the large deviation theory and
  quite similar to that one used in  statistical mechanics.
Such an approach is called  multifractal~\cite{paladin87,bohr98,beck97}:
the starting point  consists in grouping all the boxes having the same singularity index 
$\alpha$, i.e. all the $i$ such that $P_i(\ell) \sim \ell^{\alpha}$.
Let $d N(\alpha, \ell)$  be the number of boxes with crowding index in the infinitesimal interval 
 $[\alpha, \alpha + d \alpha]$,
 we can rewrite the quantity (\ref{C2}) as an integral over $\alpha$
\begin{equation}
{\cal M}_q(\ell) \sim \int \ell^{\alpha q}  d N(\alpha, \ell), 
\label{C3}
\end{equation}
where we have used the scaling relation 
$P_i(\ell) \sim \ell^{\alpha}$.
Let us now   introduce the multifractal spectrum of singularities, i.e. the fractal dimension
 $f(\alpha)$  of the subset with singularity $\alpha$:
 in the limit $\ell \to 0$   the number of boxes  with crowding index $\alpha$, scales as
 \begin{equation}
  d N(\alpha, \ell) \sim  \ell^{-f(\alpha)} d \alpha.
 \label{C4}
 \end{equation}
 Therefore we have
 \begin{equation}
 {\cal M}_q(\ell) \sim \int_{\alpha_m}^{\alpha_M} \ell^{[q \alpha -f(\alpha)]}  d \alpha,
 \label{C5}
 \end{equation}
 where $\alpha_m$ and $\alpha_M$ are the minimum and maximal  value of $\alpha$ respectively, 
 corresponding to the strongest ($\alpha_m$) and weakest ($\alpha_M$) singularity.
 In the limit  $\ell \to 0$  one has 
 $$
d_q={ 1 \over q-1} min_{\alpha}\{ q \alpha -f(\alpha) \}=
{ 1 \over q-1}  [q \alpha_*(q) -f(\alpha_*(q))],
$$
where   $\alpha_*(q)$  is  the  solution of
$$
{d \over d \alpha} [  q \alpha -f(\alpha)] =0.
$$
It is easy to see that  for $q \gg 1$ one has $\alpha_*(q)\simeq  \alpha_m$, i.e. the strongest singularity,
while for $q \ll- 1$ the weakest singularity is selected $\alpha_*(q)\simeq  \alpha_M$.
At varying $q$, from  the value $d_q$  we are able to understand 
the weight of   the singularities $\alpha$.

\subsubsection{Fluctuation of the finite time Lyapunov exponent}

A  quite similar approach has been used to describe the statistical features 
of the   fluctuations of the finite time Lyapunov exponent~\cite{benzi85,paladin87,beck97}.
Let us introduce   the fluctuating quantity $\gamma$ defined as
$$
\gamma(\tau, t)= {1 \over \tau} \ln \left[ { |{\bf w}(t+\tau)| \over | {\bf w}(t)|} \right],
$$
where  ${\bf w}(t)$ is the tangent vector in phase space (basically 
the difference between two very close trajectories when such a difference is very small);
 $\gamma(\tau, t)$  indicates the  local, growth rate of the tangent vectors within the time interval 
$[t, t+\tau]$.
\\
The large fluctuations of $\gamma$ can be quantitatively characterised 
in terms of  the so called generalised exponents $L(q)$:
$$
\langle e^{q \gamma \tau} \rangle \sim e^{L(q) \tau},
$$
where $\tau$ is large.
It is easy to show that
the Lyapunov exponent can be expressed in terms of $L(q)$:
$$
\lambda= \lim_{\tau \to \infty}  \langle  \ln \gamma \rangle= {d L(q) \over d q}\Big|_{q=0}.
$$
The large deviation theory suggests the shape, at large $\tau$, of the probability distribution density:
\begin{equation}
P(\gamma, \tau) \sim e^{-S(\gamma) \tau},
\label{C6}
\end{equation}
where $S(\gamma)$ is a Cramer's function (see Section 2.2), with
$S(\lambda)=0, \, S(\gamma \ne \lambda) >0$ and $d^2 S/d\gamma^2 >0$.
Using  Eq.~(\ref{C6})  we have
$$
\langle e^{q \gamma \tau}  \rangle =
\int e^{\tau q \gamma}   P(\gamma, \tau) d \, \gamma 
\sim \int e^{\tau [q \gamma -S(\gamma)]} d \, \gamma,
$$
and the Laplace's method shows how the generalised Lyapunov exponents are given by a Legendre transformation~\cite{benzi85,paladin87,fujisaka83,fujisaka87}
$$
L(q)=max_{\gamma} [q \gamma -S(\gamma)].
$$
Of course for $L(q)$ and $S(\gamma)$ we can repeat in a straightforward way the same
remarks previously discussed for $d_q$ and $f(\alpha)$.

\subsubsection{Analogies with the equilibrium statistical mechanics}

Let us now briefly remind that from the statistical mechanics
of a system of $N \gg 1$,
we can write the partition function ${\cal Z}$ as
$$
{\cal Z}(\beta, N)=\int e^{ -N[ \beta e - s(e)]} d \, e =  e^{ - N \beta F(\beta)}\,\, ,
$$
being $e=E/N$  the energy for particle, $s(e)$ the microcanonical entropy
and $F(\beta)$ the free energy per particle.
We have   $F(\beta)=e_* - T s(e_*)$,  where $e_*$ depends on the temperature,
and is selected by the minimum of  $e - T s(e)$.
We can say that varying the temperature $T$, we can "explore" the phase space of
the Hamiltonian.

It is not difficult to realize the strong analogies among the multifractal description
of singular measure, fluctuations of Lyapunov exponent and equilibrium statistical
mechanics~\cite{paladin87,fujisaka87,beck97}.
The following table summarises the correspondences.
\\
\\
\begin{tabular}{ccc}
\hline
\multicolumn{3}{c}{{\bf Correspondences among multifractal measures (MM), fluctuations}}\\
\multicolumn{3}{c}{{\bf of the Lyapunov exponent (FL), and statistical mechanics (SM)}}\\
\hline
$ {\bf MM}$  & $\,\,\,\,\,\,\,\,\,\,\,\,\,\,\,\,\,\,\,\,\,\,\,\,\,\,\,\,\,\,\ {\bf FL}$  & ${\bf SM}$\\
 $\ell \to 0$   &  $\,\,\,\,\,\,\,\,\,\,\,\,\,\,\,\,\,\,\,\,\,\,\,\,\,\,\,\,\,\,\  \tau \to \infty$  &  $N \to \infty$   \\
 $\alpha$    &  $\,\,\,\,\,\,\,\,\,\,\,\,\,\,\,\,\,\,\,\,\,\,\,\,\,\,\,\,\,\,\ \gamma$  & $e$ \\
 ${\cal M}_q(\ell)$  & $\,\,\,\,\,\,\,\,\,\,\,\,\,\,\,\,\,\,\,\,\,\,\,\,\,\,\,\,\,\,\  e^{L(q)\tau}$   & ${\cal Z}(\beta, N)$ \\
 $f(\alpha)$  &   $\,\,\,\,\,\,\,\,\,\,\,\,\,\,\,\,\,\,\,\,\,\,\,\,\,\,\,\,\,\,\ S(\gamma)$  & $s(e)$ \\
  $q$ &  $\,\,\,\,\,\,\,\,\,\,\,\,\,\,\,\,\,\,\,\,\,\,\,\,\,\,\,\,\,\,\ q$ & $\beta$ \\
 $(q-1)d_q$  &\,\,\,\,\,\,\,\,\,\,\,\,\,\,\,\,\,\,\,\,\,\,\,\,\,\,\,\,\,\,\  $L(q)$  & $\beta F(\beta)$ \\
\hline
\end{tabular}
\\
\\

We conclude this subsection noting that in MM (as well in FL) $q$ is a
free parameter, which does not appear in $\mu ({\bf x})$ (as well as in
$P(\gamma, t)$), and can be changed in order to explore the typical
statistical features (for $|q| < O(1)$), or extreme ones (for $|q| \gg
1$).  Since in SM $\beta$ appears in the probability distribution of
$e$, one could conclude that the above analogies are just formal.

On the contrary  even in SM it is possible,  following a procedure quite similar to that one 
used in MM (as well as FL), to investigate the statistical features at any $\beta $
computing the mean value of suitable observables
at a given $\beta_0$.
For instance it is possible  to compute  the free energy per particle $F(\beta)$ performing an average
with the canonical probability density  at $\beta_0$.
For $\beta=\beta_0 + \Delta \beta$ a trivial computation gives
$$
e^{-\beta N F(\beta)}=\int e^{-\beta H({\bf x})} d \, {\bf x}=\int e^{-\beta_0 H({\bf x})} 
e^{-\Delta \beta H({\bf x})} d \, {\bf x}
=e^{-\beta_0 N F(\beta_0)} \langle  e^{-\Delta \beta H({\bf x})} \rangle_{\beta_0} \, ,
$$
where $\langle (\,\, ) \rangle_{\beta_0}$ indicates the average according to the canonical distribution
at $\beta_0 $:
$$
\rho_{\beta_0}({\bf x})= e^{\beta_0 N F(\beta)} \,  e^{- \beta_0 H({\bf x})} \,\, .
$$ 
The idea of the  previous result is  the so called {\it data reweighting} technique  
used in Monte Carlo numerical computations~\cite{peli14}.

 \subsection{Statistical Hydrodynamics}

In Section 3 we already discussed the statistical mechanics of a
system of point vortices in $2d$ incompressible fluids.  Such a system
is Hamiltonian, therefore, although the Hamiltonian has not the usual
shape (\ref{Usual1}), it is enough to use the standard statistical
mechanics. 
  
\subsubsection{Perfect fluids}
Usually, apart from a few special cases, a fluid (even in the absence of viscosity) does not obey Hamiltonian equations.
In spite of this, one can easily build an equilibrium statistical mechanical approach 
to the Euler equation: it is enough to follow straightforwardly  the path used to obtain
 the  micro-canonical formalism~\cite{rose78,kraichnan80,bohr98}.

Let us consider a $3d$ perfect fluid, i.e. with zero viscosity,
and without external forcing, in a cube of edge $L$ with periodic boundary
conditions, so that the velocity field ${\bf u}({\bf x},t)$  can be expanded in Fourier
series as
\begin{equation}
 u_j({\bf x},t)=\frac{1}{L^{3/2}}\sum_{n_1,n_2,n_3} e^{i(k_1x+k_2y+k_3z)} v_j({\bf k},t),
\label{eqno7}
\end{equation}
where
$$
{\bf k}={2\pi \over L}(n_1,n_2,n_3),
$$
being  $n_j$ integer numbers.
In addition we introduce an ultraviolet truncation $v_j({\bf k})=0$ for
$|{\bf k}|>K_M$, being $K_M$ the maximum allowed wave vector. 

Because of the incompressibility condition ($\nabla \cdot {\bf u}=0$)
and the fact that the velocity field is real, the variables $\{
v_j({\bf k},t) \}$ are not independent, and indeed one has
\begin{eqnarray}
\sum_{j=1}^3 k_jv_j({\bf k}, t) = 0 & {\rm and} & v_j(-{\bf k},t) = \left[v_j({\bf k}, t)\right]^\star,
\end{eqnarray}
where $\star$ denotes complex conjugation. Therefore it is useful to  
introduce a new set of real variables
$\{ X_n(t) \}$ which replace $\{ v_j({\bf k},t) \}$, and obey an
ordinary differential equation:
\begin{equation}
{d X_n \over dt}=\sum_{m,\ell}M_{n,m,\ell}X_m X_{\ell}, 
n=1,2,...,N.
\label {Flu1}
\end{equation}
From Euler's equation we have the following properties:
 $M_{n,m,\ell}=M_{n,\ell,m}$  and
$M_{n,m,\ell}+ M_{m,\ell,n}+ M_{\ell,n,m}=0$: for details see~\cite{kraichnan80}.
Because of the introduction of the ultraviolet truncation,
we have a finite system of equations, and therefore one
avoids the infinite energy problems of the classical field
theory.

Since Eq.~(\ref{Flu1}) conserves the volume in the phase space
(Liouville theorem)
\begin{equation}
\sum_n{\partial \over \partial X_n} {d X_n \over dt}=0,
\label{eqno9}
\end{equation}
and in addition one has the (energy) conservation law
$$
{1 \over 2}\sum X_n^2=E.
$$
It is straightforward, following the usual approach of 
equilibrium statistical mechanics, i.e. assuming that ergodicity holds, to obtain
 the microcanonical distribution:
\begin{equation}
P_{mc}(\{ X_n \}) \propto \,\,\, \delta\left( {1 \over 2}\sum X_n^2-E \right).
\label{eqno10}
\end{equation}
In addition, the $N \to \infty$ limit yields the canonical distribution
\begin{equation}
P_{c}(\{ X_n \}) \propto  \exp \left[- \left( {\beta \over 2}\sum X_n^2 \right)\right],
\label{eqno11}
\end{equation}
and therefore
\begin{equation}
\langle X_n^2\rangle={2 E \over N}= {1 \over \beta}.
\label{eqno12}
\end{equation}
Let us note that,
 although the system is not Hamiltonian, we
have a well defined temperature (which can be
identified with $1/\beta$), which, as in the usual statistical mechanics
for Hamiltonian systems,
is the relevant parameter for the probability distribution in phase space.

It is  remarkable   that  for perfect fluids one has a rather simple fluctuation-dissipation
relation. From the general result discussed in Section 4, one has
$$
\overline{ \delta X_n(t)}=\sum_j R_{n,j}(t) \delta X_j(0),
$$
and, in the limit $N \gg 1$,
\begin{equation}
R_{n,j}(t)= {\langle X_n(t) X_n(0) \rangle \over \langle X_n^2 \rangle } \delta_{nj}.
\label{FRPF}
\end{equation}
Let us stress  an important technical aspect:
 although the probability density distribution for
$\{ X_n \}$ is Gaussian, the dynamics is not linear and the shape of each
correlation  $\langle X_n(t) X_n(0) \rangle$ is not trivially exponential~\cite{biferale02,BPRV08}.

The previous procedure,  used for $3d$ perfect fluids,
can be easily generalized to the
two-dimensional case.
In such a system, in addition to the energy, there is a second conserved quantity,
the enstrophy:
\begin{equation}
\Omega={1 \over 2} \sum_n k_n^2 X_n^2.
\end{equation}
Therefore the microcanonical distribution should be defined on the
surface in which both energy and enstrophy are constant:
\begin{equation}
P_{mc}(\{ X_n \}) \propto \,\,\, \delta\Big( {1 \over 2}\sum X_n^2-E \Big) \,
\delta\Big( {1 \over 2}\sum k_n^2 X_n^2-\Omega \Big),
\label{2D}
\end{equation}
and, in the large $N$ limit, we have the canonical distribution
\begin{equation}
P_{c}(\{ X_n \}) \propto  \exp \left[- \Bigl(
 {\beta_1 \over 2}\sum X_n^2 +  {\beta_2 \over 2}\sum k_n^2 X_n^2   
\Bigr)\right],
\label{eqno13}
\end{equation}
with
\begin{equation}
<X_n^2>={1 \over \beta_1+ \beta_2 k_n^2}.
\label{eqno14}
\end{equation}
Now, due to  the presence of two conservation laws,
we have two ``temperatures'', $1/\beta_1$ and $1/\beta_2$.

Let us open a short parenthesis on the possibility to have ``more than
one temperature'': the presence of a unique temperature in the typical
cases of the statistical mechanics is due to the fact that usually one
considers systems with only one conservation law (energy).  However in the presence of 
another conservation law, a ``second temperature'' appears in
a quite natural way~\cite{grad52}.  For instance in a system with
central forces among particle pairs in a cylindrical vessel with elastic
walls, the energy and the angular momentum along the axis of the
cylinder are constant, resulting in a microcanonical distribution
similar to Eq.~(\ref{2D}).

Beyond $3d$ and $2d$ Euler equations, in hydrodynamics there are other
interesting systems described by inviscid ordinary differential
equations, such as Eq.~(\ref{Flu1}), with quadratic invariants, for
which the Liouville theorem holds (e.g. in magnetohydrodynamics and
geostrophyc flows).  Detailed numerical simulations show that such
systems are ergodic and mixing if $N$ is large: arbitrary initial
distributions of $\{ X_n \}$ evolve towards the
Gaussian~(\ref{eqno11}) or~(\ref{eqno13}), and the statistical
mechanics predictions are in perfect agreement with the actual
results, see~\cite{kraichnan80}.

\subsubsection{Viscous fluids}

In the presence of viscosity $\nu >0$,  Eq.~(\ref{Flu1}) must be modified:
\begin{equation}
{d X_n \over dt}=\sum_{m,\ell}M_{n,m,\ell}X_m X_{\ell} -\nu k^2_n X_n + f_n,
\qquad n=1,2,\ldots, N.
\label {TU1}
\end{equation}
The external forcing $f_n$ is necessary in order to have a statistical
steady state.  Apparently in the limit $\nu \to 0$, and $f_n \to 0$,
the system (\ref{TU1}) can sound rather similar to the perfect fluids.
The actual situation is rather different: as paradigmatic example we
can remind that in $3d$ the limit $\nu \to 0$ of the Navier-Stokes
equations, is singular and cannot be interchanged with the limit $K_M
\to \infty$.  Therefore the statistical mechanics of an inviscid fluid
has a rather limited relevance for the Navier-Stokes equations at
very high Reynolds numbers~\cite{frisch95}.

Let us briefly remind the basic phenomenology of the fully developed
turbulence: in the $\nu \to 0$ limit and in the presence of forcing at
large scale (small $k$), one has an intermediate range, called the
inertial range, where practically there are neither pumping nor
dissipation.  In addition there is a strong departure from
equipartition~(\ref{eqno12}): instead of $\langle X_n^2 \rangle =
const.$ one has $\langle X_n^2 \rangle \sim k_n^{-\gamma}$ where
$\gamma \simeq 11/3$, the value $\gamma=11/3$ corresponding to the
Kolmogorov spectrum.  

Let us stress that while the statistical mechanics of perfect fluids
has an equilibrium behavior, the statistical features of turbulence
belongs to the non equilibrium realm.  There is an energy flux
(cascade in the turbulent jargon) from large scale, i.e. small $k$
where the energy is injected, to small scale, where the energy is
dissipated by the molecular viscosity~\cite{frisch95,monin75,bohr98}.

Because of viscosity, the system~(\ref{TU1}) is dissipative, and therefore
the measure on the attractor is singular, see~\cite{bohr98}.  On the
other hand the above point is not the main difficulty: one can avoid
the mathematical trouble of the singular measure simply by adding a
small noise.  With the introduction of a noisy term, which is quite
natural from a physical point of view, we have a Langevin equation,
and therefore a Fokker-Planck equation whose stationary solution for
the PdF is surely non singular and continuous with respect to the
Lebesgue measure, but, unfortunately its shape is not known.

Although from the generalised fluctuation-dissipation relation, see
Section 4, we can say that there exists a relation linking response
and a (suitable) correlation, i.e.
$$
R_{n,j}(t)=  \langle X_n(t) F_j({\bf X}(0) \rangle,
$$ 
where the functions $F_j(\,)$ depend on the shape of the stationary
PdF, as a matter of fact the statistical mechanics of turbulence is
still a difficult open problem and also the existence of an FDR is not
particularly useful at a practical level.

\subsubsection{Difficulties for the introduction of temperature}

Let us note that in all the cases where  we are able   to introduce,
in a consistent way, the temperature, such a quantity   plays   the role of the  (unique) parameter
appearing in the probability distribution, with a known specific  functional shape.
For equilibrium  Hamiltonian systems one has
$$
\rho({\bf x})={1 \over Z} e^{-\beta H({\bf x})} \,\, .
$$
In a similar way in  systems ruled by a Langevin equation with a gradient drift $-\nabla V({\bf x})$,
\begin{equation}
\frac{dx_n}{dt}= -\frac{\partial V}{\partial x_n} +\sqrt{\frac{2}{\beta}} \eta_n \, ,
\label{LG}
\end{equation}
one has
$$
\rho({\bf x})={1 \over Z} e^{-\beta V({\bf x})} \,\,.
$$
The possibility to introduce a temperature from the comparison
between response function and correlation function
is a consequence of the specific shape of the PdF,
e.g. for Langevin  equation (\ref{LG}) one has
\begin{equation}
R_{n,j}(t)=  -\beta \frac{ d}{d t} \langle x_n(t) x_j(0) \rangle=
 -\beta \left\langle x_n(t) {\partial V({\bf x}(0)) \over \partial x_j(0)} \right\rangle    \,\, .
\label{TU2}
\end{equation}
\\
On the contrary, for  turbulence the introduction of the concept of temperature does not appear as
an easy task even assuming a simple relation between response functions and correlations.
For  instance in the  direct-interaction-approximation (DIA) developed by 
Kraichnan~\cite{kraichnan59,kraichnan00,kiyani04} one assumes:
\begin{equation}
R_{n,n}(t)= \frac{ \langle X_n(t) X_n(0) \rangle} {\langle X_n^2 \rangle}\,,
\label{TU3}
\end{equation}
as a consequence, in such an approximation, the PdF is a Gaussian function.
In spite of  the fact that the above relation appears identical to (\ref{FRPF}) for the perfect fluid, 
there is an important difference.
For the perfect fluids $ {\langle X_n^2 \rangle}=1/\beta$, so $\beta$ is the unique parameter
in the (Gaussian) PdF.
On the contrary  in turbulent flows the scenario is rather different, even assuming the validity of the DIA,
we have a Gaussian PdF for the $\{ X_n \}$ where 
 ${\langle X_n^2 \rangle}$ depend on  $k_n$, e.g. in the inertial range
 ${\langle X_n^2 \rangle} \sim k_n^{-\gamma}$.
 \\
 Therefore also   the validity of a fluctuation-response relation and a Gaussian PdF does not imply the possibility
 to introduce a temperature-like quantity.
 Of course we can introduce a sort of  "effective temperature"  $T_n$ for each $n$, simple 
 writing the relation (\ref{TU3})  in the form
 \begin{equation}
R_{n,n}(t)= \frac{ \langle X_n(t) X_n(0) \rangle} {T_n}\,,
\label{TU4}
\end{equation}
in other words $\langle X_n^2 \rangle = T_n$, now the ``temperatures'' $T_n$ have a rather trivial role:
they are the variance of the $\{X_n \}$. 
 \\
Let us stress that the above difficulty does not originate from the ``complexity'' of the
turbulence.
This is quite clear noting  that
the same difficulty is present also in the case of a generic linear Langevin equation,
i.e. different from Eq. (\ref{LG}),
 whose PdF is gaussian but  the relation (\ref{TU2}) does not hold, and therefore
 it is  not possible to introduce a temperature, see the discussion in Section 4.
 In such a system the PdF is a multivariate Gaussian
 $$
 P({\bf X})= \frac{1} {(2 \pi)^{N/2} \sqrt{Det \, \sigma}} \, 
 e^{ - \frac{1}{2} \sum_{ij} (\sigma^{-1})_{ij} x_i x_j } \,\,,
 $$
and, since the matrix ${\bf \sigma}$ is symmetric we can perform a linear transformation:
 $$
 Z_j=\sum_{i=1}^N C_{ji} X_i
 $$ 
in  such a way  that the PdF is
 $$
 P({\bf Z})= \frac{1} {(2 \pi)^{N/2} \sqrt{Det \, \sigma}} \, 
 e^{ -  \sum_{ij} \frac{ Z_i^2} {2 a_i} } \,\,
 $$
 where $\{ a_i \}$ are the eigenvalues of ${\bf \sigma}$.
 Of course Eq. (\ref{TU4}) holds, and one as that each $a_i$  can be (naively)
 considered as the "effective temperature", but we are not able   to see any physical meaning
in the above result.

\subsection{Granular systems}   
\label{sub51}

Granular materials are important systems which
appear in our everyday life as well as in many industrial applications,
posing interesting and partly unanswered questions to statistical
physics, geophysics and technology~\cite{JN92,andreotti13,puglio15,dearcangelis2016}. One of those questions is
the possibility of defining a suitable and meaningful concept of
``granular temperature''~\cite{BBDLMP05}. 

A granular medium is an ensemble of ``grains'', which are macroscopic
objects interacting among each other, and with the surroundings,
through non-conservative forces. Being almost undeformable, a granular
particle is fairly described by a solid of mass $m$, for instance a
sphere of radius $r$, and fully characterised by the position and
velocity of its center of mass, and (possibly) angular
momentum. In a collision, a fraction of the kinetic
energy associated with such macroscopic variable is transferred into internal
degress of freedom and is irreversibly lost. Several orders of
magnitude separate the average energy of internal thermal fluctuations at room temperature
- $k_B T \sim 5 \cdot 10^{-21} J$ - and the macroscopic energy of a
grain: for instance $m g r \sim 10^{-5} J$ for a steel sphere with
$r=2 mm$,  $g$ being the gravity acceleration. For this reason, until a few
decades ago, granular matter was commonly described as athermal. Recently a different point of view has been accepted, where
fluctuations have a role and can be characterized by some kind of
temperature, which is however very different  - in nature and in
value - from the temperature associated with a reservoir at equilibrium,
such as the surrounding environment (air, walls of the container,
etc.).

Granular media can be in very different ``phases'',
depending on boundary conditions, including external forcing: under violent shaking and with
enough allowed volume, one may realise a granular ``gas'', but when
allowed volume and/or the intensity of shaking are reduced, the
observed regime is much closer to dense liquids or even slowly
deforming solids~\cite{JNB96b}. Temperature can be defined in different ways, or may have different meaning, depending
upon the dominant regime: for this reason we separate the discussion
in three sections.

\subsubsection{Granular gases}

A granular gas is realized when the packing fraction (percentage of
container volume which is occupied by grains) is small, typically of
the order of $1\%$ or less. In such a situation one can assume for the interactions
instantaneous inelastic binary collisions. In the most common model
in a collision between particles of mass $m_1$ and $m_2$,
the velocities ${\bf v}_1$ and ${\bf v}_2$ are changed into ${\bf
  v}_1'$ and ${\bf v}_2'$ with the rule
\begin{align} \label{colrule}
{\bf v}_1' &= {\bf v}_1 - m_2\frac{1+\alpha}{m_1+m_2}[(\mathbf{v}_1-\mathbf{v}_2) \cdot \hat{\mathbf{n}}] \hat{\mathbf{n}}\\
{\bf v}_2' &= {\bf v}_2 + m_1\frac{1+\alpha}{m_1+m_2}[(\mathbf{v}_1-\mathbf{v}_2) \cdot \hat{\mathbf{n}}] \hat{\mathbf{n}},
\end{align}
where $\hat{\mathbf{n}}$ is the unit vector joining the center of
particle $1$ to the center of particle $2$ and $\alpha \in [0,1]$ is
the restitution coefficient, which is $1$ in the limit of elastic interactions~\cite{puglio15}.

Since experiments are usually done under gravity, in order to keep the
packing fraction small everywhere in the allowed volume it is
necessary to shake the container with accelerations much larger than
gravity~\cite{c90,poeschel,puglio15}. Many theoretical models (in
simulations or analytical calculations, for instance kinetic theories)
of granular gases exist. Here we list three
main categories:

1) cooling granular gases, which are obtained by initialising the
velocities and positions of the particles as in a gas (for instance at
equilibrium) and leaving the total kinetic energy dissipate under the effect of inelastic
collisions~\cite{h83,BMC96,NE98};

2) boundary driven granular gases, obtained by considering the
presence of at least one wall which injects energy into the gas (e.g. in
the fashion of a thermostat, or by a periodic vibration, etc.);

3) bulk driven granular gases, where each particle is constantly in
contact with some source of energy, for instance a monolayer of grains
hopping above a vibrating rough plate~\cite{NETP99,PLMPV98}.

The first category is in the class of non-stationary states (the
ultimate fate is a gas of non-colliding particles, possibly at rest)
and is considered very difficult to be observed in a laboratory. The
second category is close to many experimental realisations with shaken
containers, and has the peculiarity of leading to steady states which
are non-homogeneous in the spatial distribution of positions and velocities of grains,
with interesting emerging patterns~\cite{meerson2,lohse07,PGVP16}. The
third category results in spatially homogeneous steady states and has
been recently realised in experiments~\cite{OU98,puglisi11,puglisi12}.

In all those examples the lack of a Hamiltonian structure (due to the presence
of non-conservative interactions) is the main obstacle to the
realisation of an equilibrium-like description and therefore of a statistical mechanics 
definition of temperature. The easiest and effective replacement is the kinetic
temperature~\cite{ogawa80,kumaran98,BBDLMP05,gold08}, often called
``granular temperature''
\begin{equation}
k_B T_g = \frac{m \langle |{\bf v}|^2 \rangle }{d},
\end{equation}
with ${\bf v}$ the velocity of each particle, $d$ the dimensionality
of space and $k_B$ is usually replaced with $1$, as we do in the following. The average can have
several possible meanings, depending on the considered case: in an
experimental steady state it can be the empirical average cumulating
observations made upon many particles and many different instants; in
a theoretical study it can be realised with a theoretical probability distribution
which can be time-dependent (as in cooling cases) or steady (as in
driven systems), and in the latter case is naturally replaced by an
average over time. 

The concept of kinetic granular temperature is
grounded in kinetic theory, which gives a description of granular gases
through the inelastic Boltzmann equation~\cite{NE98,TPNE01} and its
hydrodynamic limit (slow variations of the fields in space and time)~\cite{BDKS98}. The study of inelastic Boltzmann equation
shows that deviations from a Maxwellian are inevitable in the presence
of inelastic collisions, but are small in many practical situations: therefore one may
characterize the statistics of velocities through a second moment (proportional to granular temperature $T_g$) and a
kurtosis excess (or second Sonine coefficient) which is small in many
regimes~\cite{NE98,TPNE01}. Granular hydrodynamics introduces a field
$T_g({\bf x},t)$ which is assumed to change slowly in space and time (with
respect to mean free path and mean free time) obeying an equation
which is similar to the analogous equation for molecular (elastic)
gases, but with two crucial differences: 1) a sink term representing
the loss of kinetic energy due to collisions, and 2) a conduction term
which is proportional to the {\em density} gradient (in addition to
the usual Fourier term)~\cite{soto99}. While energy is globally
conserved in molecular gases, the same is not true in granular gases,
and for this reason the assumption of slow variation of the
temperature field has a smaller range of validity, restricting the
application of granular hydrodynamics to small inelasticities.

Another challenge to the concept of granular
temperature, even in granular gases, is posed by the breakdown of the validity of the
equipartition of energy among degrees of freedom. This has been
observed comparing the granular temperature of different species in a
granular mixture, as well as comparing the kinetic energy along
different Cartesian components in a single species gas under
non-isotropic
conditions~\cite{MP99,FM02,BT02,MP02,PMP02,MP02b}. Primarily for this
reason, it is not believed that granular temperature contains more information
than just a description of the (local) average kinetic energy. In the
following we discuss some exception to this general view.

The role of granular temperature in the description of energy fluctuations
has solicited further questions about its role in linear response
relations~\cite{PBL02,DB02,BLP04,G04,SBL06,PBV07,BGM08,VPV08,VBPV09,GPSV14,naert17}. While
 in cooling granular gases linear response is not
described by the equilibrium Fluctuation-Dissipation
relation~\cite{DB02,BGM08}, the case is considerably simpler in driven
granular gases, where such a relation is satisfied, provided that the
canonical temperature is replaced with granular temperature $T_g$~\cite{PBL02,BLP04,G04,VPV08,VBPV09}.
For instance, a granular tracer under the action of a weak perturbing
force in a dilute driven granular system satisfies the Einstein
relation
\begin{equation}
R_{vF}(t) = \frac{\langle v(t)v(0)\rangle}{T_g}.
\end{equation}
Such a result is quite interesting as, on the basis of the generalized
Fluctuation-Response relation (see Section~\ref{sub42}) and of the
non-Gaussian distribution of velocities, one would expect a correction
to it. Nevertheless, in many different dilute cases, such corrections are not
observed or - in certain solvable models - can even be proven to
vanish~\cite{VBPV09}. A possible explanation to such a general result
is the following. In the dilute limit, {\em molecular chaos} is likely
to be valid, which - for statistical purposes (and provided that the
gas is in a bulk-driven steady state) - is equivalent to say that a
particle $1$ meets particle $2$ only once. As a matter of fact
the inelastic rule in Eq.~\eqref{colrule}, if restricted only to
particle $1$ (that is, disregarding the fate of particle $2$) is
equivalent to an {\em elastic} collision with effective
masses $m_1'$ and $m_2'$ such that~\cite{PVTW06}
\begin{equation} \label{effmass}
\frac{m_1'}{m_2'}=\frac{2}{1+\alpha}\left(\frac{m_1}{m_2}+1\right)-1.
\end{equation}
This is equivalent to say that - if the
information about the second particle is lost - then for the first
particle it is irrelevant if the collision is elastic or not. 

A phenomenology which is consistent with the above observation has
been seen studying the dynamics of a massive intruder~\cite{sarra10},
i.e.  a (spherical) granular particle with mass $M$ diffusing in a
stochastically driven granular gas of particles of mass $m \ll M$. It
is not important the exact mechanism of driving, provided that it
guarantees in a finite time a relaxation to a dilute and spatially
homogeneous steady state of the gas at granular temperature $T_g$, as
it occurs for instance in the model introduced in~\cite{PLMPV98}. For
simplicity we assume that the intruder does not interact directly with
the driving mechanism (in the more realistic case of interaction with
the bath, additional terms appear in the following formula). In
appropriate limits (large number of granular bath particles and
dilute limit) the massive tracer does not
influence appreciably the statistics of the surrounding granular gas
and therefore its statistics is governed by a Lorentz-Boltzmann
equation. Such an equation may be further simplified in the limit of
large intruder's mass, i.e. basically truncating an expansion in
powers of (small) $m/M$~\cite{vk}. This procedure, which has given results
in fair agreement with numerical simulations, has shown that the
dynamics of the intruder's velocity ${\bf V}(t)$ is fairly described (in the large mass limit)
by the following Ornstein-Uhlenbeck process:
\begin{equation} \label{gbm}
M \dot{\bf V}(t) = -\Gamma {\bf V}(t) + \mathcal{E}(t),
\end{equation}
with $\mathcal{E}$(t) an unbiased white Gaussian noise and
\begin{align}
\Gamma = (1+\alpha) m f_c,\\
\langle \mathcal{E}_i(t)\mathcal{E}_j(t') \rangle = 2 \Gamma \frac{1+\alpha}{2} T_g \delta_{ij}\delta(t-t'),
\end{align}
where $\alpha$ is the restitution coefficient for the
intruder-particle collisions, $f_c$ is the frequency of collisions
experienced by the intruder and $T_g$ is the granular temperature of
the surrounding gas. Both $f_c$ and $T_g$ are determined by the
average number density of the gas, the dimension of the particles and
the parameters of the external driving
mechanism~\cite{sarra10}. Note that Eq.~\eqref{gbm} leads to
\begin{equation} \label{equi1}
T_{tr}=M \frac{\langle |V|^2 \rangle}{d} = \frac{1+\alpha}{2} T_g,
\end{equation}
where we have defined the tracer granular temperature $T_{tr}$. Interestingly, formula~\eqref{equi1} is equivalent to {\em equipartition}
\begin{equation}
M' \frac{\langle |V|^2 \rangle}{d} = T_g,
\end{equation}
with $M'= \frac{2}{1+\alpha} M$ which is the large $M$ limit of formula~\eqref{effmass} (with replacing $m_1 \to M$ and $m_2 \to m$).

Another observation that gives relevance and usefulness to the
concept of granular temperature, again in the dilute limit, comes from
the study of ``granular ratchets''. Changing the shape of the
intruder, one can break a spatial
symmetry (for instance $x \to -x$, which also requires some constraint
to prevent rotation of the object), 
a large mass expansion
shows the appearance of an additional constant force in
Eq.~\eqref{gbm}, which is known as motor or ratchet
effect~\cite{CPB07,CB07}. The role of granular temperatures is evident in
the observation that the motor force is proportional to $T_{tr}-T_g =
\frac{1-\alpha}{2} T_g$~\cite{CPB07,puglisi13,SGP13}.

The departure from the dilute limit changes many of the above results
and make deviations from an equilibrium description emerge: as a
consequence the effective meaning of granular temperature is
drastically reduced.

\subsubsection{Granular liquids}

The first experiment focusing on a Brownian-like description of a
large intruder in a granular liquid, i.e. a shaken granular material
in a non-dilute but rapidly evolving regime, is discussed in
Ref.~\cite{DMGBLN03}. The validity of an equilibrium-like
Fluctuation-Dissipation relation was observed, with an effective
temperature which - as orders of magnitude - was ``related to the
granular temperature'', a fact which is reminiscent of the
observations discussed in the previous paragraphs, but cannot be fully evaluated since the real granular temperature was not measured. Most recent
studies, both theoretical~\cite{PBV07,VPV08,VBPV09,sarra10b,sarra12}
and experimental~\cite{GPSV14}, have shown that when the granular is a
liquid and not a gas, deviations from the equilibrium
Fluctuation-Dissipation relation are observed: most importantly,
such deviations are not well described by an effective
temperature. In granular liquids, as a matter of fact, granular
temperature is much less useful than in gases, and cannot be replaced
by some other temperature for the purpose of an effective description.

\begin{figure}
\includegraphics[width=7cm,clip=true]{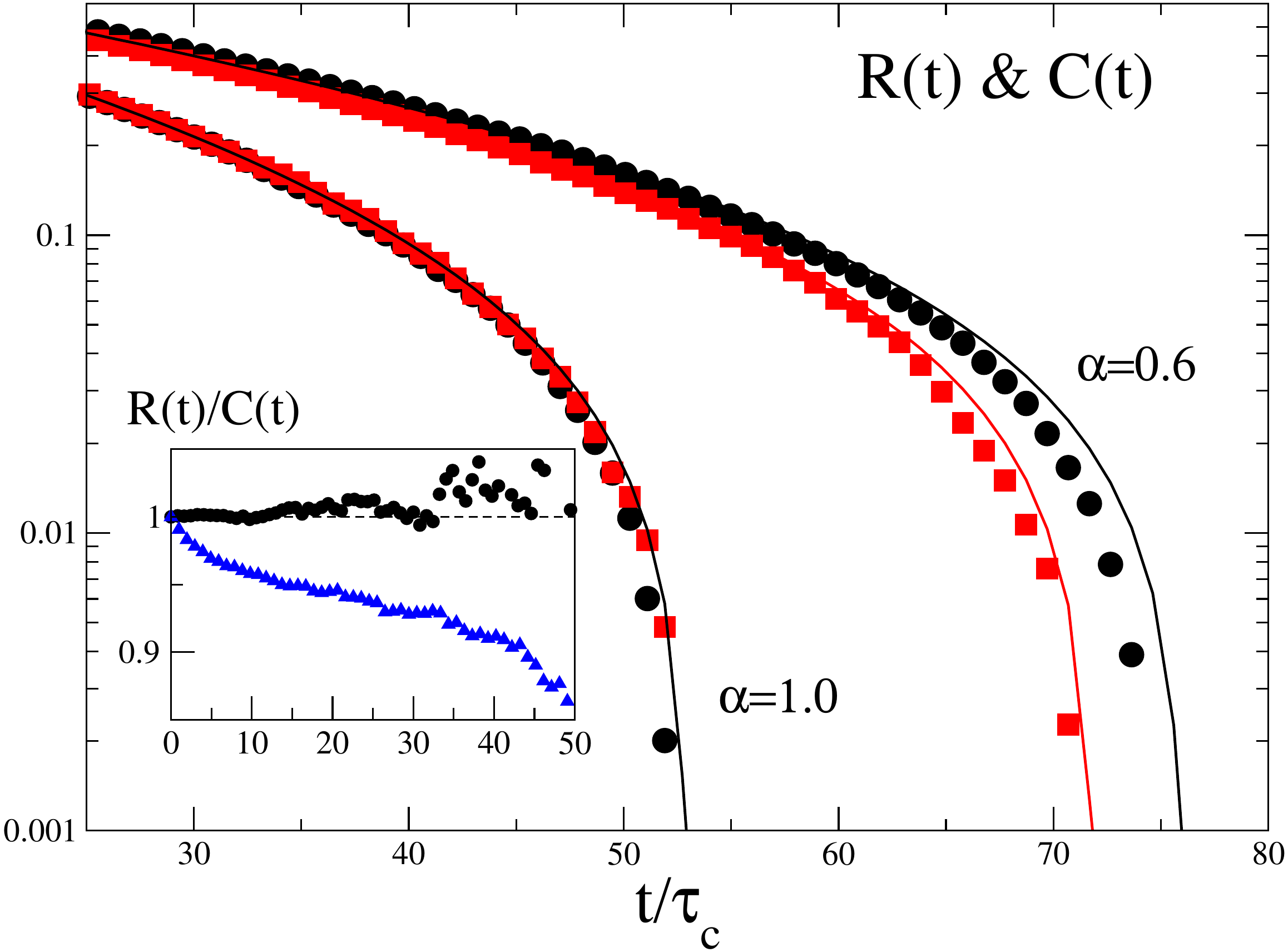}
\includegraphics[width=7cm,clip=true]{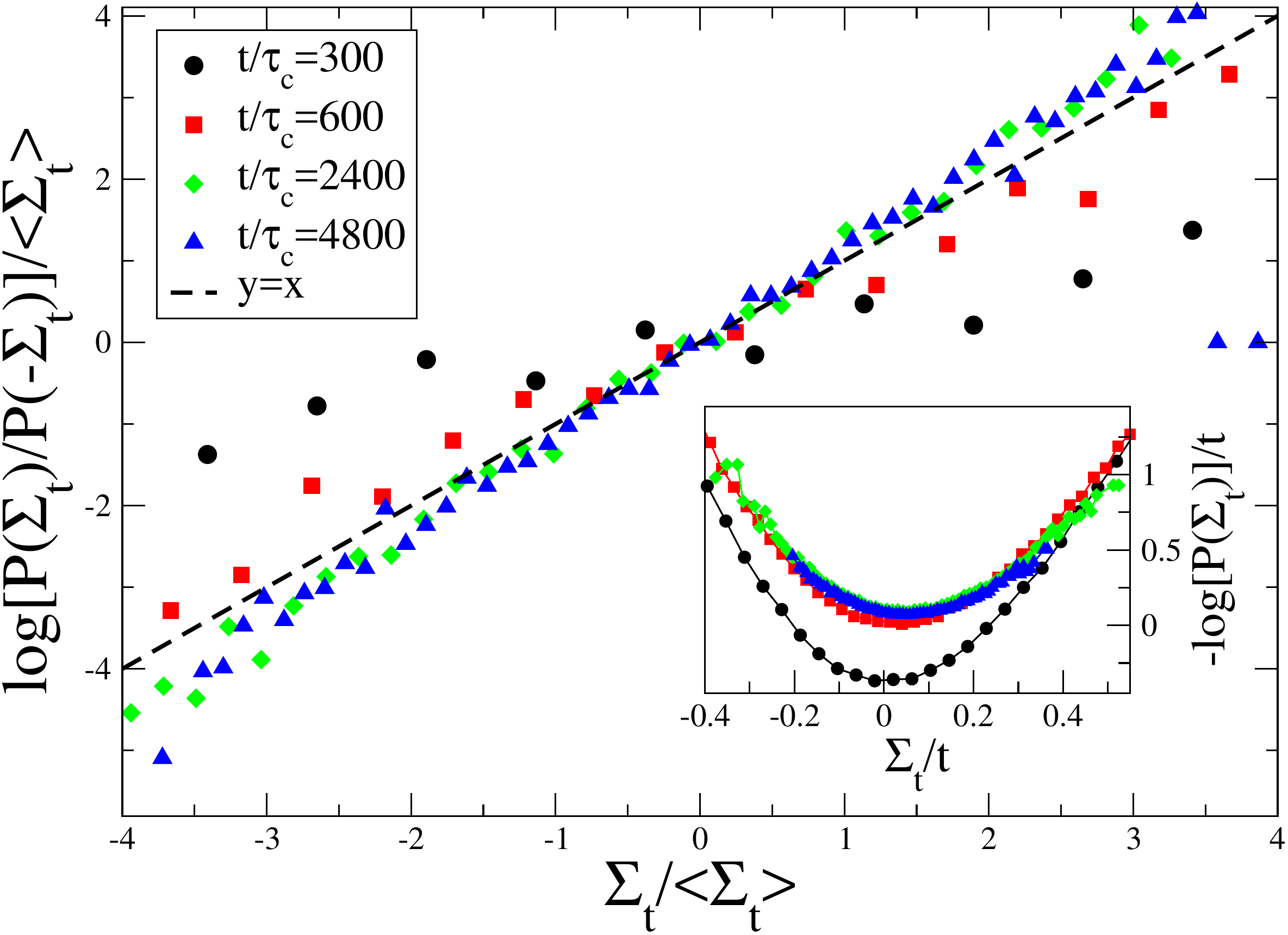}
\caption{Left: response function $R_{vF}(t)$ and auto-correlation
  function $C(t)=\langle V(t)V(0)\rangle/\langle V(0)V(0)\rangle$ as a
  function of time, measured in molecular dynamics simulations of a
  system composed of a massive intruder interacting with a driven
  granular fluid. Right: Fluctuation Relation satisfied at large times
  by the fluctuating entropy production measured in the same system,
  using the definition~(\ref{entrprod}).}
\label{fig_rc}
\end{figure}

A simple example is provided, again, by the case of a massive
intruder $M \gg m$~\cite{sarra10b,sarra12}. For the purpose of
describing the velocity autocorrelation of the tracer and its linear
response, the model in Eqs.~\eqref{twovar} has revealed to give an excellent
description. A particularly clear re-formulation of that model, using the notation of the present section, and restricting to a single dimension for the purpose of clarity, reads
\begin{subequations} \label{grintr}
\begin{align} 
M \dot{V}(t)= - \Gamma [V(t)-U(t)] + \sqrt{2 \Gamma T_{tr}} \mathcal{E}_v(t)\\
M' \dot{U}(t) = -\Gamma' U(t) - \Gamma V(t) + \sqrt{2 \Gamma' T_b} \mathcal{E}_U(t),
\end{align}
\end{subequations}
where $\Gamma'$ and $M'$ are parameters to be determined, for instance
from the measured autocorrelation function, and $T_b$ is the value of
$T_g$ in the elastic case: for instance if the steady state is
guaranteed by an external ``bath'', it is the bath
temperature~\cite{PLMPV98} (a more detailed discussion is given
below). The dilute limit, Eq.~\eqref{gbm}, can be obtained with
$\Gamma' \sim M' \to \infty$, which implies small $U$. In the elastic
limit ($T_{tr} = T_b=T_g$), on the other side, the coupling with $U$
can still be important (as a first approximation for the breakdown of
Molecular Chaos), but the equilibrium Fluctuation-Dissipation relation
is recovered, as seen in Section 4.1. The model in Eqs.~\eqref{grintr}
is, of course, an approximation which is seen to work for packing
fractions smaller than $40\%$: its advantage is in its particular
simplicity and analytical solvability. Interestingly, numerical
simulations have shown that the auxiliary field $U(t)$ is a local
average of the velocities of the particles surrounding the intruder
(in an {\em persistence} radius which has been discussed
in~\cite{sarra10b}). A striking confirmation of the usefulness of
Eqs.~\eqref{grintr} came also from a measurement of the fluctuating
entropy production, which is proportional to the work done by the
force $\Gamma U(t)$ upon the intruder,
\begin{equation}\label{entrprod}
\Sigma_t \approx \Gamma \left(\frac{1}{T_{tr}}-\frac{1}{T_b} \right)\int_0^t ds V(s)U(s),
\end{equation}
and which satisfies very well the Fluctuation
Relation~\cite{sarra10b}, see Fig.~\ref{fig_rc}. The latter is quite a
remarkable result, in particular if compared with previous attempts to
define and measure a granular entropy
production~\cite{FM04,PVBTW05,VPBTW05,VPBTW06,PVTW06,VPBWT06}.

A few words are in order about the temperatures $T_{tr}$ and $T_b$
appearing in the ``thermostats'' of the $V(t)$ and $U(t)$ fields in
Eqs.~\eqref{grintr}. As discussed above, in the dilute limit $T_{tr}
\to \frac{1+\alpha}{2}T_g$. On the contrary, when the density
increases numerical simulations suggest $T_{tr} \to T_g$, likely due
to a reduction of effective inelasticity in recolliding particles. The
appearance of $T_b$ is also interesting: the fact that the “temperature” of
the local velocity field $U$ is equal to the bath temperature comes as
a consequence of the conservation of momentum in collisions, implying
that the average velocity of a group of particles is not changed by
collisions among themselves and is only affected by the external bath
and a (small) number of collisions with outside
particles. Summarizing, model~\eqref{grintr} suggests that in a granular liquid -
at some level of approximation - {\em two} temperatures are relevant,
one related to the single particle scale and another one related to a
many-particle, or collective, scale.

\begin{figure}
\begin{center}
\includegraphics[width=7cm,clip=true]{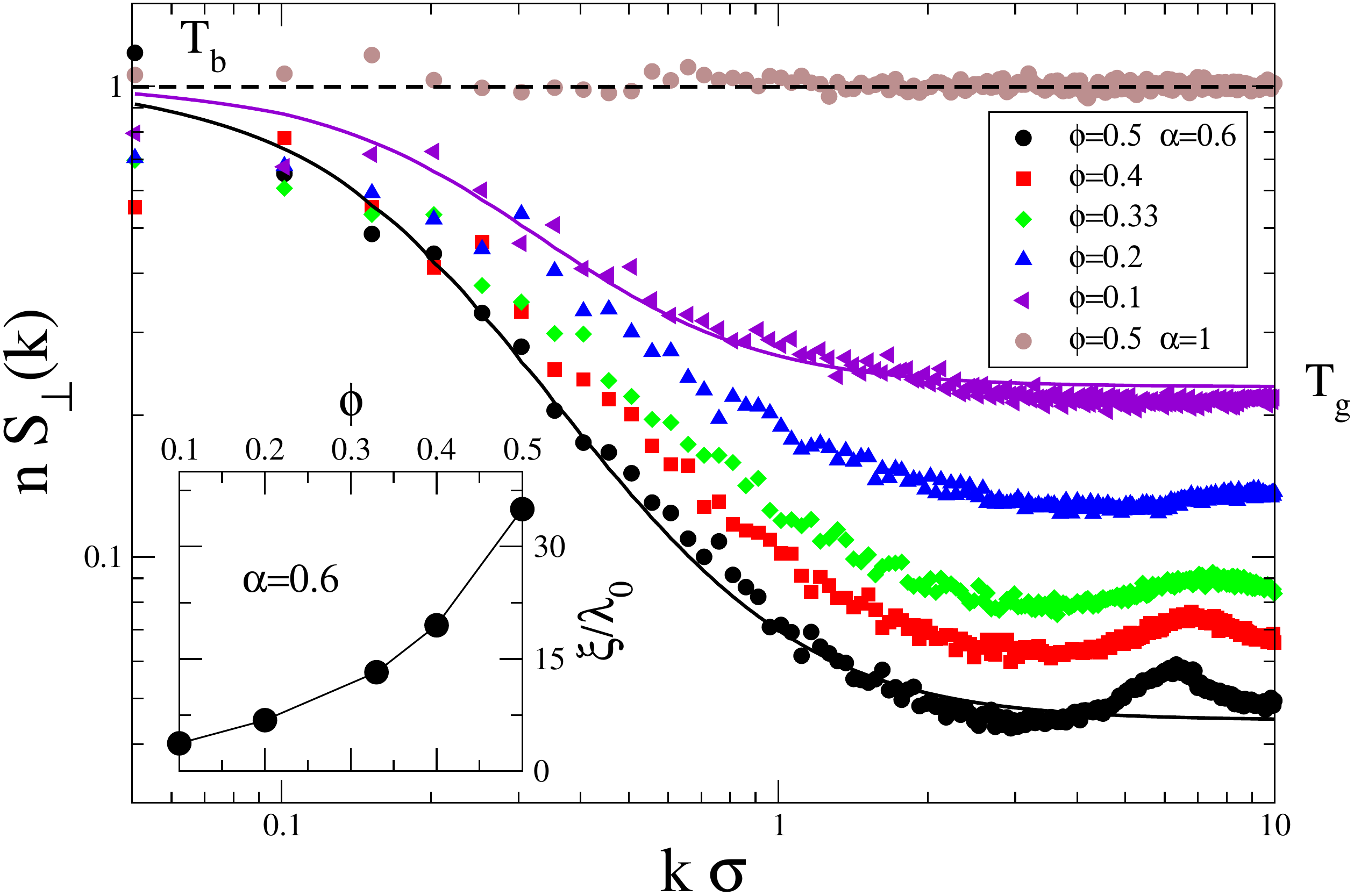}
\caption{Transverse velocity structure factor for simulations of a $2d$
  driven granular gas at different packing fractions $\phi$ and
  restitution coefficients $\alpha$~\cite{puglisi11}. In the graph
  $n=N/V$ is the average number density, $\sigma$ is the diameter of
  the disks and $\lambda_0$ is the mean free path. \label{struct}}
\end{center}
\end{figure}

Such a conclusion is consistent with a series of recent results
about spatial velocity correlations, typically measured as structure
factors of the velocity field. The question about velocity structure
factors has been initially addressed - through numerical studies and
models of fluctuating hydrodynamics~\cite{BMG09,trizac09} - for the
case of cooling granular gases~\cite{NEBO97,BMP02,BMP02b} or bulk-driven gases with
an infinite relaxation time~\cite{NETP99}. In those studies the large
scale behavior of velocity structure factors appeared to be unbounded
and therefore useless for the definition of a large-scale granular
temperature. More recently, the velocity correlations for a
bulk-driven model of granular particles interacting at a finite
frequency, proportional to $\gamma_b$, with the external bath~\cite{PLMPV98} has been studied under
some theoretical approximation and in
experiments~\cite{puglisi11,gradenigo11,puglisi12,PLH12,GCR13}. The key finding is
represented by a study of the steady transverse velocity structure factor, which can be
defined - for instance - as
\begin{equation}
S_\perp(k)=\sum_{n,j=1}^N v_{n,y} v_{j,y} e^{-i k \pi (x_n-x_j)},
\end{equation}
where $x_n$ and $v_{n,y}$ are the $x$-component of the position  and the $y$-component of the velocity of the $n$-th particle respectively.
The structure factor was found to be given by the following expression for a large range of packing fractions in the liquid regime ($0-40\%$)
\begin{equation}
S_\perp(k) \approx \frac{\gamma_b T_b + \nu k^2 T_g}{\gamma_b + \nu k^2},
\end{equation}
where $\nu$ is the kinematic viscosity of the granular fluid.
Such an expression clearly shows the dependence upon the two temperatures which appear at
large scales, $\lim_{k \to 0} S_\perp(k) = T_b$, and at small scale
$\lim_{k \to \infty} S_\perp(k) = T_g \le T_b$. The structure factor
becomes flat, as it should, at equilibrium ($T_g =
T_b$). Out-of-equilibrium, on the contrary, a finite correlation
length $\xi \sim \sqrt{\nu/\gamma_b}$ appears.  In the numerical
studies~\cite{puglisi11,gradenigo11}, where full control of $T_b$ was available,
it was possible to put in evidence that (at fixed $T_b$) the value of
$T_g$ decreases with the packing fraction: this implies a larger
contrast $T_b/T_g=S_\perp(k \to 0)/S_\perp(k \to \infty)$, which is
equivalent to more pronounced correlations between velocities, see Fig.~\ref{struct}. A
strict analogy with the several intruder's examples discussed above is
recovered~\cite{PBV07,sarra10b}.

\subsubsection{Dense granular materials}

When packing fraction is further increased, the granular liquid
typically falls in a more complicate regime which is associated with a
very slow relaxation dynamics, often known as ``glassy'' granular
state. A dynamics can still be observed, for instance in the presence
of low-frequency tapping, but on time-scales much longer than in the
previously considered collisional or liquid states. Schemes of
deduction of the constitutive relations starting from the microscopic
dynamics can hardly be applied to such a regime, unless at the cost of
uncontrolled approximations. Phenomenological
descriptions~\cite{andreotti13} are more common and useful, at least
as a descriptive tool or for industrial
applications. In these models,
see~\cite{barnes89,cavagna09,andreotti13} for reviews, macroscopic
stresses are related to strains and/or strain rates through empirical
formulae inspired by viscoelasticity (e.g. the so-called Maxwell model
or the Kelvin-Woigt model) or Bingham materials (e.g. mayonnaise,
which is characterised by the presence of a yield - or threshold -
stress), or Herschel-Bulkeley models, which are fully non-linear in
the strain/shear rate, or to more recent friction-like
theories~\cite{gdr04,cruz05,jop06,forterre08,gnoli16}. A noticeable
exception to the phenomenological approach is the so-called Edwards
theory of compactivity, which is an elegant attempt to build a theory
of dense granular states, formally equivalent to equilibrium
statistical
mechanics~\cite{mehta89,edwards89c,edwards94,nowak98,bcl02,makse04,RNDRB05,ciamarra06}.

The essence of Edwards' idea is to reproduce the observables attained
dynamically by a sort of {\em ergodicity hypothesis}. In the theoretical approach one fixes the density of
the system (which in real tapping experiment is not constant), and
then calculates the value of the observables in an ensemble consisting
of all the {\em ``blocked'' configurations} at that density. The blocked
configurations are defined as those in which every grain is unable
to move. 

It is clear that configurations with low mobility are relevant in a jammed
situation, while it is a much stronger hypothesis  that the
configurations reached dynamically are {\em the typical ones} of given
energy and density. Had we restricted averages to blocked
configurations having {\em all} macroscopic observables coinciding
with the dynamical ones, the construction would exactly, and
trivially, reproduce the dynamic results. The fact that conditioning
averages to the observed energy and density suffices to give (even if
maybe only as an approximation) other dynamical observables is highly
non-trivial~\cite{ciamarra06}.

This ``Edwards ensemble'' leads naturally to the definition of an
entropy $S_{Edw}$, given by the logarithm of the number of blocked
configurations of given volume, energy, etc.  Associated with this entropy are
state variables which play a role similar to the usual thermodynamics quantity (as temperature), e.g. ``compactivity''
\begin{equation}
\frac{1}{X_{Edw}}=\frac{\partial}{\partial V}S_{Edw}(V),
\end{equation}
and ``Edwards temperature''
\begin{equation}
\frac{1}{T_{Edw}}=\frac{\partial}{\partial E}S_{Edw}(E).
\end{equation}

A series of interesting results helped to clarify and support Edwards'
proposal. In particular several points of contact emerged with linear
response theory of glassy systems, which is quite a different
framework with respect to Edwards' approach. As discussed in
Subsection~\ref{sub43}, in the presence of a large separation of
timescales, it is meaningful to define an effective temperature
$T_{dyn}$ as the one replacing the bath temperature in the Equilibrium
Fluctuation-Dissipation Relation, for instance in the Einstein-like
relation between diffusion and mobility~\cite{CR03}.  In several
mean-field models of disordered systems exhibiting glassy dynamics, it
was possible to identify all the energy minima (the blocked
configurations in a gradient descent dynamics), and calculate
$1/T_{Edw}$ as the derivative of the logarithm of their number with
respect to energy. Explicit computations showed that $T_{Edw}$
coincides with $T_{dyn}$ obtained from linear response of the out of
equilibrium dynamics of the models, i.e. during a slow aging transient
in contact with an almost zero temperature
bath~\cite{remi,jamming,Theo,Frvi,Felix,biroli,nicodemi99,BB11}.
Moreover, given the energy $E(t)$ at long times, the value of any
other macroscopic observable is also given by the flat average over
all blocked configurations of energy $E(t)$. In some simplified models
of dense granular media the Edwards configurational entropy can be
computed and a negative temperature can arise in certain ranges of
packing fraction~\cite{cc08}.  Within the same approximation, one can
also treat systems that, like granular matter, present a non-linear
friction and different kinds of energy input, and the conclusions
remain the same~\cite{jorge-trieste}, despite the fact that there is
no thermal bath temperature. Edwards' scenario then seems to be
correct within mean-field schemes and for very weak vibration or
forcing. The problem that remains is to what extent it carries through
to more realistic models~\cite{BL00}.  Recently, this approach has
been revitalised due to the development of efficient algorithms for
computing granular entropy~\cite{APF14,MSS16}.

\subsection{Active matter}   
\label{sub53}

The study of active matter embraces systems composed by many
elementary constituents with a common fundamental ingredient: self
propulsion~\cite{ram10,marchetti2013}. Typical examples are found in
biology, ranging from aggregates of macroscopic living
beings~\cite{cavagna12} (e.g. bird flocks, insect swarms, fish
schools, etc.) to ensembles of microscale entities (e.g. interacting
cells or bacteria) and, even smaller, sub-cellular unities such as
actuated biopolymers (actin filaments, kinesin molecules
etc.). Another important category of active matter is that of
non-living - often man-made - systems, for instance self-propelled
colloids at the micro-scale (typical example are the Janus particles,
which propel due to asymmetric interactions with the solvent induced
by inhomogeneous chemical properties of their surfaces), driven
granular particles at the scale of millimiters~\cite{dauchot13}
(asymmetrically manufactured granular objects may crawl due to
friction asymmetries along a vibrating surface), and - at larger
scales - swarming robots~\cite{dileo16,dileo16b}.

The most salient aspect of self-propulsion is an intrinsic lack
of thermodynamic equilibrium with the surrounding environment, with a
net transfer of energy from the ``active reservoir'' (a source of
self-propulsion energy) to the ``thermal reservoir'' (the environment,
usually a liquid solvent or air)~\cite{speck16,fodor16,umb17}. Moreover the number of elementary
constituents of an active system is large but not too much: hundreds or thousands are the most common
figures. This makes active systems very interesting for the study of
fluctuations, which are rarely negligible~\cite{ginelli06,menon2007}.

There is a plethora of collective phenomena observed in real active
systems as well as in theoretical models. Two major
ordering modes in active systems - which can also combine together -
are phase separation~\cite{cates15} and swarming~\cite{vicsek95,toner05}: the former is
typically a dilute-dense transition (often simply called clustering,
or in the most extreme cases, jamming and crystallisation), the latter
consists in a macroscopic fraction of the particles moving in the same
direction. In the presence of polar or chiral particles, the number of
symmetries to break increases and so do the observable ordered
phases~\cite{marchetti2013}. Turbulent phases with whirls, jets and vortices have also been
observed~\cite{bratanov15}. Other rich fields of study are that of confined active
systems, where the interaction between self-propelled particles and
complex geometries are considered, and that of active-passive
mixtures, where the transport of passive particles is affected by the
presence of self-propelled particles~\cite{dileo16}.

\subsubsection{Modelling active particles}

The motion of an active ``particle'' is usually modelled trough simple
dynamical equations, i.e. neglecting the details of the self-propulsion
mechanism.  Given the huge variety of systems under the label ``active'', a
homogeneous and agreed classification is lacking. According to a recent review~\cite{dileo16}, a major distinction
concerns the interaction between the self-propelled agent and its
surrounding fluid: with respect to that, one can distinguish between

1) active Brownian particles (ABP), which feel just viscous damping as
a ``simplified'' effect of their solvent, and

2) microswimmers (MS), which are more sophisticate models taking into
account the modification of the surrounding fluid due to momentum
transfer: indeed the self-propelled particle acts as a localised
source of momentum in the fluid.

This difference of course reverberates into the complexity of
particle-particle interactions: ABP interact through short-range
forces (unless an additional long-range interaction of external origin
is considered, e.g. in the case of electrically charged particles); MS
are subject to hydrodynamic interactions, which - in incompressible
fluids - are typically long-range. Apart from the fluid-particle mode
of interaction, another discriminating property is the shape of the
particle or of its self-propulsion mechanism. Shape may induce some
kind of chirality, which is relevant for the possible observable
collective behaviors.

Irrespective of their range - short or long - particle-particle
interactions for active systems can be placed into two main
categories: aligning and non-aligning. Non-aligning forces include all
the kinds which are already known to act in passive (equilibrium)
systems, such as excluded volume, charge repulsion, liquid-like
interactions (e.g. Lennard-Jones), depletion forces, etc. Aligning
forces usually involve some dissipative mechanism, the most common
being a force which depends upon the velocity of the interacting
particles which tends to reduce the misaligned component. The ancestor
of aligning self-propelled particles is the Vicsek model, where the
aligning mechanism is collective: at each time-step all particles in a
given radius take as new orientation (which determine the direction
of motion) their average orientation plus a small amount of noise~\cite{vicsek95}.

Notwithstanding the simplicity of the models, emergent collective
behaviors can be surprisingly rich.  The theoretical tools most often
adopted to analyse such collective behaviors are numerical simulations
and coarse-graining techniques. Interestingly, a role of a certain
relevance has been played by {\em equilibrium-like} or {\em
  equilibrium-inspired} models, where the intrinsic non-equilibrium
condition of active systems has been neglected or drastically
reduced~\cite{cavagna10,cates15}. This occurred for two main reasons:
1) equilibrium is deeply rooted in statistical mechanics and offers
well-established approaches to explain complex phase diagrams in the
presence of non-trivial broken symmetries (active particles, such as
microswimmers, spermatozoa, bacteria, etc., but also birds in flocks,
are elongated ``polar'' particles which can display many kinds of
order, in analogy - for instance - with liquid crystals); 2) in
several cases - for instance in low-Reynolds situations, e.g. in
viscous solvents and at the micro-scale - the lack of thermodynamic
equilibrium affects the dynamics of the self-propelled particles only
at short time-scales, while on long time scales the difference between
equilibrium and nonequilibrium is much harder to be
observed~\cite{fodor16,umb17}.

\subsubsection{Toward a temperature for active matter}

The simplest way of defining a temperature for active systems is
considering active diffusivity $D_{eff} = \lim_{t \to \infty} \langle |\mathbf{X}|^2(t)\rangle/(2d t)$, where $\mathbf{X}$ indicates the particle's position in $d$ dimensions,  and the
viscous damping $\gamma$ exherted on the particle by the surrounding
solvent:
\begin{equation}
k_B T_{eff}=\gamma D_{eff}.
\end{equation}
This is perfectly analogous to the definition of effective
temperatures based upon linear response theory (see Section 4.1).

For several basic models of non-interacting ABPs, e.g. the so-called ``active Brownian
motion'', ``run-and-tumble'' and the ``Gaussian colored noise'' model
(also known as Ornstein-Uhlenbeck ABP), the mean-squared displacement
(considering $d=2$ for the sake of simplicity) reads~\cite{dileo16}
\begin{equation} \label{eq:msd_abp}
  \langle |\mathbf{X}|^2(t)\rangle = [4 D_T + v^2 \tau_R] t + \frac{v^2}{2}\tau_R^2 \left[e^{-2 t/\tau_R}-1 \right],
\end{equation}
where $D_T = k_B T/\gamma$ is the passive translational diffusion
coefficient due to Brownian motion at the thermal temperature $T$, $v$
is the typical velocity of self-propulsion, and $\tau_R$ is the mean
re-orientation time, related to rotational diffusion (the rotational
diffusion coefficient is $\tau_R^{-1}$). Even without entering into
the details of the mentioned ABP models, the expression in
Eq.~\eqref{eq:msd_abp} already gives an idea of the difference between
ABP and passive Brownian motion: in the latter the rotational
diffusion is not coupled to translational diffusion and therefore
$\tau_R$ does not appear in the expression of $\langle |\mathbf{X}|^2(t)\rangle$. The
translational movement of an ABP, on the contrary, occurs
preferentially along a direction parallel to its orientation, for
instance marked by its flagellum, and therefore the rotational
dynamics affects translations. Eq.~\eqref{eq:msd_abp} also shows that
at small times $t \ll \tau_R$ (of course still larger than the very
small average time between collisions with the fluid-molecules) one
has $\langle |\mathbf{X}|^2(t)\rangle \approx 4 D_T t$, i.e. the short-time behavior is
equivalent to the thermal diffusion of passive Brownian particles. 

As a consequence of
Eq.~\eqref{eq:msd_abp}, the effective (long-time) translational diffusion
coefficient of an ABP reads
\begin{equation}
  D_{eff} = D_T + \frac{1}{4}v^2 \tau_R
\end{equation}
 and therefore its ``effective temperature'' takes the expression
\begin{equation}
k_B T_{eff}= k_B T + \frac{\gamma \tau_R}{4} v^2.
\end{equation}
Typical velocities $v$ of biological or artificial micro-scale active
particles are in the range of $1-100 \mu m s^{-1}$, while their
translational and rotational diffusivities are (in water) $D_T \sim
0.1 \mu m^2 s^{-1}$ and $\tau_R \sim 5 s$. For such particles, the
ratio $T_{eff}/T$ is commonly of the order of $10^3$ or larger. This
is the reason why ABP may be considered as ``hot colloids'', a
scenario which is also observed in some experiments. However, when
interactions or external (inhomogeneous) fields become important, this
simple picture based upon a single equilibrium-like temperature breaks down.

Numerical studies of the linear-response-theory-based effective
temperature in active models with relevant particle-particle
interactions (e.g. active liquids) have been performed
in~\cite{mossa08,marchetti12,bialke12,kurchan13}. These results,
however, are not conclusive. Some of them~\cite{mossa08,kurchan13}
indicate a certain robustness in the definition of $T_{eff}$, which
can also be measured by coupling ``thermometers'' (e.g. tracer
particles whose mass selects time-scales and therefore degrees of
freedom) or which can be relevant for a liquid-glass transition. The
measured $T_{eff}/T$ is however of order $1$ (typically between $1$
and $2$), which suggests that the adopted models are
not very close to real active systems. Other
studies~\cite{marchetti12,bialke12}, apparently enforcing stronger
self-propulsions, indicate that a temperature based upon linear
response or diffusivity/mobility ratios is not a useful concept.

Recent theoretical and experimental studies have also tested the
concept of effective temperature in different active systems, such as
hair bundle's spontaneous oscillations~\cite{MHJ01}, biopolymers in an
active medium~\cite{KEKK09}, red-blood-cell membranes~\cite{BIP11},
DNA harpins~\cite{DC15}, and active colloid suspensions~\cite{palacci10} and
mixtures~\cite{HYGL16}.

\subsubsection{Effective temperature in the Gaussian colored noise model}

An interesting alternative to effective temperatures based upon
response has emerged, recently, in the study of the Gaussian colored
noise model for active systems~\cite{umb15b,umb16a}. Even the simple one-dimensional case
with no particle-particle interactions, but only in the presence of an
external inhomogeneous field, is interesting~\cite{umb15a}. In this case each
active particle obeys the following equation of motion:
\begin{subequations} \label{eq:gcn}
\begin{align}
  \dot{x}(t)&=v(t)\\
  m\dot{v}(t)&=-\gamma v(t) + \sqrt{2 \gamma T} \xi(t) - U'(x) + \gamma v_a(t) \label{gcnv}\\
  \dot{v}_a(t)&=-\frac{v_a(t)}{\tau_R} + \sqrt{2 \frac{D_{eff}}{\tau_R^2}} \xi_a(t),
\end{align}
\end{subequations}
where $\gamma$, $T$, $\tau_R$ and $D_{eff}$ have the same qualitative
meaning as above, $m$ is the particle's mass, $U(x)$ is an external
potential and both $\xi(t)$ and $\xi_a(t)$ are Gaussian white noises
with zero average and unitary variance. The model in
Eq.~\eqref{eq:gcn} has three main timescales:

1) $\tau_m=m/\gamma$,
which is the momentum relaxation timescale due to interaction with
fluid molecules;

2) $\tau_R$, which is the persistence timescale of the
active force;

3) $\tau_U$, the characteristic time needed by the
particle to see the variations of the potential $U(x)$: for instance,
it can be defined as $\tau_U=\Delta x/V$, with $\Delta x$ a
characteristic length-scale (e.g. periodicity) of the external
potential and $V$ the typical particle's velocity, which in the
presence of activity is of the order of $\sqrt{D_{eff}/\tau_R}$ while
in the passive case is $\sqrt{T/m}$.
Apart from atypical choices of the potential (e.g. $U(x)$ varying over
very small lengthscales) and in the low-Reynolds regime ($Re \sim
\tau_m/\tau_R$), the order of the three timescales is $\tau_m \ll
\tau_R \ll \tau_U$.

In the literature, the model is usually presented in its overdamped
limit~\cite{umb15a,umb15b,umb16a}, i.e. disregarding the time-scales of order $\tau_m$ or smaller,
that is equivalent to consider $\dot{x} \approx \frac{x(t+\Delta
  t)-x(t)}{\Delta t}$ with $\tau_R \gg \Delta t \gg\tau_m$. The
overdamped equation reads:
\begin{subequations} \label{eq:gcn_ov}
\begin{align}
  \dot{x}(t)&= \sqrt{2 \frac{T}{\gamma}} \xi(t) - \frac{U'(x)}{\gamma} + v_a(t) \label{ov1}\\
  \dot{v}_a(t)&=-\frac{v_a(t)}{\tau_R} + \sqrt{2 \frac{D_{eff}}{\tau_R^2}} \xi_a(t). \label{ov2}
\end{align}
\end{subequations}
In view of the quite large difference between
thermal and active diffusivities, it is reasonable to neglect thermal
noise, the first term in the r.h.s. of Eq.~\eqref{ov1}. This
has the advantage of leading to a differentiable dynamics for
$\dot{x}(t)$. Indeed, upon time-differentiation of Eq.~\eqref{ov1} (and few more passages), one gets
\begin{equation} \label{eq:gcn3}
\dot{v}(t)=-\frac{1}{\tau_R}\left(1+\frac{\tau_R}{\gamma}U''(x)\right)v(t)-\frac{1}{\tau_R
  \gamma}U'(t)+\sqrt{\frac{2D_{eff}}{\tau_R^2}} \xi_a(t).
\end{equation}
Of course this equation is different from Eq.~\eqref{gcnv}, because
of the overdamped approximation considered to derive it: basically
$v(t)$ in Eq.~\eqref{eq:gcn3} is not the same velocity as in
Eq.~\eqref{gcnv}. Interestingly, a second level of coarse-graining
is available, that is looking at larger time-scales $\Delta t$, with
$\tau_U \gg \Delta t \gg \tau_R$. This is the so-called uniform colour
noise approximation (UCNA)~\footnote{A rigorous discussion of the
  UCNA, see for instance~\cite{hanggi95}, shows that it can be
  obtained in the limit $\tau_R \to 0$ but also in the opposite limit
  $\tau_R \to \infty$, provided that one focuses in the regions of
  stable potential $U''>0$.}, which reads
\begin{equation} \label{eq:ucna}
\dot{x}(t)=-\frac{1}{\gamma \Gamma(x)} U'(x) + \sqrt{\frac{2D}{\Gamma^2(x)}} \xi_a(t),
\end{equation}
where the Stratonovich convention is assumed and $\Gamma(x)=1+(\tau_R/\gamma)U''(x)$. The stationary distribution for Eq.~\eqref{eq:ucna} is known and reads
\begin{equation}
p_{ucna}(x) = \mathcal{N} |\Gamma(X)| \exp \left( -\frac{U(x)}{\gamma D_{eff}}-\frac{\tau_R}{2\gamma^2 D_{eff}} [U'(x)]^2\right).
\end{equation}
Most importantly, a joint position-velocity stationary distribution
\begin{equation} \label{eq:fullucna}
  p(x,v)=p_{ucna}(x) \sqrt{\frac{\Gamma(x) \tau_R}{2 \pi D_{eff}}}\exp\left( -\frac{\Gamma(x)\tau_R}{2 D_{eff}} v^2\right)
\end{equation}
is an approximate solution of the stationary Fokker-Planck
equation associated with Eq.~\eqref{eq:gcn3}. Among the properties of
the joint probability density function in Eq.~\eqref{eq:fullucna}, a
remarkable one is that it satisfies a coarse-grained form of detailed
balance: more precisely, the first three velocity moments of the
irreversible probability current are zero~\cite{umb17}.

The above considerations suggest an interesting interpretation of the
quantity $T_a(x)=D_{eff}/(\tau_R \Gamma(x)) \equiv T_{eff}/(\tau_R \gamma \Gamma(x))$ as the local 
temperature of the active bath. Two additional arguments support such
an interpretation:

1) an equation of state of the kind $p(x) \approx \rho(x) T_a(x)$
(with local density $\rho(x)=\int dv p(x,v)$) fairly explains
anomalous density fluctuations due to activity in the presence of
confining potentials~\cite{umb15a};

2) a generalised Clausius inequality can be demonstrated involving a
local heat flux normalised by $T_a(x)$~\cite{umb17}:
\begin{equation} \label{newcla}
\int dx \frac{\dot{q}(x)}{T_a(x)} \le 0,
\end{equation}
where $\dot{q}(x)$ represents the local density of heat flux from the system to the active bath:
\begin{equation} \label{tildeq}
  \dot{q}(x,t) = \frac{\Gamma(x)}{\tau_R}\left[T_a(x) \rho(x,t)-\int dv v^2 p(x,v,t)\right ].
\end{equation}
The interpretation of $T_a(x)$ as a temperature with some
thermodynamic meaning is still a conjecture. Nevertheless it has
interesting fascinating consequences, including a possible connection
with the issue of negative temperatures.  

Another open question strictly connected to the previous
considerations on heat fluxes is that of entropy production. Only few
studies have addressed such a problem, adopting - to represent
self-propulsion - the Gaussian colored noise
model~\cite{fodor16,umb17} or a velocity-dependent
force~\cite{chaudhuri14}. The latter approach has been seen to raise
ambiguities in the equivalence between zero entropy production and
equilibrium~\cite{CP15}.

\section{Conclusions}
\label{sec:concl}

In this review we have revisited the notion of temperature, focusing
on some difficult aspects that arise when one tries to extend such a
concept beyond the standard cases covered by thermodynamics and
equilibrium statistical mechanics.  The need for such a critical
analysis is motivated also by new experimental achievements, allowing
us to study the non-equilibrium statistical properties of ``small
systems'', at the micro and nano-scales, such as macromolecules (like
DNA, proteins and molecular motors), colloidal suspensions, granular
media and active matter, just to name a few examples.  Our main
purpose was to provide the reader with the basic aspects and the open
issues related to a mindful use of the concept of temperature, from
different perspectives.

We have followed a line of reasoning that, starting from the
two possible definitions of temperature (Boltzmann \emph{vs} Gibbs) in
the well-established framework of equilibrium statistical mechanics,
led us to clarify some debated points, such as the problem of
``temperature fluctuations'' and the existence of negative
temperatures, related to cases of systems with few degrees of freedom
or with a limited phase space, respectively.

Our journey, then,
crossed the realm of non-equilibrium systems, where a notion of
effective temperature can be introduced via the generalization of the
fluctuation-dissipation relation. We critically discussed its physical
meaning and usefulness in the description of non-equilibrium states,
focusing on some specific examples of non-Hamiltonian systems, such as
statistical hydrodynamics, granular media and active matter.

In particular, we have discussed how, in dilute granular fluids, an
effective temperature can be safely introduced, while, in dense
regimes, a good description cannot be obtained with a single
macroscopic parameter, due to the presence of not well-separated
time-scales in the system.  On the contrary, a statistical approach
based on few macroscopic parameters, directly inspired by the
equilibrium theory, seems to be effective for describing the slow
dynamics of very dense granular systems, as suggested by Edwards.  In
the context of active matter, we have focused our attention on a
specific model characterized by coloured noise, where a local
temperature of the active bath can be introduced, consistently with
the thermodynamic requirements.  Another rich field of research where
similar issues naturally arise, is represented by systems
characterized by aging dynamics, such as glassy and disordered models,
or colloidal suspensions, for reviews of such aspects see~\cite{CR03,cuglirev}.

In conclusion, some questions related to the notion of temperature, at
least in the context of equilibrium statistical mechanics, seem to
have been by now sufficiently clarified.  Others remain largely open,
in particular in the field of out of equilibrium and non-Hamiltonian systems. Here, the
search for general theories based on few macroscopic parameters, in
analogy with the equilibrium approach, is far from being
completed. Despite some important recent results, such as the
Fluctuation Relations or the extensions of the fluctuation-dissipation
theorem~\cite{ECM,GC,bsgj01}, general methods to treat non-equilibrium dynamics are
not available. From this perspective, the generalization of the notion
of temperature to characterize non-equilibrium behaviors seems to play
a central role. To pursue such an ambitious program, further efforts
have to be devoted, in particular, to the study of the effects of
typical non-equilibrium features, such as coarse-graining, memory
terms, or mixed time-scales in the system.

\section*{Acknowledgements}

We warmly thank Luca Cerino for many fruitful discussions and close
collaboration on the topics of this review. We are grateful to
M. Baldovin, M. Falcioni, L. Peliti and H. H. Rugh for remarks,
suggestions and correspondence. We also thank the Authors of
Refs.~\cite{Braun2013} and~\cite{sdh14} for allowing us to use their
figures.


\section*{References}

\bibliographystyle{elsarticle-num} 
\bibliography{biblio}

\end{document}